\documentclass[12pt,notitlepage]{revtex4-1}
\usepackage{amsmath}
\usepackage{amsfonts}
\usepackage{amssymb}
\usepackage{graphicx}
\usepackage{epstopdf}
\usepackage{bbm}

\usepackage{color}
\usepackage{xcolor}
\usepackage{dsfont}

\font\msytw=msbm9 scaled\magstep1 
scaled\magstep1

\let\a=\alpha \let\b=\beta  \let\g=\gamma  \let\d=\delta \let\e=\varepsilon
  \let\h=\eta     \let\l=\lambda
\let\m=\mu    \let\n=\nu         \let\p=\pi    \let\r=\rho
\let\s=\sigma \let\t=\tau    \let\c=\chi
   \let\o=\omega
 \let\D=\Delta  \let\L=\Lambda 
         
\let\O=\Omega 

\def\EE{{\cal E}} \def\VV{{\cal V}}
 \def\HHH{{\cal H}}\def\WW{{\cal W}}
\def\TT{{\cal T}}\def\NN{{\cal N}} \def\BBB{{\cal B}}
  
\def\DD{{\cal D}}

   \def\qq{{\bf q}}
  \def\PP{{\bf P}} \def\pp{{\bf p}}
 \def\xx{{\bf x}} \def\yy{{\bf y}} \def\zz{{\bf z}}

\def\kk{{\bf k}}

\def\RRR{\hbox{\msytw R}}

\def\\{\hfill\break}
\def\={:=}
\let\io=\infty
\let\0=\noindent
\def\media#1{{\langle#1\rangle}}
\let\dpr=\partial

\def\const{{\rm const}}
\def\tende#1{\,\vtop{\ialign{##\crcr\rightarrowfill\crcr\noalign{\kern-1pt
    \nointerlineskip} \hskip3.pt${\scriptstyle #1}$\hskip3.pt\crcr}}\,}
\def\otto{\,{\kern-1.truept\leftarrow\kern-5.truept\to\kern-1.truept}\,}

\def\to{\rightarrow}

\def\qed{\hfill\raise1pt\hbox{\vrule height5pt width5pt depth0pt}}

\def\V#1{{\bf#1}}
\def\be{\begin{equation}}
\def\ee{\end{equation}}
\def\bea{\begin{eqnarray}}
\def\eea{\end{eqnarray}}
\def\nn{\nonumber}

\def\lb{\label}

\def\Tr{\mathrm{Tr}}

\newtheorem{lemma}{Lemma}[section]
\newtheorem{prp}{Proposition}[section]
\newtheorem{theorem}{Theorem}[section]
\newtheorem{cor}{Corollary}[section]
\newtheorem{oss}{Remark}

\def\uu{{\bf u}}
\def\zz{{\bf z}}

\numberwithin{equation}{section}

\begin{document}

\title{Universality of the Hall conductivity\\ in interacting electron systems}

\author{Alessandro Giuliani}
\affiliation{University of Roma Tre, Department of Mathematics and Physics, L.go S. L. Murialdo 1, 00146 Roma, Italy}
\author{Vieri Mastropietro}
\affiliation{University of Milano, Department of Mathematics ``F. Enriquez'',\\
 Via C. Saldini 50, 20133 Milano, Italy}
\author{Marcello Porta}
\affiliation{University of Z\"urich, Mathematics Department, Winterthurerstrasse 190, 8057 Z\"urich, Switzerland}

\begin{abstract} We prove the quantization of the Hall 
conductivity for general weakly interacting gapped fermionic systems on 
two-dimensional periodic lattices. The proof is based on fermionic cluster expansion techniques combined with lattice Ward identities,
and on a reconstruction theorem that allows us to compute the Kubo conductivity as the 
analytic continuation of its imaginary time counterpart. 
\end{abstract}
 \maketitle

\section{Introduction}

Two-dimensional condensed matter systems often present remarkable transport properties. A famous 
example is the {\it Integer Quantum Hall Effect} (IQHE): the Hall conductivity of thin samples at very low 
temperatures, exposed to strong transverse magnetic fields, is equal to an integer times the von 
Klitzing constant $e^{2}/h$, \cite{Klitzing}. This measurement is amazingly sharp: the observation of the 
Hall plateaux is by now used to measure the fine structure constant, at a very high level of accuracy.
In view of the complexity of the underlying microscopic Hamiltonian, which depends on 
several material- and sample-dependent parameters,
the universality of the Hall conductivity is a very remarkable phenomenon.

The quantization of the Hall conductivity for non-interacting fermions 
has a deep topological interpretation \cite{AS2,TKNN}, and the intrinsic robustness of a topological 
quantity offers a natural qualitative explanation of the observed universality. 
The universality of the Hall conductivity in the presence of disorder has later been established in full 
mathematical rigor in \cite{Bellissard, AS2index, AG}. 

A similar universality property 
is expected to be valid in the presence of many-body interactions, as well.
However, while in the non-interacting case the features of the many-body problem can be deduced from the single-particle Schr\"odinger operator,
in the interacting case one needs to consider the 
full $N$-particle Schr\"odinger equation, which is much harder to study. 
This explains why a mathematical proof 
of the quantization of the Hall conductance for {\it interacting} electrons remained 
open \cite{OpenProblems} for many years. 

Effective field theories \cite{BFr,W, FrK, FZ, Z, FrS, FrST},
have been used for explaining a possible topological mechanisms 
underlying both the integral and the fractional QHE {in interacting electron systems}. 
However, they are 
based on certain phenomenological assumptions, such as the {\it incompressibility} of the ``quantum 
Hall fluid'', which may be very hard to check from first principles in concrete models.

More recently, the quantization of the Hall conductivity
has been rigorously proved \cite{HM}.
The proof of \cite{HM} is based on the hypothesis that the interacting ground state is non-degenerate and, as in the 
effective field theory approach, incompressible, which amounts to say that the interacting ground state is gapped, uniformly in the system size. This assumption is 
unproven in most physically relevant cases, {at least in the context of interacting fermions.}
As far as we know,  the only cases for which it is proved
are perturbations of  ``topologically trivial'' classical reference states 
\cite{DFF, DFFRB}, {or of ``frustration free'' Hamiltonians 
\cite{BHM,BH,MZ,BNY}, that is of Hamiltonians 
that can be written as sums of projectors geometrically localized around the sites of the underlying lattice. }

In this work, by using a different approach, we prove 
the quantization of the Hall conductivity for general interacting 
fermionic systems, under the assumptions that the reference non-interacting system is gapped,
and that the interaction is weak and short-ranged.  
In particular, our result applies to the interacting versions of the Hofstadter \cite{AEG,Hof}
and Haldane \cite{H1} models. See also \cite{R2} and \cite{R1} 
for numerical and experimental results on the interacting Haldane model. We stress that 
our proof does not require any a priori assumption on the interacting spectrum of the system. 
It is based on constructive cluster expansion  techniques combined with lattice Ward Identities. 
We write a {\it convergent} power series expansion for 
the conductivity, defined in terms of the Kubo formula, 
and we show that all the interaction-dependent corrections vanish exactly, in the 
infinite volume and zero temperature limits. 

The idea that the universality of Hall conductance follows from Ward Identities 
is well known \cite{IM,CH}. However, their implementation was so far limited 
to continuum effective quantum field theory models plagued by ultraviolet divergences,
and their use was combined with formal manipulations of 
non-convergent Feynman graph expansions.
Here, we consider lattice Hamiltonian models, and we develop a strategy similar to \cite{IM,CH},
based on {\it lattice} Ward Identities.
The convergence of the perturbative series is achieved by re-summing the usual Feynman diagram 
expansions in the form of  a suitable determinant expansion, which admits improved combinatorial 
estimates. Similar techniques, combined with an infrared Renormalization Group analysis, 
were used earlier for constructing the ground state of several low-dimensional interacting Fermi 
systems, and for proving universality relations among critical exponents, amplitudes and 
conductivities \cite{BGPS, BMdensity, BMchiral, BM, BFM0, BFM1, GMP1,GMP2,Mbook}. In this paper 
we apply these ideas for the first time to the study of the transverse (Hall) conductivity.

An informal statement of our main result is the following. 

\bigskip

{\it Consider a fermionic system on a two-dimensional periodic lattice, with grand canonical Hamiltonian $H_0+UV$, where $H_0$ is a quadratic gapped Hamiltonian, $V$ is a 
density-density interaction, decaying faster than any power at large distances, and $U$ is its strength. 
If $U$ is small enough, then the interacting correlation functions are analytic in $U$ and decay faster 
than any power at large distances, uniformly in the system size and in the temperature. The 
conductivity matrix, defined by the Kubo formula, 
is analytic as well, and its infinite volume and zero temperature limit is independent of $U$. In 
particular, the longitudinal conductivity is zero, while the transverse one is quantized.}

\bigskip

The rest of the paper is devoted to the proof of this result. In section \ref{sec:modelgen} we define 
the general class of Hamiltonians we consider, we define the 
current observable and the conductivity, and state our main theorem in a mathematically precise way. 
In section \ref{sec:3}, we introduce the imaginary-time counterpart of the conductivity and state 
a ``reconstruction theorem'' that guarantees its equivalence with the standard (real-time) Kubo conductivity. In section \ref{sec:proofuniv}, we prove the quantization of the conductivity, under the assumption of analyticity and smoothness of the multipoint current correlations at imaginary times. 
The key ingredient in the proof is the use of Ward Identites, 
which are nothing but the
restatement of the continuity equation for the {density} at the level of correlation functions. 
In section \ref{app:analyt} we prove the analyticity and smoothness of the imaginary-time/Matsubara frequency correlations, by using 
multiscale fermionic cluster expansion techniques. Strictly speaking, the content of section \ref{app:analyt} is a straightforward adaptation of previous results, but we include it here for the sake of 
self-containedness. In section \ref{sec:reconstr} we prove the reconstruction theorem
stated in section \ref{sec:3}, thus concluding the proof of our main result. 
In the appendices we collect some auxiliary results: in appendix \ref{app:kubo}
we reproduce the well known result that the non-interacting Kubo conductivity 
is equal to the Chern number of the filled Bloch bands; 
in appendix \ref{sec:haldane} we apply our main result to the interacting Haldane model, and show that it displays a non-trivial topological phase diagram;
in appendix \ref{app:LR} we discuss the existence of the infinite volume dynamics, required in the proof of the reconstruction theorem. 

\section{The model and the main result}\label{sec:modelgen}

In this section, we give a mathematically precise formulation of the class of models we consider.
First, we introduce the periodic lattice and the fermionic operators associated with its sites. Next,
we define the grand-canonical Hamiltonian and state our main assumptions on 
its quadratic and interaction parts, including the gap condition for the non-interacting theory. 
We proceed by  introducing the current and conductivity observables, and finally we state our main result. 

\subsection{Lattice fermionic operators} We let $\L=\big\{n_{1}\vec\ell_{1} + n_{2}\vec \ell_{2},\; n_{i} \in \mathbb{Z}\}$ be the Bravais lattice generated by the 
two linearly independent vectors $\vec \ell_{1},\, \vec \ell_{2} \in \mathbb{R}^{2}$. 
Given $L\in \mathbb{N}$, we also let $\L_{L}=\L/ L\L$ be the corresponding finite torus of side $L$, which can be thought of as the set 
\be
\L_L= \big\{ \vec x \mid \vec x = n_{1}\vec\ell_{1} + n_{2}\vec \ell_{2},\; n_{i} \in \mathbb{Z},\; 0\leq n_{i} < L \big\},\label{eq:2.1}
\ee
with periodic boundary conditions (i.e., endowed with the euclidean distance on the torus, denoted by $|\vec x-\vec y|_L=\min_{\vec n\in\mathbb Z^2}|\vec x-\vec y+n_1\vec \ell_1 L+n_2\vec\ell_2 L|$).
The number of sites of $\Lambda_{L}$ is denoted by $|\L_{L}| = L^{2}$. With each site $\vec x\in \L_L$, we associate 
fermionic creation and annihilation operators $\psi^{\pm}_{\vec x,\s}$, with $\s \in I$, and $I$ a finite set of 
indices, which can be thought of as ``color'' labels, possibly corresponding to the spin, 
or to different sublattices. In particular, the fermion labeled by $\s$ can be thought of as 
living on a physical lattice
obtained by translating $\L_L$ by a fixed amount $\vec r_\s\in\mathbb R$ (possibly equal to $\vec 0$, in the case that, e.g., $\s$ is a spin index). 

The fermionic operators satisfy the usual canonical anticommutation relations:
\be
\{ \psi^{\e}_{\vec x,\s}, \psi^{\e'}_{\vec y,\s'} \} = \delta_{\e,-\e'}\,\delta_{\vec x,\vec y}\,\delta_{\s,\s'}\;, 
\ee
where $\e,\e' = \pm$, $\vec x,\vec y\in \L_L$, $\s,\s' \in I$, and $\delta_{\cdot,\cdot}$ is the Kronecker delta. Consistently with the periodic boundary conditions,
we identify the fermionic operators obtained by translating $\vec x$ by an integer multiple of $L \vec \ell_i$. 
We let $\vec G_{1}$, $\vec G_{2}$ be a basis of the reciprocal lattice $\L_{L}^{*}$ of $\L$, i.e., 
$\vec G_i\cdot\vec \ell_j=2\pi \d_{i,j}$, and we define the finite-volume Brillouin zone as
\be
\mathcal{B}_{L} := \Big\{ \vec k \mid \vec k = \frac{n_{1}}{L} \vec G_{1} + \frac{n_{2}}{L}\vec G_{2},\; n_{i}\in \mathbb{Z},\; 0\leq n_{i}< L \Big\}\,.
\ee
We will also denote $\BBB=\BBB_\infty$. 
We let the Fourier transforms of the fermionic operators be:
\be
\psi^{\pm}_{\vec x,\s} = \frac{1}{ L^{2}}\sum_{\vec k\in \mathcal{B}_{L}} e^{\pm i \vec k\cdot \vec x} \hat \psi^{\pm}_{\vec k,\s}\;,\quad \forall \vec x\in \L_{L}\,,\qquad \Longleftrightarrow\qquad 
\hat \psi^{\pm}_{\vec k,\s} = \sum_{\vec x\in \L_{L}} e^{\mp i\vec k\cdot x} \psi_{\vec x,\s}^\pm\;,\quad 
\forall \vec k\in \mathcal{B}_{L}\;.
\ee
Note that, with this definition, the fermionic operators in momentum space are periodic over the first Brillouin zone, that is $\hat \psi^{\pm}_{\vec k,\s} = \hat \psi^{\pm}_{\vec k + \vec G_{i},\s}$, $i=1,2$.
Moreover, 
\be
\{\hat \psi^{\e}_{\vec k,\s}, \hat \psi^{\e'}_{\vec k',\s'}\} = L^{2}\delta_{\e,-\e'}\delta_{\vec k,\vec k'}\delta_{\s,\s'}\;.
\ee

\subsection{The Hamiltonian and the Gibbs state}\label{sec2.b} The grand-canonical Hamiltonian of the system is assumed to be of the form:
\be 
\mathcal{H}_{L}-\m \mathcal N_L = \mathcal{H}^{(0)}_{L} + U\mathcal{V}_{L}-\m \mathcal N_L\;,
\label{eq:2.6}\ee
with 
\bea && \mathcal{H}^{(0)}_{L} = \sum_{\vec x,\vec y\in \L_{L}}\sum_{\s,\s'\in I} \psi^{+}_{\vec x,\s} H^{(0)}_{\s\s'}(\vec x - \vec y) \psi^{-}_{\vec y,\s'}\;,\nn\\
&&\mathcal{V}_{L} = \sum_{\vec x,\vec y\in \L_{L}} \sum_{\s,\s'\in I} n_{\vec x}^{\s}\, v_{\s\s'}(\vec x - \vec y) \,n^{\s'}_{\vec y}\;,\quad {\rm where}\quad n_{\vec x}^{\s} = \psi^{+}_{\vec x,\s}\psi^{-}_{\vec x,\s}\;,
 \label{eq:defham}\\
&& {\rm and} \qquad \mathcal N_L=\sum_{\vec x\in \L_L}\sum_{\s\in I}n^\s_{\vec x}\;.\nn
\eea
The operator $\mathcal{H}^{(0)}_{L}$ is called the {\it free Hamiltonian}, while 
$U\mathcal{V}_{L}$  is the {\it many-body interaction}, and $U$ plays the role of the interaction strength. The constant $\m$ is the {\it chemical potential}, or Fermi level. 

We assume the {\it hopping function} $H^{(0)}_{\s\s'}(\vec x)=H^{(0)}_{L;\s\s'}(\vec x)$ to be a function on the torus (i.e., a periodic function on $\L_{L}$), 
such that $H^{(0)}_{\s\s'}(\vec x)=\sum_{\vec n\in\mathbb Z^2}H^{(0)}_{\infty;\s\s'}(\vec x+n_1\vec \ell_1 L+n_2\vec \ell_2 L)$, and 
$H^{(0)}_{\s\s}(\vec 0)=0$. In order for the free Hamiltonian to be self-adjoint, 
we require $\big[H^{(0)}_{\s\s'}(\vec x)\big]^* = H^{(0)}_{\s'\s}(-\vec x)$. Moreover, we assume that $H^{(0)}_{\infty;\s\s'}$ 
decays faster than any power at large distances, so that: 
\be \|H^{(0)}(\vec x)\|\le \frac{C_N}{1+|\vec x|_L^{N}}\,,\quad \forall N\ge 0\;.\label{eq:2.8}\ee
As a consequence of these assumptions, we see that the
{\it Bloch Hamiltonian}
\be
\hat H^{(0)}(\vec k) := \sum_{\vec x\in \L_{L}} e^{i\vec k\cdot \vec x} H^{(0)}(\vec x)\;,
\ee
is a self-adjoint matrix, so that the spectrum 
$\s(\hat H^{(0)}(\vec k)) = \{\e_{\sigma}(\vec k)\}_{\sigma\in I}$ is real. The functions $\vec k \mapsto 
\e_{\s}(\vec k)$ are called the {\it energy bands}. We let 
\be\label{eq:h0}\mathfrak e_0=\sup_{\vec k\in\BBB}||\hat H^{(0)}(\vec k)||,\ee which sets the 
energy scale. Note also that the infinite volume limit of $\hat H^{(0)}(\vec k)$ is infinitely differentiable 
in $\vec k$.

Concerning the interaction, we assume, similarly, that 
$v_{\s\s'}(\vec x)$ is periodic on $\L_{L}$, equal to the sum over the images of its infinite volume limit, such that
$v_{\s\s}(\vec 0)=0$, $v_{\s\s'}(\vec x-\vec y)=v_{\s'\s}(\vec y-\vec x)$ and 
\be \|v(\vec x)\|\le \frac{C_N}{1+|\vec x|_L^{N}}\,,\quad \forall N\ge 0\;.\label{eq:intdef}
\ee
In particular, the infinite volume limit of \be
\hat v_{\s\s'}(\vec p) =\sum_{\vec x\in \L_{L}} e^{i\vec p\cdot \vec x} v_{\s\s'}(\vec x)
\ee
is infinitely differentiable 
in $\vec p$.

Finally, concerning the choice of the Fermi level, we assume the following {\it gap condition}:
\be \d_{\m}= \inf_{\vec k\in \mathcal{B}}\text{dist}(\mu, \s(\hat H^{(0)}(\vec k)))>0\;.
\label{gap}\ee
One of the important implications of the gap condition is that the projector over the ```filled bands'',
i.e., over the bands with energy smaller than $\mu$, is smooth in $\vec k$: more precisely, the operator
$P_-(\vec k)=\sum_{\a:\ \e_\a(\vec k)<\mu}P_\a(\vec k)$, with $P_\a(\vec k)$ the projector over the $\a$-th energy band, is a projector itself and, in the infinite volume limit, it is infinitely differentiable in $\vec k$. 

\medskip

{\bf Remark.} The assumptions that $H^{(0)}(\vec x)$ and $v(\vec x)$ 
decay faster than any power are 
far from being optimal. It is easy to adapt the following discussion to the case of sufficiently fast 
power-law decay of the hopping matrix and of the interaction. Since we are not interested in optimal 
bounds, for simplicity we illustrate our method in the case of faster-than-any-power decay. 
Note also that, if the hopping matrix decays exponentially fast at large distances, 
then the Bloch Hamiltonian and the projector $P_-(\vec k)$ are analytic in $\vec k$, rather than just 
infinitely differentiable.

\bigskip


In the following, we construct the 
grand-canonical {\it Gibbs state} associated with \eqref{eq:2.6}, which is characterized by its correlation
functions, defined as follows. 
Given an {\it observable} $O$, that is a self-adjoint operator on the fermionic Fock space $\mathcal{F}$, 
its expectation value is:
\begin{equation}\label{eq:gibbs}
\langle O \rangle_{\beta,\mu,L} :=\Tr_{\mathcal{F}}\, \r_{\b,\m,L}\,O\;,\qquad 
\r_{\b,\m,L}=
 \frac{e^{-\beta (\mathcal{H}_{L} - \mu \mathcal{N}_{L})} }{\Tr_{\mathcal{F}}\, e^{-\beta (\mathcal{H}_{L} - \mu\mathcal{N}_{L})}}\;,
\end{equation}
where $\mathcal{F}$ is the fermionic Fock space. The chemical potential $\mu$ should be thought 
of as being fixed once and for all, so that \eqref{gap} is verified. Therefore, for notational convenience, 
we shall drop the label $\mu$ from the symbol for the Gibbs state and for the density matrix: $\langle \cdot \rangle_{\beta,\m,L} \equiv \langle \cdot \rangle_{\beta,L}$ and $\r_{\b,\m,L}\equiv \r_{\b,L}$.

\subsection{The current and the conductivity}\label{sec:linear}

Let us preliminarily define the current observable on the infinite lattice $\L$. 
Given $\vec x\in \L$, we let $\vec x_{\s} := \vec x + \vec r_\s$ be the location of the fermion labeled 
by $\s\in I$, see the discussion following \eqref{eq:2.1}. The total position operator is defined as 
$\vec X=\sum_{\s\in I}\sum_{\vec x\in \L} \vec x_\s n^{\s}_{\vec x}$, while the d.c. current is
\be\label{eq:currdefgen}
\vec J := i\big[\mathcal{H}, \vec X\big]\;.\ee
where $\mathcal H$ is the formal infinite volume limit of $\mathcal H_L$.
Note that $\vec J = i\big[\mathcal{H}^{(0)}, \vec X\big]$, because $\mathcal{V}$ commutes with 
$\vec X$. Moreover, $\vec X$ can be naturally decomposed as $\vec X=\vec X^{(1)}+\vec X^{(2)}$,
where $\vec X^{(1)}=\sum_{\vec x,\s}\vec x\, n^\s_{\vec x}$ represents the position of the ``centers''
of the cells of the Bravais lattice, and $\vec X^{(2)}$ the displacement with respect to the centers. 
The current $\vec J$ inherits this decomposition. As we shall see below, the ``displacement current'' 
$\vec J^{(2)}$ does not contribute to the conductivity, in the infinite volume and zero temperature limits. 

By using the definition of $\mathcal H$, we can rewrite $\vec J$ more explicitly as
\be\label{eq:currbond}
\vec J = \frac12\sum_{\vec x,\vec y\in \L} \sum_{\s,\s' \in I} (\vec y_{\s'} - \vec x_\s) J_{\vec x\,\vec y}^{\s\s'}\;,\ee
where 
\be\label{eq:defbondcurr}
J_{\vec x\, \vec y}^{\s\s'}= i\big[\psi^{+}_{\vec x,\s} H^{(0)}_{\s\s'}(\vec x-\vec y) \psi^{-}_{\vec y, \s'} - \psi^{+}_{\vec y,\s'} H^{(0)}_{\s'\s}(\vec y-\vec x) \psi^{-}_{\vec x, \s}\big]
\ee
is the {\it bond current} flowing from $\vec x_\s$ to $\vec y_{\s'}$. 
The bond current satisfies a natural {\it continuity equation}, which is reviewed in the next section. The finite volume current (to be still denoted by $\vec J$ in the following, with some abuse of notation) is defined by an expression analogous to \eqref{eq:currbond}, with the sums over 
$\vec x$ and $\vec y$ restricted to $\L_L$, and the vector $(\vec y_{\s'}-\vec x_\s)$
to be interpreted as $(\vec y-\vec x)_L+\vec r_{\s'}-\vec r_\s$, where, if $\vec y-\vec x=(n_1\vec \ell_1,n_2\vec\ell_2)$, then $(\vec y-\vec x)_L=(\{n_1\}_L\vec \ell_1,\{n_2\}_L\vec \ell_2)$, with $\{n\}_L=n-L\lfloor\frac{n}L+\frac12\rfloor$. 

\bigskip

The conductivity matrix in the infinite volume and zero temperature limits 
is defined via the Kubo formula \cite{Ku57} 
as:
\be \s_{ij}(U) =\frac1A \lim_{\omega\to 0^{+}} \frac{1}{\o }\Big( i\int_{-\infty}^{0} dt\, e^{\o t}\, \langle\!\!
\langle \big[e^{i\mathcal{H}t} J_{i}e^{-i\mathcal{H}t}, J_{j}\big] \rangle\!\!\rangle_{\infty} -\langle
\!\!\langle\big[[\mathcal H,X_i],X_j\big]\rangle\!\!\rangle_\infty\Big)\;,\label{eq:kubo0}\ee
where $A=|\vec\ell_{1} \wedge \vec\ell_{2}|$ is the area of the fundamental cell, and 
$ \langle\!\!\langle O
\rangle\!\!\rangle_{\infty}=\lim_{\b\to\infty}\lim_{L\to\infty}L^{-2} \langle O_L
\rangle_{\b,L}$. The second term in parentheses is known as the diamagnetic term, or Schwinger term.
This formula describes the response of the system at $t=0$ after adiabatically switching on at $t=-\infty$ a time-dependent external field, whose amplitude is damped by a factor $e^{\o t}$, see \cite[Eq.(A.7)]{AG}.

\subsection{The main result}\label{sec:main}

We are finally in the position of stating our main result in a mathematically precise way. 

\begin{theorem}{{\bf [Universality of the conductivity matrix]}}\label{thm:main}
Let $\s_{ij}(U)$ be the conductivity matrix of the model with Hamiltonian \eqref{eq:2.6}, 
as defined in \eqref{eq:kubo0}. 
Under the assumptions on the Hamiltonian spelled out after 
\eqref{eq:defham} (see \eqref{eq:2.8}, \eqref{eq:intdef} and \eqref{gap}), 
there exists $U_0>0$ such that
\be
\sigma_{ij}(U) = \sigma_{ij}(0)\;,\qquad \forall i,j = 1,2\;,
\ee
as long as $U\in(-U_0,U_0)$.
\end{theorem}

{\bf Remarks.}
\begin{itemize}
\item The non-interacting conductivity $\s_{ij}(0)$ is well-known to be quantized \cite{AG,AS2,TKNN}.
The proof that the non-interacting conductivity is equal to the first Chern number is reproduced for 
completeness in appendix \ref{app:kubo}.
Therefore, theorem \ref{thm:main} tells us that, if $|U|\le U_0$,
$\s_{ij}(U)\in \mathbb Z/(2\pi)$, which is the usual quantization formula of the Hall 
conductivity, in units where $e=\hslash=1$. 
Moreover, the computation in appendix \ref{app:kubo} shows that $\vec J^{(2)}$,
which was defined after \eqref{eq:currdefgen}, gives a vanishing contribution to $\s_{ij}(0)$ and, 
therefore, in light of theorem \ref{thm:main}, it gives vanishing contribution to the interacting 
conductivity, as well. 
\item The infinite volume and zero temperature current correlations entering the definition of conductivity are defined by first sending the volume to infinity and then 
the temperature to zero. However, in the situation we are considering, the two limits can be interchanged, thanks to the gap condition. 
\end{itemize}

\section{Imaginary times and Matsubara frequencies}\label{sec:3}

In this section we introduce the notion of Euclidean (imaginary-time) correlation functions, which are 
the most natural class of correlations that can be studied by the many-body thermodynamic formalism.
We also introduce their Fourier transform, which are known as the correlations 
at imaginary (or ``Matsubara'') frequencies. Next, we define the imaginary time/Matsubara 
frequency counterpart of the Kubo conductivity, and state a ``reconstruction theorem''
about the equivalence between the two definitions of conductivity, to be proved in section \ref{sec:reconstr}. 

\subsection{Euclidean correlation functions}

Given an observable $O$, we let 
 $O_{t}$ with $t\in[0,\b)$ be its imaginary time evolution,
 namely
\begin{equation}\label{eq:17}
O_{t} := e^{t(\mathcal{H}_{L} - \mu \mathcal{N}_{L})} O e^{-t (\mathcal{H}_{L} - \mu\mathcal{N}_{L})}\;.
\end{equation}
Given $n$ observables $O_{t_{1}}^{(1)},\ldots, O_{t_{n}}^{(n)}$, each of which can be written as
a polynomial in the time-evolved  creation and annihilation operators $\psi^\pm_{(t,\vec x),\s}=
e^{t(\mathcal{H}_{L} - \mu \mathcal{N}_{L})} \psi^\pm_{\vec x,\s}e^{-t(\mathcal{H}_{L} - \mu \mathcal{N}_{L})} $,
we define their time-ordered average as:
\begin{equation}\label{eq:18}
\langle {\bf T}\, O^{(1)}_{t_1}\cdots O^{(n)}_{t_{n}} \rangle_{\beta,L} := \frac{\Tr_{\mathcal{F}}\, e^{-\beta (\mathcal{H}_{L} - \mu\mathcal{N}_{L})} \mathbf{T} \big\{ O_{t_{1}}^{(1)}\cdots O_{t_{n}}^{(n)} \big\}  }{\Tr_{\mathcal{F}}\, e^{-\beta (\mathcal{H}_{L} - \mu\mathcal{N}_{L})}}\;,
\end{equation}
where the (linear) operator $\mathbf{T}$ is the fermionic time-ordering, acting
on a product of fermionic operators as:
\begin{equation}
\mathbf{T} \big\{ \psi^{\e_{1}}_{(t_1,\vec x_1),\s_1}\cdots \psi^{\e_{n}}_{(t_n,\vec x_n),\s_n} \big\} = \text{sgn}(\pi) \psi^{\e_{\pi(1)}}_{(t_{\p(1)},\vec x_{\p(1)}),\s_{\p(1)}}\cdots \psi^{\e_{\pi(n)}}_{
(t_{\p(n)},\vec x_{\p(n)}),\s_{\p(n)}}
\;,
\end{equation}
where $\pi$ is a permutation of $\{1,\ldots, n\}$ with signature $\text{sgn}(\pi)$
such that $t_{\pi(1)}\geq \ldots \geq t_{\pi(n)}$. If some operators are evaluated at the same time, 
the ambiguity is solved by normal ordering. 

Moreover, we denote by $\langle {\bf T}\, O^{(1)}_{t_{1}}\,;\, \cdots \,;\, O^{(n)}_{t_{n}}\rangle_{\beta,L}$ the time-ordered {\it truncated} correlation function, or {\it cumulant}, of $O_{t_{1}}^{(1)},\ldots, O_{t_{n}}^{(n)}$. If the observables are all even, i.e., if they are linear combinations of even monomials  in the creation and annihiliation operators, the cumulant is defined as follows \cite{Araki}:
\be\label{eq:defcumulant}
\langle {\bf T}\, O^{(1)}_{t_1}\,;\, O^{(2)}_{t_2}\,;\, \cdots \,;\, O^{(n)}_{t_n} \rangle := \frac{\partial^n}{\partial \l_{1}\cdots \partial \l_{n}} \log \Big\{ 1 + \sum_{I\subseteq \{1,2, \ldots, n\}} \l(I) \langle {\bf T}\, O(I) \rangle \Big\}\Big|_{{\boldsymbol{\l}}=\V0}\;,
\ee
where $\l(I) = \prod_{i\in I}\l_i$ and $O(I) = \prod_{i\in I}O^{(i)}_{t_i}$.
For $n=1$, this definition reduces to $\langle O^{(1)}_{t_1} \rangle=\media{O^{(1)}}$. For $n=2$ one gets $\langle {\bf T}\, O^{(1)}_{t_{1}}\,;\, O^{(2)}_{t_{2}} \rangle = \langle {\bf T}\, O^{(1)}_{t_{1}}O^{(2)}_{t_{2}} \rangle - \langle O^{(1)}_{t_1} \rangle\langle O^{(2)}_{t_{2}} \rangle$, and so on. A similar definition of time-ordered truncated expectation is valid in the case that $O^{(i)}_{t_i}$
is replaced by an operator depending on multiple times, e.g., by $O^{(i)}_{t_i}\tilde O^{(i)}_{t_i'}$, with $\tilde O^{(i)}$ another even observable. 

We also introduce the  notion of Fourier transform with respect to the 
imaginary time. Consider, again, the case of $n$ even observables $O_{t_1}^{(1)},\ldots, O^{(n)}_{t_n}$, with $t_i\in[0,\b)$. We denote by $\widehat{O}_{\o_i}^{(i)} := \int_{0}^{\beta} d t\, e^{-i
\o_i t}\, O^{(i)}_{t}$ their Fourier transforms, where $\o_i \in \frac{2\pi}{\beta}\mathbb{Z}$ are 
called {\it Matsubara frequencies}. By using the definition of time-ordered correlations 
and the cyclicity of the trace, it is straightforward to 
check that 
\bea\label{eq:corrFdef}
&&\int_0^\b dt_1\cdots\int_0^\b dt_n\, e^{-i\o_1 t_1\cdots-i\o_nt_n} \langle {\bf T}\, O^{(1)}_{t_1}\,;\, \cdots \,;\, O^{(n)}_{t_n}\rangle_{\beta,L}= \nn\\
&&\qquad\qquad  = \delta_{\o_1+\cdots+\o_n,0}\langle {\bf T}\, \widehat{O}^{(1)}_{\o_1}\,;\, \cdots \,;\, \widehat{O}^{(n-1)}_{\o_{n-1}}\,;\, \widehat{O}^{(n)}_{-(\o_1+\cdots+\o_{n-1})}\rangle_{\beta,L}\;,
\eea
which sets our convention on the Fourier transform of the truncated correlations. 

\subsection{The continuity equation}

As anticipated in section \ref{sec:linear}, the (imaginary-time) evolution of the bond current satisfies
a natural continuity equation, which reads as follows: 
 if $n^\s_{(t,\vec x)}$
and $\big(J^{\s\s'}_{\vec x\,\vec y}\big)_t$ are the imaginary time evolutions of $n^\s_{\vec x}$ and of $J^{\s\s'}_{\vec x\,\vec y}$, respectively, then 
\be\label{eq:cont} \partial_t n^\s_{(t,\vec x)}=i\sum_{\vec y\in \L_L}\sum_{\s'\in I}
\big(J_{\vec x\,\vec y}^{\s\s'}\big)_t\;.\ee
The continuity equations \eqref{eq:cont} for different values of $\s$ can be conveniently 
combined in a single equation, by letting 
\be\label{eq:J0}
\tilde J_{0,(t,\vec p)} := \sum_{\vec x\in \L_{L}}\sum_{\s\in I} e^{-i\vec p\cdot \vec x_\s} n^{\s}_{(t,\vec x)}\;, 
\ee
with $\vec p\in \{ \vec k :  \vec k = \frac{n_{1}}{L} \vec G_{1} + \frac{n_{2}}{L}\vec G_{2},\; n_{i}\in \mathbb{Z} \}$. 
The observable $\tilde J_{0,(t,\vec p)}$ 
satisfies 
\be \partial_t\tilde J_{0,(t,\vec p)}+ \vec p\cdot \vec{\tilde{ J}}_{(t,\vec p)}=0\;,
\label{eq:contcompact}\ee
where 
\be \vec{\tilde{ J}}_{(t,\vec p)}=
\frac12\sum_{\vec x,\vec y\in\L_L} \sum_{\s,\s'\in I}
e^{-i\vec p\cdot\vec x_\s}(\vec y_{\s'}-\vec x_\s)\eta_{\vec x\,\vec y}^{\s\s'}(\vec p)
\big(J^{\s\s'}_{\vec x\,\vec y}\big)_t
\label{eq:Ji}\ee
and 
\be\label{eq:etap}
\eta_{\vec x\,\vec y}^{\s\s'}(\vec p)=\frac{1-e^{-i\vec p\cdot (\vec y_{\s'}-\vec x_\s)}}
{i\vec p\cdot (\vec y_{\s'}-\vec x_\s)}
\;,\ee
with the understanding that $\eta_{\vec x\,\vec y}^{\s\s'}(\vec 0)=1$, if $\vec y_{\s'}\neq \vec x_\s$,
and $\eta_{\vec x\,\vec y}^{\s\s'}(\vec p)=0$, if $\vec y_{\s'}=\vec x_\s$. 
Note that, in general, $\tilde J_{0,(t,\vec p)}$ and $\vec{\tilde J}_{(t,\vec p)}$ are not periodic 
in $\vec p$ over the Brillouin zone.

\subsection{The conductivity at imaginary frequency. A reconstruction theorem}

The natural counterpart of the Kubo conductivity \eqref{eq:kubo0} at imaginary time/Matsubara frequency is defined as follows (cf., e.g., \cite[Eqs.(3.388) to (3.391)]{Mah} and \cite[Eqs.(6) to (10)]{SPG}):
\be\label{eq:kubo}
\bar\sigma_{ij}(U):= -\lim_{\omega \to 0^{+}} \frac{1}{A}\frac{1}{\o}\big[ \widehat K_{ij}(\o,\vec 0) - \widehat K_{ij}(0,\vec 0) \big]\;,
\ee
where 
\be \widehat K_{ij}(\o,\vec p)=\lim_{\b\to\infty}\lim_{L\to\infty}\frac1{\b L^2}
\media{{\bf T} \hat J_{i,(\omega,\vec p)};\hat J_{j,(-\omega,-\vec p)}}_{\b,L}\;,
\label{2.29}\ee
and $\hat J_{i,(\o,\vec p)}=\int_0^\b dt\, e^{-i\o t}\tilde J_{i,(t,\vec p)}$
is the Fourier transform of the current-current correlation. Note that the labels $i,j\in\{1,2\}$ in the previous two equations refer to the basis $\hat e_{1} = (1,\, 0)$, $\hat e_{2} = (0,\, 1)$. 
Now, {\it if} the infinite-volume current-current correlation function $\widehat K_{ij}(\o,\vec 0)$ is differentiable at $\o = 0$, then \eqref{eq:kubo} reduces to:
\be\label{eq:2.30}
\bar\sigma_{ij}(U) = -\frac{1}{A}\frac{\partial\widehat K_{ij}}{\partial \o} (0,\vec 0)\;,\qquad i=1,2\;.
\ee
This is in fact the case for the class of systems we are considering, as we shall prove
on the basis of fermionic cluster expansion methods, by taking advantage of the 
gap condition \eqref{gap} on the non-interacting spectrum. See proposition \ref{prp:analyticity} below.

Remarkably, for the class of gapped systems we are considering, the Kubo conductivity at imaginary
frequency, in the limit of zero frequency, is the {\it same} as its real-time counterpart $\s_{ij}(U)$ 
defined in \eqref{eq:kubo0}. This is summarized in the following proposition. 

\begin{theorem}{{\bf [Reconstruction of the real-time Kubo formula]}}\label{thm:realkubo} 
Under the same assumptions as theorem \ref{thm:main}, 
\be \bar\s_{ij}(U)=\s_{ij}(U)\;,\ee
for all $U\in(-U_0,U_0)$.
\end{theorem}

This theorem will proved in section \ref{sec:reconstr}. It allows us to study the Kubo conductivity via its 
imaginary time counterpart, which is more directly accessible to constructive many-body techniques,
to be described in the following two sections. 

\section{Universality and Ward Identities}
\label{sec:proofuniv}

In this section, we prove theorem \ref{thm:main}, by combining the use of exact lattice Ward Identities
with the information that the multipoint density and current correlations are analytic in $U$ and smooth in the momenta. More in detail, we first introduce the definition of multipoint density and current correlations, and state a result, to be proved in section \ref{app:analyt}, concerning the regularity of these correlations. Next, we introduce the notion of Ward Identities, and prove an important consequence thereof, in the form of an identity relating certain correlations to the derivatives of other correlation functions. We then proceed to prove the so-called Schwinger-Dyson equation for the 
correlation functions. Finally, we put together all these ingredients and 
prove that $\bar\s_{ij}(U)=\bar\s_{ij}(0)$, for $U$ small enough. In light of theorem \ref{thm:realkubo},
this implies theorem \ref{thm:main} for $\s_{ij}(U)$.

\subsection{Multipoint density and current correlation functions}\label{sec:correlations}

For $n\ge 2$, we let $\pp_{i} = (\o_i, \vec p_{i}) \in (2\pi/\beta)\mathbb{Z}\times \mathcal{D}_{L}$
and $\a_i\in \{0,1,2\}\cup I$, with $\DD_L= \{ \vec k :  \vec k = \frac{n_{1}}{L} \vec G_{1} + \frac{n_{2}}{L}\vec G_{2},\; n_{i}\in \mathbb{Z} \}$ and
$i=1,\ldots, n-1$, and we define:
\be\label{eq:Kdef1}
\widehat K^{\beta, L}_{\a_{1},\ldots, \a_n}(\pp_{1},\ldots, \pp_{n-1}):=\frac{1}{\beta  L^{2}}
 \langle {\bf T}\, \hat J_{\a_{1},\pp_{1}}\,;\, \cdots \,;\, \hat J_{\a_{n-1},\pp_{n-1}}\,;\, \hat J_{\a_n,-\pp_{1} - \ldots - \pp_{n-1}} \rangle_{\beta,L}\;,
\ee
where, if $\a=\m\in\{0,1,2\}$, then $\hat J_{\m,\pp} = \int_{0}^{\beta} dt\, e^{-i\o t} \tilde J_{\m,(t,\vec p)}$,
with $\tilde J_{\m,(t,\vec p)}$ as in \eqref{eq:J0} and \eqref{eq:Ji}, while, if
$\a=\s\in I$, then $\hat J_{\s,\pp}=\hat n_{\pp}^{\s} = \int_{0}^{\beta} dt\, e^{-i\o t}\, 
\tilde n_{(t,\vec p)}^{\s}$, in which case the vector $\vec p\in \DD_L$ should be identified with its image in $\BBB_L$ modulo vectors in $\L_L^*$.
If $n=1$, we introduce the one-point correlation as
\be\label{eq:Kdef_1}
\widehat K^{\beta, L}_{\a}:=\frac{1}{\beta  L^{2}}
 \langle \hat J_{\a,\V0}\rangle_{\beta,L}\;.
\ee
Moreover, for $\pp_{i} \in \mathbb{R}^2$, we let 
\be\label{eq:Kdef2}
\widehat K_{\a_{1},\ldots, \a_{n}}(\pp_{1},\ldots, \pp_{n-1})  := \lim_{\beta\to\infty}\lim_{L\to\infty} 
\widehat K^{\beta, L}_{\a_{1},\ldots, \a_{n}}(\pp_{1},\ldots, \pp_{n-1})\;.
\ee
Note that these correlations are invariant under the exchange of the indices $(\a_i,\pp_i)$ 
with $(\a_j,\pp_j)$ (and, if either $i$ or $j$ are equal to $n$, $\pp_n$ should be interpreted as 
$-\pp_2-\cdots-\pp_{n-1}$).

A crucial fact for the following is that the infinite volume and zero temperature correlations functions are analytic in $U$ and infinitely differentiable in the momenta, as summarized in the following proposition. 
\begin{prp}{{\bf [Existence and regularity of the interacting correlations]}}\label{prp:analyticity}
There exists $U_{0}>0$, independent of $\beta$ and $L$, such that, for $|U|<U_{0}$
and $\b,L$ sufficiently large, 
the correlations $\widehat K^{\beta, L}_{\a_{1},\ldots, \a_{n}}
(\pp_{1},\ldots, \pp_{n-1})$ are analytic in $U$, uniformly in $\b,L$ and in their arguments. 
Moreover, the infinite volume and zero temperature limits of the correlations in \eqref{eq:Kdef2} exist, 
and define a sequence of functions $\widehat K_{\a_{1},\ldots, \a_{n}}
(\pp_{1},\ldots, \pp_{n-1})$ that are analytic in $U$ in $|U|\le U_0$, and are $C^{\infty}$ in their arguments. 
\end{prp}

The proof is given in section \ref{app:analyt}. The $C^\io$ regularity of 
$\widehat K_{\a_{1},\ldots, \a_{n}}(\pp_{1},\ldots, \pp_{n-1})$ 
could be improved to analyticity, in the case that both $H^{(0)}(\vec x)$ and $v(\vec x)$ decay 
exponentially at large distances. 

\subsection{Ward Identities} The components of $\hat J_{\a,\pp}$ with $\a=\m\in\{0,1,2\}$ are related 
among each other by the continuity equation \eqref{eq:contcompact}, which is an exact 
identity at finite volume and temperature. If plugged into the definition of correlation functions, 
this equation implies exact relations among the correlations, known as Ward Identities. 

\begin{prp}{{\bf [Ward Identities]}}\label{prop:WI}
Under the same hypotheses as proposition \ref{prp:analyticity}, if $n\ge 2$ and $\pp_i=(\o_i,\vec p_i)\in(2\p/\b)\mathbb Z\times\DD_L$, $\forall i=1,\ldots, n-1$, the following identity holds:
\be\label{eq:WI1prop}
\sum_{\m=0}^2 (i)^{\d_{\m,0}}(\pp_1)_\m \widehat K^{\beta,L}_{\m,\a_2,\ldots,\a_n}(\pp_1,\pp_2, \ldots, \pp_{n-1}) = \sum_{j=2}^n \widehat S^{\beta,L}_{\a_j;\widehat{\underline\a}_j}
(\pp_{1}, \ldots, \pp_{n-1})
\ee
where $\widehat{\underline\a}_j$ denotes the sequence $\underline\a=(\a_2,\ldots,\a_n)$ with the
element $\a_j$ removed, and
\bea &&\label{eq:D}
\widehat S^{\beta,L}_{\a_j;\widehat{\underline\a}_j}(\pp_{1}, \pp_{2}, \ldots, \pp_{n-1}) =\\
&&\quad =\frac{1}{\b L^{2}}\big\langle {\bf T}\,\hat \D_{\a_j}(\pp_1,\pp_j) \, ; \, \hat J_{\a_2,\pp_2}\, ; \, \cdots \,;
 \, \hat{J}_{\a_{j-1},\pp_{j-1}}\, ;\, \hat{J}_{\a_{j+1},\pp_{j+1}}\, ;\,\cdots\,; \, \hat J_{\a_n,-\pp_1\cdots-\pp_{n-1}}\big\rangle_{\beta,L}\;.\nonumber
\eea
with 
\be \hat\D_{\a}(\pp_1,\pp_2)=\int_0^\b dt\, e^{-it(\o_1+\o_2)}
\big[ \tilde J_{0, \vec p_{1}}, \tilde J_{\a,\vec p_{2}} \big]_t\,.\label{eq:3.6}\ee 
The identity \eqref{eq:WI1prop} is also valid for the infinite volume and zero temperature limits 
of the correlation functions.
\end{prp}

{\bf Remarks.}
\begin{itemize}
\item In the right side of \eqref{eq:D}, if $j=n$, then the vector $\pp_n$ in  the 
argument of $\hat\D_{\a_n}$ should be interpreted as $-\pp_1-\cdots-\pp_{n-1}$. In \eqref{eq:3.6}, 
$\tilde J_{\a,\vec p}=\tilde J_{\a,(t,\vec p)}\big|_{t=0}$, and $\big[ \tilde J_{0, \vec p_{1}}, \tilde J_{\a,\vec p_{2}} \big]_t$ denotes the imaginary time evolution of $\big[ \tilde J_{0, \vec p_{1}}, \tilde J_{\a,\vec p_{2}} \big]$.
\item The term in the right side of (\ref{eq:WI1prop}) is called the 
{\it Schwinger term}. It is, of course, absent if the observables $\tilde J_{0,\vec p_1}$ and $\tilde J_{\a,\vec p_2}$ commute, i.e., for $\a=0$ or $\a\in I$. 
\end{itemize}

\medskip

{\bf Proof of proposition \ref{prop:WI}.}
By integrating by parts with respect to $t_1$, we find (noting that the boundary terms cancel)
\be\label{eq:proofWI1}
i\o_1 \widehat K^{\beta,L}_{0,\underline\a}(\pp_{1}, \ldots, \pp_{n-1}) = 
\frac{1}{\beta L^{2}} \int_{0}^{\beta} dt_1\, e^{-i\o_1 t_1} \partial_{t_{1}}\big\langle {\bf T}\, \tilde J_{0,(t_1,\,\vec p_{1})}\,;\, \hat J_{\a_2,\pp_{2}} \,;\, \cdots \,;\, \hat J_{\a_n,\pp_{n}} \big\rangle_{\beta,L}\;,
\ee
where $\pp_n=-\pp_1-\cdots-\pp_{n-1}$.
Recalling the definition of time-ordered correlation function, we see that the derivative with respect to 
$t_1$ can act either on the observable $\tilde J_{0,(t_1,\,\vec p_{1})}$, in which case we can apply 
the continuity equation \eqref{eq:contcompact}, or on the characteristic functions 
entering the definition of time-ordering. Therefore, \eqref{eq:proofWI1} can be rewritten as
\bea%
&&i\o_1 \widehat K^{\beta,L}_{0,\underline\a}(\pp_{1}, \ldots, \pp_{n-1}) = 
-\frac1{\b L^2}\vec p_1\cdot \big\langle {\bf T}\, \vec{\hat J}_{\pp_1}\,;\, \hat J_{\a_2,\pp_{2}} \,;\, \cdots \,;\, \hat J_{\a_n,\pp_{n}}\big\rangle_{\beta,L}+\frac1{\b L^2}\int_0^\b dt_1\sum_{j=2}^n\times\nn\\
&& \times e^{-i(\o_1+\o_j)t_1}\big\langle {\bf T}\, \big[\tilde J_{0,\vec p_{1}}\,;\, \hat J_{\a_j,\vec p_{j}}\big]_{t_1} \,;\, 
\hat J_{\a_2,\pp_2};\cdots;\hat J_{\a_{j-1},\pp_{j-1}};\hat J_{\a_{j+1},\pp_{j+1}};\cdots;\hat J_{\a_n,\pp_n}\big\rangle_{\beta,L}\,\label{eq:WIderivative} \eea
where, for any collection of even observables $O^{(1)},  \ldots, O^{(n)}$, 
\bea \big\langle {\bf T}\, \big[O^{(1)};O^{(2)}\big]_{t_1};O^{(3)}_{t_3};\cdots;O^{(n)}_{t_n}\big\rangle_{\beta,L}&=&
\big\langle {\bf T}\, O^{(1)}_{t_1};O^{(2)}_{t_2};O^{(3)}_{t_3};\cdots;O^{(n)}_{t_n}\big\rangle_{\beta,L}\big|_{t_2=t_1+0^-}\nn\\
&-&
\big\langle {\bf T}\, O^{(2)}_{t_2};O^{(1)}_{t_1};O^{(3)}_{t_3};\cdots;O^{(n)}_{t_n}\big\rangle_{\beta,L}
\big|_{t_2=t_1+0^+}\;.\label{3.9}
\eea
Now, a straightforward implication of the definition of time-ordered truncated expectations is that 
\bea\label{eq:decomp}
&&\langle {\bf T}\, O^{(1)}_{t_1}\,;\, O^{(2)}_{t_2}\,;\, \cdots \,;\, O^{(n)}_{t_n} \rangle_{\b,L} =
\langle {\bf T}\, O^{(1)}_{t_1} O^{(2)}_{t_2}\,;\, \cdots \,;\, O^{(n)}_{t_n} \rangle_{\b,L} 
\nn\\
&& \quad - \sum^{*}_{\substack{ \{i_{1},\ldots, i_{p}\} \\ \{j_{1},\ldots j_{q}\} }} \langle {\bf T}\, O^{(1)}_{t_1}\,;\, O^{(i_{1})}_{t_{i_1}}\,;\, \cdots \,;\, O^{(i_{p})}_{t_{i_p}}\rangle_{\b,L}\langle {\bf T}\, O^{(2)}_{t_2}\,;\, O^{(j_{1})}_{t_{j_1}}\,;\, \cdots \,;\, O^{(j_{q})}_{t_{j_q}}\rangle_{\b,L} 
\eea
where the sum in the second line is over all the partitions of $\{3,\ldots , n-1\}$ in two disjoint subsets, $\{i_{1},\ldots, i_{p}\}$ and $\{ j_{1},\ldots, j_{q} \}$. By plugging this identity in the right side of \eqref{3.9},
we obtain that 
\be \big\langle {\bf T}\, \big[O^{(1)};O^{(2)}\big]_{t_1};O^{(3)}_{t_3};\cdots;O^{(n)}_{t_n}\big\rangle_{\beta,L}=\big\langle {\bf T}\, \big[O^{(1)},O^{(2)}\big]_{t_1};O^{(3)}_{t_3};\cdots;O^{(n)}_{t_n}\big\rangle_{\beta,L}\;.\label{3.11}\ee
(Note the comma between $O^{(1)}_{t_1}$ and $O^{(2)}_{t_2}$ in the right side, instead of the semicolon). Finally, by using \eqref{3.11} in the second line of \eqref{eq:WIderivative}, we
obtain \eqref{eq:WI1prop}. By taking the limit as the volume goes to infinity and the temperature to zero, 
and using the existence and analyticity of the limiting correlations stated in proposition 
\ref{prp:analyticity}, we obtain that \eqref{eq:WI1prop} is also valid for the limiting correlations. 
\qed

\bigskip

The Ward identities have important consequences on the momentum-dependence of the current-current correlations. The following corollary will play a crucial role in the proof of our main result.

\begin{cor}\label{cor:CH} Under the same hypotheses as proposition \ref{prp:analyticity},
the infinite volume and zero temperature correlations satisfy the following identities:\\
(1) If $n\ge 2$, $j\in\{1,2\}$ and $\underline\s=(\s_2,\ldots,\s_n)\in I^{n-1}$, 
\be\label{eq:CH1}
\widehat K_{j,\underline{\s}}((\o, \vec 0),\pp_{2}, \ldots, \pp_{n-1}) = - i\omega \frac{\partial
\widehat K_{0,\underline{\s}}}{\partial p_{1,j}}((\o,0),\pp_{2}, \ldots, \pp_{n-1})
\;.\ee
\0(2) If $n\ge 3$, $j,j'\in\{1,2\}$ and $\underline\s=(\s_3,\ldots,\s_n)\in I^{n-2}$,  
\bea
&&\widehat K_{j,j',\underline{\s}}((\o_1,\vec 0), (\o_2,\vec 0),\pp_{3},\ldots, \pp_{n-1}) - \frac{\partial
 \widehat S_{j';\underline{\s}}}{\partial p_{1,j}}((\o_1,\vec 0),(\o_2,\vec 0),\pp_{3},\ldots, \pp_{n-1})=
 \label{eq:CH2}\\
&& \hskip6.truecm=-\o_1\o_2 \frac{\partial^2  \widehat K_{0,0,\underline{\s}}}{\partial p_{1,j}\partial p_{2,j'}}(
(\o_1,\vec 0),(\o_2,\vec 0),\pp_{3},\ldots, \pp_{n-1})\;.\nn
\eea
\end{cor}

\medskip

{\bf Remarks.}
\begin{itemize}
\item These equations are just two special examples of relations among the correlations and their derivatives that can be obtained from the Ward Identities, by using the differentiability of the 
correlation functions stated in proposition \ref{prp:analyticity}. We limit ourselves to stating these two equations, because they are only the ones playing a role in the proof of our main result.
\item Similar consequences of the Ward Identities have been used by Coleman and Hill \cite{CH} to prove that all the contributions to the topological mass of QED$_{2+1}$ beyond one-loop vanish exactly.
\end{itemize}

{\bf Proof of corollary \ref{cor:CH}.} In order to prove \eqref{eq:CH1}, consider the limit as $\b,L\to\infty$
of \eqref{eq:WI1prop} with $(\a_2,\ldots,\a_n)=\underline\s$, which reads
\be i\o_1\widehat K_{0,\underline\s}(\pp_1,\ldots,\pp_{n-1})+\sum_{i=1}^2p_{1,i}\widehat K_{i,\underline\s}(\pp_1,\ldots,\pp_{n-1})=0\;.\ee
Recall that these correlations are differentiable, by proposition \ref{prp:analyticity}. 
Therefore, we can derive this equation with respect to $p_{1,j}$. If we do so, and then compute it at $\pp_1=(\o,\vec 0)$, we obtain \eqref{eq:CH1}. 

In order to prove \eqref{eq:CH2}, let us proceed as follows. Consider the $\b,L\to\infty$
of \eqref{eq:WI1prop} with $(\a_2,\ldots,\a_n)=(j',\underline\s)$, which reads
\be i\o_1\widehat K_{0,j',\underline\s}(\pp_1,\ldots,\pp_{n-1})+\sum_{i=1}^2p_{1,i}\widehat K_{i,j',\underline\s}(\pp_1,\ldots,\pp_{n-1})=\widehat S_{j';\underline\s}(\pp_1,\ldots,\pp_{n-1})\;.\ee
By deriving it with respect to $p_{1,j}$, we obtain 
\bea\label{eq:WI3}
&&\widehat K_{j,j',\underline{\s}}(\pp_{1},\ldots, \pp_{n-1}) - \frac{\partial}{\partial p_{1,j}} \widehat S_{j';\underline{\s}}(\pp_{1},\ldots, \pp_{n-1})= \nn\\
&& = -i\o_1 \frac{\partial}{\partial p_{1,j}} \widehat K_{0,j',\underline{\s}}(\pp_{1},\ldots, \pp_{n-1}) - \sum_{i=1}^2 p_{1,i} \frac{\partial}{\partial p_{1,j}} \widehat K_{i,j',\underline{\s}}(\pp_{1},\ldots, \pp_{n-1})
\eea
Similarly, consider (\ref{eq:WI1prop}), with $(\a_2,\ldots,\a_n)=(0,\underline\s)$. By using the 
invariance of $\widehat K_{\a_1,\ldots,\a_n}(\pp_1,\ldots,\pp_{n-1})$ under the exchange of 
$(\a_1,\pp_1)$ with $(\a_2,\pp_2)$, we obtain
\be i\o_2\widehat K_{0,0,\underline\s}(\pp_1,\ldots,\pp_{n-1})+\sum_{i=1}^2p_{2,i}\widehat K_{0,i,\underline\s}(\pp_1,\ldots,\pp_{n-1})=0\;,\ee
and by deriving this with respect to $p_{2,j'}$, we find:
\bea\label{eq:WI4}
\widehat K_{0,j',\underline{\s}}(\pp_{1},\ldots, \pp_{n-1})&=& -i\o_2\frac{\partial}{\partial p_{2,j'}}\widehat K_{0,0,\underline{\s}}(\pp_{1},\ldots, \pp_{n-1})\nn\\
&& - \sum_{i=1,2} p_{2,i} \frac{\partial}{\partial p_{2,j'}} \widehat K_{0,i,\underline{\s}}(\pp_{1},\ldots, \pp_{n-1})\;.
\eea
By plugging this equation into \eqref{eq:WI3} and then setting $\pp_1=(\o_1,\vec 0)$ and $\pp_2=(\o_2,\vec 0)$, we obtain \eqref{eq:CH2}.
\qed

\subsection{Schwinger-Dyson equation}

In this section we derive an equation relating the two-point current-current correlation, which enters the 
definition of conductivity, with higher-point correlations, known as the Schwinger-Dyson equation. 
The equation will be expressed order by order in perturbation {theory}, which is not a limitation, since,
in light of proposition \ref{prp:analyticity}, the correlations are analytic in $U$, if $|U|\le U_0$. 

The starting point is the convergent perturbative expansion of the current-current correlation. 
By using the Duhamel's formula, one can easily prove that 
\be\label{eq:cumulantK}
\widehat K^{\beta,L}_{i,j}(\pp) = \sum_{k\geq 0} (-1)^k\frac{U^{k}}{k!} \widehat K^{\beta,L,(k)}_{i,j}(\pp)
\;,\ee
where, if $\widetilde\VV_L=\int_0^\b dt\, (\VV_L)_t$ is the integral of the imaginary-time evolution of the interaction, 
\be\label{eq:Kkdef}
\widehat K^{\beta,L,(k)}_{i,j}(\pp) = \frac{1}{\beta L^{2}} \langle {\bf T}\,\hat J_{i,\pp}; \hat J_{j,-\pp}; \widetilde{\VV}_L^{\,; k}\rangle_{\beta,L}^{(0)}\;.
\ee
Here $\widetilde{\mathcal{V}}^{\,; k}_{L}$ is a shorthand notation for 
$\underbrace{\widetilde{\mathcal{V}}_{L}\,;\, \widetilde{\mathcal{V}}_{L}\,;\, \cdots \,;\, 
\widetilde{\mathcal{V}}_{L}}_{\text{$k$ times}}$,
and the superscript ${}^{(0)}$ is a shorthand for $\big|_{U=0}$: this means that both the Gibbs state
and the time evolution of the operators in \eqref{eq:Kkdef} are computed at $U=0$, i.e., 
with respect to the grand canonical  Hamiltonian $\HHH^{(0)}_L-\m \NN_L$. 

Note that $\widetilde\VV_L$ can be conveniently rewritten in momentum space as
\be\label{eq:deftildeV}
\widetilde{\mathcal{V}}_{L} = \frac{1}{\beta L^{2}} \sum_{\qq\in \frac{2\pi}{\beta}\mathbb{Z}\times \mathcal{B}_{L}} \sum_{\s,\s'\in I}\hat n^{\s}_{\qq} \hat v_{\s\s'}(\vec q)\, \hat n^{\s'}_{-\qq}\;.
\ee
Now, plugging \eqref{eq:deftildeV} in 
\eqref{eq:Kkdef}, we obtain that, for all $k\ge 1$, 
\be\label{eq:dec0}
\widehat K^{\beta,L,(k)}_{i,j}(\pp) = \frac{1}{(\beta L^{2})^{2}}\sum_{\qq\in \frac{2\pi}{\beta}\mathbb{Z}\times \mathcal{B}_{L}}\sum_{\s,\s' \in I} \hat v_{\s\s'}(\vec q)\langle {\bf T}\, \hat J_{i,\pp}\,;\, \hat J_{j,-\pp}\, ; \, \widetilde{\mathcal{V}}_{L}^{\,;\, k-1}\,;\, \hat n^{\s}_{\qq}\, \hat n^{\s'}_{-\qq} \rangle^{(0)}_{\beta,L}\;.
\ee
By using the combinatorial identity \eqref{eq:decomp}, we can further rewrite the average on the right side as
\bea
&&\langle {\bf T}\, \hat J_{i,\pp}\,;\, \hat J_{j,-\pp} \,;\, \widetilde{\mathcal{V}}_{L}^{\,;\, k-1}\,;\, \hat n^{\s}_{\qq}\, \hat n^{\s'}_{-\qq}  \rangle^{(0)}_{\beta,L} = \langle {\bf T}\, \hat J_{i,\pp}\,;\, \hat J_{j,-\pp} \,;\, \widetilde{\mathcal{V}}_{L}^{\,;\, k-1}\,;\, \hat n^{\s}_{\qq}\,;\, \hat n^{\s'}_{-\qq}  \rangle^{(0)}_{\beta,L}+ \sum_{m=0}^{k-1} {k-1 \choose m}\times \nn\\
&&\quad  \times \Big[\langle {\bf T}\, \hat J_{i,\pp}\,;\, \hat J_{j,-\pp} \,;\, \widetilde{\mathcal{V}}_{L}^{\,;\, m}\,;\, \hat n^{\s}_{\qq} \rangle^{(0)}_{\beta,L}\cdot
\langle {\bf T}\, \widetilde{\mathcal{V}}_{L}^{\,;\, k-1-m}\,;\, \hat n^{(\s')}_{-\qq}  \rangle^{(0)}_{\beta,L}
+ \langle {\bf T}\, \hat J_{i,\pp}\,;\, \widetilde{\mathcal{V}}_{L}^{\,;\, m}\,;\, \hat n^{\s}_{\qq} \rangle_{\beta,L}^{(0)}\cdot\nonumber\\
&&\qquad \cdot \langle {\bf T}\, \hat J_{j,-\pp}\,;\, \widetilde{\mathcal{V}}_{L}^{\,;\, k-1-m}\,;\, \hat n^{\s'}_{-\qq} \rangle_{\beta,L}^{(0)}
+ \text{terms obtained by replacing $n_\qq^\s \otto n_{-\qq}^{\s'}$}\Big]\;.\label{eq:dec}
\eea
The translation invariance of the Gibbs state implies that (denoting $\qq=(\o',\vec q)$):
\bea
\langle {\bf T}\, \hat J_{i,\pp}\,;\, \hat J_{j,-\pp} \,;\, \widetilde{\mathcal{V}}_{L}^{\,;\, m}\,;\, \hat n^{\s}_{\qq} \rangle^{(0)}_{\beta,L} &=& \delta_{\o',0}\delta_{\vec q,\vec 0}\langle {\bf T}\, \hat J_{i,\pp}\,;\, \hat J_{j,-\pp} \,;\, \widetilde{\mathcal{V}}_{L}^{\,;\, m}\,;\, \hat n^{\s}_{\V0} \rangle^{(0)}_{\beta,L} \nn\\
\langle {\bf T}\, \hat J_{i,\pp}\,;\, \widetilde{\mathcal{V}}_{L}^{\,;\, m}\,;\, \hat n^{\s}_{\qq} \rangle_{\beta,L}^{(0)} &=& \delta_{\o'+\o,0}\delta_{\vec q+\vec p,\vec 0}
 \langle {\bf T}\, \hat J_{i,\pp}\,;\, \widetilde{\mathcal{V}}_{L}^{\,;\, m}\,;\, \hat n^{\s}_{-\pp} \rangle_{\beta,L}^{(0)}\label{eq:dec1}\\
\langle {\bf T}\, \widetilde{\mathcal{V}}_{L}^{\,;\, m}\,;\, \hat n^{\s}_{-\qq}  \rangle^{(0)}_{\beta,L} &=& 
\delta_{\o',0}\delta_{\vec q,\vec 0}\langle {\bf T}\, \widetilde{\mathcal{V}}_{L}^{\,;\, m}\,;\, \hat n^{\s}_{\V0}  \rangle^{(0)}_{\beta,L}\;.\nn
\eea
If we now substitute \eqref{eq:dec} and \eqref{eq:dec1} into \eqref{eq:dec0}, we obtain 
the following remarkable identity, summarized here as a proposition.

\begin{prp}{{\bf [Schwinger-Dyson equation]}}\label{prp:SD}
For all $k\ge 1$, $i,j\in\{1,2\}$, and
$\pp \in \frac{2\pi}{\beta}\mathbb{Z}\times \mathcal{D}_{L}$, the following identity holds:
\bea\label{eq:SD}
\widehat K^{\beta,L,(k)}_{i,j}(\pp) &=& \frac{1}{\beta L^{2}} \sum_{\qq \in \frac{2\pi}{\beta}\mathbb{Z}\times \mathcal{B}_{L}}\sum_{\s,\s' \in I}\hat v_{\s\s'}(\vec q) \widehat K^{\beta,L,(k-1)}_{i,j,\s,\s'}(\pp,-\pp,\qq) \nn\\
&& + 2\sum_{m=0}^{k-1} {k-1 \choose m} \sum_{\s,\s' \in I} \hat v_{\s\s'}(\vec 0) \widehat K^{\beta,L,(m)}_{i,j,\s}(\pp,-\pp) \widehat K^{\beta,L,(k-1-m)}_{\s'}\label{eq:schwingerdyson}\\
&& + 2\sum_{m=0}^{k-1} {k-1 \choose m} \sum_{\s,\s' \in I} \hat v_{\s\s'}(-\vec p)\widehat K^{\beta,L,(m)}_{i,\s}(\pp) \widehat K^{\beta,L,(k-1-m)}_{j,\s'}(-\pp)\;.\nn
\eea
\end{prp}

Note that $\hat v_{\s\s'}(-\vec p)=\hat v_{\s'\s}(\vec p)$, and the argument of $\hat v_{\s\s'}$ should be identified with its image in $\BBB_L$ modulo vectors in $\L_L^*$.

\subsection{Proof of theorem \ref{thm:main}}

We are finally in the position of proving our main result, theorem \ref{thm:main}. The proof is based on 
a combination of the three results discussed in the previous subsections, namely: the analyticity of the 
correlation functions (proposition \ref{prp:analyticity}), the Ward identities (proposition \ref{prop:WI}), 
and the Schwinger-Dyson equation (proposition \ref{prp:SD}).

\medskip

{\bf Proof of theorem \ref{thm:main}.} 
In light of theorem \ref{thm:realkubo}, it is enough to prove that 
$\bar\s_{ij}(U)=\bar\s_{ij}(0)$, for $U$ small enough. 
First of all, by the analyticity of the current correlations stated in proposition \ref{prp:analyticity},
we have that $\bar\s_{ij}(U)$ is analytic in $U$ as well, as long as $|U|\le U_0$. In this domain, 
$\bar\s_{ij}(U)$ can be written in convergent perturbation series as 
\be\label{eq:sij}
\bar\sigma_{ij}(U)=\bar\sigma^{(0)}_{ij} + \sum_{k\geq 1}(-1)^k\frac{U^{k}}{k!}\bar\sigma^{(k)}_{ij}\;,
\ee
where
\be\label{eq:sijk}
\bar\sigma_{ij}^{(k)} = -\frac{1}{A}\lim_{\o\to 0} \frac{\partial}{\partial\o}\widehat K^{(k)}_{ij}(\o,\vec 0)\;.
\ee
Since the series in \eqref{eq:sij} is convergent, in order to prove theorem \ref{thm:main} it 
sufficies to show that:
\be\label{eq:vanishing}
\bar\sigma^{(k)}_{ij} = 0\;,\qquad \text{for all $k\geq 1$.}
\ee
This will be proved by showing that the derivative of  $\widehat K^{(k)}_{ij}(\o,\vec 0)$ with respect to 
$\omega$ vanishes linearly as $\o\to0$.
To this aim, we plug the Schwinger-Dyson equation \eqref{eq:schwingerdyson}
into \eqref{eq:sijk}, thus getting, for all $k\ge 1$, 
$\bar\sigma_{ij}^{(k)} =\text{I}^{(k)}+\text{II}^{(k)}+\text{III}^{(k)}$, with 
\bea\label{eq:ItoIII}
\text{I}^{(k)} &:=& - \frac{1}{A} \lim_{\o\to 0} \int_{\mathbb R\times \BBB} \frac{d\qq}{(2\pi)|\mathcal{B}|} \sum_{\s,\s' \in I}\hat v_{\s\s'}(\vec q)\frac{\partial}{\partial\o} \widehat K^{(k-1)}_{i,j,\s,\s'}\big((\o,\vec 0), (-\o,\vec 0),\qq \big) \\
\text{II}^{(k)} &:=& -\frac{2}{A}\lim_{\o \to 0} \sum_{m=0}^{k-1} {k-1 \choose m}  \sum_{\s,\s' \in I} \hat v_{\s\s'}(\vec 0) \frac{\partial}{\partial \o} \widehat K^{(m)}_{i,j,\s}\big((\o,\vec 0), (-\o,\vec 0)\big) \widehat K^{(k-1-m)}_{\s'} \nn\\
\text{III}^{(k)} &:=& - \frac{2}{A}\lim_{\o \to 0} \sum_{m=0}^{k-1} {k-1 \choose m} \sum_{\s,\s' \in I} \hat v_{\s\s'}(\vec 0) \frac{\partial}{\partial \o}\Big[\widehat K^{(m)}_{i,\s}(\o,\vec 0) \widehat K^{(k-1-m)}_{j,\s'}(-\o, \vec 0)\Big]\;.\nn
\eea
Now, by using corollary \ref{cor:CH}, we prove that the three contributions are separately zero. 

\medskip 

{\it The contribution ${\rm{I}}^{(k)}$.}
First of all, note that
\be\label{eq:proofmainI}
\frac{\partial}{\partial \o} \widehat K^{(k-1)}_{i,j,\s,\s'}\big((\o,\vec 0), (-\o,\vec 0),\qq \big) 
= \frac{\partial}{\partial \o}\Big[ \widehat K^{(k-1)}_{i,j,\s,\s'}\big((\o,\vec 0), (-\o,\vec 0),\qq \big) - \frac{\partial \widehat S^{(k-1)}_{j;\s,\s'}}{\partial p_{1,i}}\big((\o,\vec 0), (-\o,\vec 0),\qq \big)
\Big]\;,
\ee
simply because, by the very definition \eqref{eq:D}-\eqref{eq:3.6} of $\widehat S_{j;\s,\s'}(\pp_1,\pp_2,\pp_3)$, this function 
depends on $\o_1$ and $\o_2$ only upon the combination $\o_1+\o_2$, so that, in particular,
$\widehat S_{j;\s,\s'}((\o,\vec 0),(-\o,\vec 0),\qq)$ is independent of $\o$. We can now use 
\eqref{eq:CH2}, understood as an order by order identity between convergent power series in $U$,
and we thus obtain 
\bea
\text{I}^{(k)} &=& -\frac{1}{A}\lim_{\o\to 0} \int_{\mathbb R\times \BBB}  \frac{d\qq}{(2\pi)|\mathcal{B}|} \sum_{\s,\s' \in I}\hat v_{\s\s'}(\vec q) \frac{\partial}{\partial \o} \Big[ \o^{2} \frac{\partial^{2} \widehat K^{(k-1)}_{0,0,\s,\s'}}{\partial p_{1,i}\partial p_{2,j}}\big((\o,\vec 0),(-\o,\vec 0),\qq\big)\Big]\label{eq:3.30}\\
&=& -\frac{1}{A}\lim_{\o\to 0}\int_{\BBB}  \frac{d\vec q}{|\mathcal{B}|} \sum_{\s,\s' \in I}\hat v_{\s\s'}(\vec q)\Big[ 2\o F_1^{(k-1)}(\o,\vec q)+\o^2F_2^{(k-1)}(\o,\vec q)\Big]\;,\nn
\eea
where 
\bea && F_1^{(k-1)}(\o,\vec q)=\int_{\mathbb R}\frac{d\o'}{2\pi}\frac{\partial^{2}\widehat K_{0,0,\s,\s'}^{(k-1)}}{\partial p_{1,i}\partial p_{2,j}} \big(
(\o,\vec 0),(-\o,\vec 0),(\o',\vec q)\big),\label{eq:3.31}\\
&& F_2^{(k-1)}(\o,\vec q)=\int_{\mathbb R}\frac{d\o'}{2\pi}\Big[
\frac{\partial^{3}\widehat K_{0,0,\s,\s'}^{(k-1)}}{\partial \o_1\partial p_{1,i}\partial p_{2,j}} -\frac{\partial^{3}\widehat K_{0,0,\s,\s'}^{(k-1)}}{\partial \o_2\partial p_{1,i}\partial p_{2,j}} \Big]\big(
(\o,\vec 0),(-\o,\vec 0),(\o',\vec q)\big)\;.\nn\eea
Now, the key remark is that $F_1^{(k-1)}(\o,\vec q)$ and $F_2^{(k-1)}(\o,\vec q)$ are bounded uniformly 
in $\o$ and $\vec q$, for all $i=1,2,3$, which immediately implies that the second line of \eqref{eq:3.30}
is zero, as desired. In order to prove that $|F_i^{(k-1)}(\o,\vec q)|\le C$, uniformly in $\o$ and $\vec q$,
we rewrite 
\bea &&\widehat K_{0,0,\s,\s'}^{(k-1)}(\pp_1,\pp_2,\pp_3)=\sum_{\underline\s}\!\int\! d\underline t\sum_{\underline{\vec x}}\,
e^{-i\o_1 t-i\o_2 t'-i\o_3 t''}\,e^{-i\vec p_1\vec x-i\vec p_2\vec y-i\vec p_3\vec z}\,
\Big[\prod_{l=1}^{k-1} v_{\s_l,\s_l'}(\vec x_l-\vec y_l)\Big]\times\nn\\
&&\qquad  \times
\media{n^{\s_0}_{(t,\vec x)}\,;\,n^{\s_0'}_{(t',\vec y)}\,;\,n^\s_{(t'',\vec z)}\,;\,n^{\s'}_{\bf 0}\,;\,n^{\s_1}_{(t_1,\vec x_1)}n^{\s_1'}_{(t_1,\vec y_1)}\,;\,\cdots\,;\,n^{\s_{k-1}}_{(t_{k-1},\vec x_{k-1})}n^{\s_{k-1}'}_{(t_{k-1},\vec y_{k-1})}}^{\!\!(0)},
\label{eq:3.32}\eea
where:
 (i) $\underline\s$ is a shorthand for $(\s_0,\s_0',\s_1,\s_1',\ldots,\s_{k-1},\s_{k-1}')$, and  
is summed over $I^{2k}$, (ii) $\underline t$ is a shorthand for $(t,t',t'',t_1,\ldots,t_{k-1})$, and is integrated over $\mathbb R^{k+2}$, (iii) $\underline{\vec x}$ is a shorthand for $(\vec x,\vec y,\vec z,\vec x_1,\vec y_1,\ldots,\vec x_{k-1},\vec y_{k-1})$, and is summed over $\L^{2k+1}$, where 
$\L=\{\vec x : \vec x=n_1\vec \ell_1+n_2\vec \ell_2, n_i\in\mathbb Z\}$, (iv) $\media{\cdot}^{(0)}$ stands for $\lim_{\b\to\infty}\lim_{L\to\infty}\media{\cdot}_{\b,L}^{(0)}$.
By using \eqref{eq:3.32}, we can rewrite $F_1^{(k-1)}$ as
\bea && F_1^{(k-1)}(\o,\vec q)=-\sum_{\underline\s}\!\int\! d\underline t\sum_{\underline{\vec x}}\,
e^{-i\o( t-t')}\,e^{-i\vec q\vec z}\,(\vec x)_i\,(\vec y)_j\,
\Big[\prod_{l=1}^{k-1} v_{\s_l,\s_l'}(\vec x_l-\vec y_l)\Big]\times\nn\\
&&\qquad  \times
\media{n^{\s_0}_{(t,\vec x)}\,;\,n^{\s_0'}_{(t',\vec y)}\,;\,n^\s_{(0,\vec z)}\,;\,n^{\s'}_{\bf 0}\,;\,n^{\s_1}_{(t_1,\vec x_1)}n^{\s_1'}_{(t_1,\vec y_1)}\,;\,\cdots\,;\,n^{\s_{k-1}}_{(t_{k-1},\vec x_{k-1})}n^{\s_{k-1}'}_{(t_{k-1},\vec y_{k-1})}}^{\!\!(0)}.
\label{eq:3.33}\eea
The truncated expectation value of the number operators in the second line 
can be computed via the Wick rule, which is the following. Write each number operator in the form 
$n^{\r}_{(s,\vec w)}=\psi^+_{(s,\vec w),\r}\psi^-_{(s,\vec w),\r}$, and consider
all the possible pairings of the creation/annihilation operators
such that each annihilation operator $\psi^-_{(s,\vec w),\r}$ is paired with a creation operator $\psi^+_{(s',\vec w'),\r'}$, with the additional constraint that the resulting pairing $p$ is ``connected'' in the following sense. Consider the directed graph $G_p=(V,E_p)$ whose vertex set is $V=\{\vec x,\vec y,\vec z,\vec 0,\vec x_1,\vec y_1,\ldots,\vec y_{k-1}\}$, and whose edge set $E_p$ consists of the ordered pairs $(\vec x_l,\vec y_l)$, $l=1,\ldots, k-1$, as well as of the ordered pairs $(\vec w,\vec w')$ associated with the elements $\ell=(\psi^-_{(s,\vec w),\r},\psi^+_{(s',\vec w'),\r'})$ of the pairing $p$: we shall say that the pairing $p$ is connected if the graph $G_p$ is connected. 
Then associate each connected pairing $p$ with a value, given by the sign $\a_p$ of the permutation required to move every creation operator to the immediate right of 
the annihilation operator it is paired with, times the product over the pairs of the corresponding propagators, where the propagator corresponding 
to the pair $\ell=(\psi^-_{(s,\vec w),\r},\psi^+_{(s',\vec w'),\r'})$ is 
\bea&& g_\ell\equiv g_{\r,\r'}(s-s',\vec w-\vec w')=\lim_{\b\to\infty}\lim_{L\to\infty}
\media{{\bf T}\psi^-_{(s,\vec w),\r}\psi^+_{(s',\vec w'),\r'}}_{\b,L}^{(0)}=\\
&&=\int_{\BBB}\frac{d\vec k}{|\BBB|} 
e^{-i\vec k(\vec w-\vec w')}\Big[e^{-(s-s')(\hat H^{(0)}(\vec k)-\mu)}
\Big(\mathds 1(s>s') P_+(\vec k)-
\mathds 1(s\le s')\,P_-(\vec k)\Big)\Big]_{\r,\r'}\;,\nn
\eea
where $P_-(\vec k)$ is the projector over the filled bands (see the lines after \eqref{gap}), and 
$P_+(\vec k)=\mathds 1-P_-(\vec k)$. As already observed after $\eqref{gap}$, 
under the gap condition,  $P_-(\vec k)$ is 
infinitely differentiable in $\vec k$. Therefore, $g_{\s\s'}(s-s',\vec w-\vec w')$ decays exponentially in $s-s'$ 
and faster than any power in $\vec w-\vec w'$. For later convenience, if $\ell=(\psi^-_{(s,\vec w),\r},\psi^+_{(s',\vec w'),\r'})$, we denote by 
$\D \xx_\ell$ the space-time difference associated with $\ell$, namely $\D\xx_\ell=(\D t_\ell,\D\vec x_\ell)=(s-s',\vec w-\vec w')$.

On the basis of the Wick rule explained above,  \eqref{eq:3.33} can be rewritten as 
\be\label{eq:f1} F_1^{(k-1)}(\o,\vec q)=-\sum_{\underline\s}\!\int\! d\underline t\sum_{\underline{\vec x}}\,
e^{-i\o( t-t')}\,e^{-i\vec q\vec z}\,(\vec x)_i\,(\vec y)_j\,
\Big[\prod_{l=1}^{k-1} v_{\s_l,\s_l'}(\vec x_l-\vec y_l)\Big]\sum_{p\in \mathcal G_c} \a_p \prod_{\ell\in p}g_\ell\;,\ee
where $\mathcal G_c$ is the set of connected pairings. In order to bound this expression,
for each pairing $p$ we arbitrarily choose a {\it connected} tree subgraph of $G_p$, denoted by $T_p$, 
consisting of all the pairs $(\vec x_l,\vec y_l)$, with $l=1,\ldots,k-1$, and of other $k+2$ 
edges of $G_p$. 
With some abuse of notation, we shall denote by $T_p$ also the subset of $p$ whose 
pairs are graphically associated with edges of $T_p$. 
Next, we decompose $(\vec x)_i$ 
along the path $\mathcal C_p^{\vec x\to \vec 0}$ 
on $T_p$ from $\vec x$ to $\vec 0$. By walking along the path $\mathcal C_p^{\vec x\to \vec 0}$
from $\vec x$ to $\vec 0$, some 
of the edges $e\in\mathcal C_p^{\vec x\to \vec 0}$ may be oriented in the same direction as the walk, in 
which case we set $\a_e=+1$, and some others in the opposite direction, in which case we set 
$\a_e=-1$. We can then rewrite  
$(\vec x)_i=\sum_{e\in \mathcal C_p^{\vec x\to \vec 0}}\a_e(\D\vec x_{\ell_e})_i$,
where $\ell_e$ is the pair graphically associated with $e$. We use a similar decomposition for 
$(\vec y)_j$. 

In terms of these definitions, we can finally bound \eqref{eq:f1} as
\be\label{eq:f1.1} |F_1^{(k-1)}(\o,\vec q)|\le (2k+1)^{2}\sum_{\underline\s}\sum_{p\in \mathcal G_c}
\Big[\prod_{l=1}^{k-1} \|v\|_{1,2}\Big]\Big[\prod_{\ell\in p\cap T_p}\|g_\ell\|_{1,2}
\Big]\Big[\prod_{\ell\in p\setminus T_p}\|g_\ell\|_\infty\Big]\;,\ee
where $\|v\|_{1,m}=\sup_{\s,\s'}\sum_{\vec x\in \L} |\vec x|^m |v_{\s,\s'}(\vec x)|$, 
$\|g_\ell\|_{1,m}=\sup_{\s,\s'}\int_{\mathbb R} dt \sum_{\vec x\in \L} |\vec x|^m |g_{\s,\s'}(t,\vec x)|$, 
and $\|g_\ell\|_{\infty}=\sup_{\s,\s'}\sup_{t,\vec x}  |g_{\s,\s'}(t,\vec x)|$. Now, using the fact that 
$v$ and $g$ decay to zero at large distances faster than any power, as well as the fact that 
number of terms in the sums over $\underline\s$ and over $p\in\mathcal G_p$ are bounded, 
respectively,  by $|I|^{2k}$ and by $(2k+2)!$, we obtain that $|F_1^{(k-1)}(\o,\vec q)|\le (k!)^2 ({\rm const.})^k$,
where the constant depends, in general, on the gap $\d_\m$. By proceeding analogously, we see that 
$F_2^{(k-1)}$ can be bounded exactly in the same way. This concludes the proof of 
the uniform boundedness of $F_i^{(k-1)}$ and, as observed after \eqref{eq:3.31}, of the fact that I$^{(k)}=0$.

\medskip

{\bf Remark.}
The $(k!)^2$ dependence in the bounds on  
$F_1^{(k-1)}$ and $F_2^{(k-1)}$ naively suggests that the $k$-th order coefficient in the expansion \eqref{eq:sij} behaves like $\sim k!$ at large $k$ (note the extra $1/k!$ in the right side of \eqref{eq:sij}), which seems incompatible with the stated analyticity of $\bar\s_{ij}(U)$. In fact, there is a better way of bounding 
the $k$-th order coefficient of the series, which is smaller by a factor $\sim k!$, as compared to the 
bound presented above, which implies the analyticity of the series and will be discussed in the next section.

\medskip 

{\it The contributions ${\rm{II}}^{(k)}$ and ${\rm{III}}^{(k)}$.}
The proof of the fact that ${\rm{II}}^{(k)}$ and ${\rm{III}}^{(k)}$ are zero goes along the same lines as 
the proof that I$^{(k)}=0$. By using the independence of $\widehat S_{j;\s}\big((\o,\vec p_1),(-\o,\vec p_2)\big)$  on $\o$ and the identity (\ref{eq:CH2}), we find that 
\bea  \frac{\partial}{\partial\o} \widehat K^{(m)}_{i,j,\s}\big((\o,\vec 0), (-\o,\vec 0)\big) &=&
\frac{\partial}{\partial\o}\Big[ \widehat K^{(m)}_{i,j,\s}\big((\o,\vec 0), (-\o,\vec 0)\big) -
\frac{\partial \widehat S_{j;\s}^{(m)}}{\partial p_{1,i}}\big((\o,\vec 0), (-\o,\vec 0)\big)\Big]\nn\\
&=&
\frac{\partial}{\partial\o}\Big[\o^2 
\frac{\partial^2  \widehat K_{0,0,\underline{\s}}^{(m)}}{\partial p_{1,j}\partial p_{2,j'}}\big(
(\o,\vec 0),(-\o,\vec 0)\big)\Big]\;,
\eea
so that 
\be\label{eq:CHIIa}
\text{II}^{(k)} = -\frac{2}{A}\lim_{\o \to 0}  \sum_{\s,\s'\in I} \hat v_{\s\s'}(\vec 0)\Big[ 2\o F_3^{(k-1)}(\o)+
\o^2 F_4^{(k-1)}(\o)\Big]\;,
\ee
where \bea && F_3^{(k-1)}(\o)=\!\!\!\!\!\sum_{\substack{m_1,m_2:\\
m_1+m_2=k-1}}\!\!\!\!\!{k-1 \choose m_1} \frac{\partial^2  \widehat K^{(m_1)}_{0,0,\underline{\s}}}{\partial p_{1,j}\partial p_{2,j'}}\big(
(\o,\vec 0),(-\o,\vec 0)\big)\cdot\widehat K_{\s'}^{(m_2)}
\;,\\
&& F_4^{(k-1)}(\o)=\!\!\!\!\!\sum_{\substack{m_1,m_2:\\
m_1+m_2=k-1}}\!\!\!\!\!{k-1 \choose m_1} \Big[\frac{\partial^3  \widehat K^{(m_1)}_{0,0,\underline{\s}}}{\partial \o_1\partial p_{1,j}\partial p_{2,j'}}-\frac{\partial^3  \widehat K^{(m_1)}_{0,0,\underline{\s}}}{\partial \o_2\partial p_{1,j}\partial p_{2,j'}}\Big]\big(
(\o,\vec 0),(-\o,\vec 0)\big)\cdot\widehat K_{\s'}^{(m_2)}
\;.\nn\eea Similarly, using (\ref{eq:CH1}), we can rewrite
\be\label{eq:CHIIb}
\widehat K^{(m_1)}_{i,\s}(\o,\vec 0) 
= - i\o\frac{\partial \widehat K^{(m_1)}_{0,\s}}{\partial p_{i}}(\o,\vec 0)
\;,\qquad 
\widehat K^{(m_2)}_{j,\s'}(-\o, \vec 0) = i\o\frac{\partial\widehat K^{(m_2)}_{0,\s'}}{\partial p_{j}}
(-\o,\vec 0)
\;,
\ee
so that 
\be\label{eq:CHIIa_bis}
\text{III}^{(k)} = -\frac{2}{A}\lim_{\o \to 0}\sum_{\s,\s'\in I} \hat v_{\s\s'}(\vec 0)\Big[ 2\o F_5^{(k-1)}(\o)+
\o^2 F_6^{(k-1)}(\o)\Big]\;,
\ee
where 
\bea &&F_5^{(k-1)}(\o)=\!\!\!\!\!\sum_{\substack{m_1,m_2:\\
m_1+m_2=k-1}}\!\!\!\!\!{k-1 \choose m_1}    \frac{\partial \widehat K^{(m_1)}_{0,\s}}{\partial p_{i}}(\o,\vec 0)
\frac{\partial\widehat K^{(m_2)}_{0,\s'}}{\partial p_{j}}
(-\o,\vec 0)\;,\\
&& F_6^{(k-1)}(\o)=\!\!\!\!\!\sum_{\substack{m_1,m_2:\\
m_1+m_2=k-1}}\!\!\!\!\!{k-1 \choose m_1}  \Big[\frac{\partial^2 \widehat K^{(m_1)}_{0,\s}}{\partial \o\partial p_{i}}(\o,\vec 0)
\frac{\partial\widehat K^{(m_2)}_{0,\s'}}{\partial p_{j}}(-\o,\vec 0)-\frac{\partial \widehat K^{(m_1)}_{0,\s}}{\partial p_{i}}(\o,\vec 0)
\frac{\partial^2\widehat K^{(m_2)}_{0,\s'}}{\partial\o\partial p_{j}}(-\o,\vec 0)\Big].\nn
\eea
By proceeding as in the proof of \eqref{eq:f1.1}, one obtains that $F_i^{(k-1)}(\o)$ are bounded 
uniformly in $\o$, which implies that $\text{II}^{(k)}=\text{III}^{(k)}=0$, as desired.
This concludes the proof of (\ref{eq:vanishing}), and of theorem \ref{thm:main}. \qed

\section{Analyticity}\label{app:analyt}

In this section we prove proposition \ref{prp:analyticity}, concerning the analyticity in $U$ and 
the smoothness 
in $\pp$ of the multi-point current/density correlation functions. 
Roughly, the strategy will consist in: (i) reformulating the correlation functions in terms of a Grassmann 
integral, in the limit where a suitable cutoff function is removed; (ii) proving the analyticity of the 
Grassmann integral, uniformly in the cutoff parameter; (iii) using Vitali's theorem on the convergence of holomorphic functions (also known as Vitali-Porter theorem, or Weierstrass' theorem), to conclude that the correlations themselves are analytic. 
The analysis of this section is a straightforward adaptation of previous works, see, e.g., 
\cite[Appendices B,C,D]{GM} or \cite[Section 6]{Gi} for two recent reviews in the context of 
graphene with short-range interactions, and is included here just for the sake of self-
containedness. 

As we shall see, our proof of analyticity uses a multiscale analysis, which 
seems like overkill in a situation like ours where the propagator decays faster than any power in 
space and time, uniformly in $\b$ and $L$. Before we delve into the proof, which is quite technical, 
let us then explain why a naive single-scale approach fails, and what are the main ideas that 
led us to use a multiscale analysis. 

The generic order in perturbation theory 
can be expressed as a sum over connected pairings, in complete analogy with the representation 
of the second line of \eqref{eq:3.33} discussed after that equation. 
Each pairing is associated with a value, which is bounded 
uniformly in the parameters involved, in analogy with \eqref{eq:f1.1}. However, after summing over the 
possible pairings, the bound on the $k$-th order coefficient scales like $C^k k!$, where  
$k!$ should be thought of as the product of the factor $1/k!$ appearing in the Taylor expansion 
(cf., e.g., with \eqref{eq:sij}) times the number 
of possible pairings, which grows like $(k!)^2$, see the lines after \eqref{eq:f1.1}. In other words, if we 
bound pairing by pairing the contributions to the $k$-th order in perturbation theory, we 
get a contribution that is not summable over $k$, because it is off by a combinatorial factor $\sim k!$.

This problem is reminiscent of the problem of convergence of the virial (low-density/high-temperature) 
expansion in classical statistical 
mechanics, where the $k$-th order coefficient is the sum of several contributions, each of which is 
easily seen to be bounded. However, the number of contributions to the $k$-th order of the virial expansion 
is too large ($\sim C^{k^2}$) and, therefore, the convergence of the series requires the 
exhibition of cancellations among the various contributions. 

In the fermionic problem at hand, the required cancellations arise from the fermionic statistics: the 
$k$-th order coefficient in perturbation theory can be expressed in the form of a determinant, whose 
norm is in many situations much smaller than the sum of the norms of the single contributions to the determinant.
For instance, if the generic element of the matrix $A$ is expressible in the form of a scalar product, $A_{ij}=(u_i,v_j)$,
then $\det A$ can be conveniently bounded (via the so-called Gram-Hadamard inequality) as:
$|\det A|\le \prod_i\|u_i\|\cdot\|v_i\|$, where $\|\cdot\|$ is the norm induced by the scalar 
product. Such a bound is free of bad factorials and seems to make the job. Unfortunately, in the case at 
hand, the elements of the matrix of interest are propagators $g_{\s_i\s_j}(t_i-t_j,\vec x_i-\vec x_j)$, 
which are {\it not} expressible as the scalar product of two vectors on any separable Hilbert space 
\cite{PS}, due to a jump singularity in the time dependence of the propagator. The jump singularity is 
induced by the time ordering arising from the Duhamel's formula. 
There are several ways out of this problem. One is to expand the determinants in the form of 
``chronologically ordered'' determinants, each of which is free from jump singularities and can be 
bounded by the Gram-Hadamard inequality, as in \cite{PS}. This approach has the advantage that it allows to obtain constructive bounds without any 
multi-scale analysis, but in order to do so, it is crucial that the free propagator decays sufficiently fast (as it is the case, e.g., 
in gapped systems).
Recently, it has been shown that single-scale constructive bounds can also be obtained without the use of 
chronologically ordered determinants, via non-commutative H\"older inequalities \cite{BrPe2}. 
Another possibility is to impose an ultraviolet cutoff on the imaginary 
frequencies, and to re-express the regularized propagator as the sum of single-scale propagators, 
each admitting a scalar product representation, in terms of vectors of uniformly bounded norm.
In this way, we get rid of the combinatorial factor related to the $k!$ explained above, at the cost of 
analyzing a simple multiscale problem. Even if slightly more technical than the first method, 
this approach has the advantage of being adaptable to massless situations, with propagator decaying slowly (in a non-integrable way) 
at large distances, such as those studied in 
\cite{BGPS, BM, BFM0, BFM1, GM,GMP1,GMP2}. We believe that, by applying the multiscale methods 
of these papers, our theorem \ref{thm:main} could be extended arbitrarily close to the massless line 
$\d_\m=0$. In this perspective, we prefer to present here a multiscale proof of the analyticity of the 
correlations, and we plan to come back to the problem of extending it to the infrared regime 
in a future publication.

\subsection{Grassmann representation}

Let us preliminarily recall a few known facts about perturbation theory for the free energy and correlations of interacting fermionic systems, which 
we need for justifying their Grassmann representation. We first discuss 
the free energy, which is simpler. Using Duhamel's expansion, we can rewrite the (a priori formal) series expansion of the interacting partition function 
in the parameter $U$ as:
\be \frac{\Tr_{\mathcal F}e^{-\beta(\mathcal H_L-\mu\mathcal N_L)}}{\Tr_{\mathcal F}e^{-\beta(\mathcal H^{(0)}_L-\mu\mathcal N_L)}}=
1+\sum_{n\ge 1}(-U)^n\int_0^\b dt_1\cdots \int_0^{t_{n-1}}dt_n \frac{\Tr_{\mathcal F}e^{-\beta(\mathcal H^{(0)}_L-\mu\mathcal N_L)}
\VV_L(t_1)\cdots\VV_L(t_n)}{\Tr_{\mathcal F}e^{-\beta(\mathcal H^{(0)}_L-\mu\mathcal N_L)}}
\ee
where $\VV_L(t)=e^{t(\mathcal H^{(0)}_L-\mu\mathcal N_L)}\VV_L e^{-t(\mathcal H^{(0)}_L-\mu\mathcal N_L)}$ is the non-interacting ($U=0$) version 
of the imaginary time evolution of $\VV_L$, cf. with Eq.\eqref{eq:17}.
Symmetrizing over the 
permutations of $t_1,\ldots,t_n$, this can be rewritten as
\be \frac{\Tr_{\mathcal F}e^{-\beta(\mathcal H_L-\mu\mathcal N_L)}}{\Tr_{\mathcal F}e^{-\beta(\mathcal H^{(0)}_L-\mu\mathcal N_L)}}=
1+\sum_{n\ge 1}\frac{(-U)^n}{n!}\int_0^\b dt_1\cdots \int_0^{\b}dt_n \media{{\bf T}\VV_L(t_1)\cdots\VV_L(t_n)}_{\b,L}^{(0)},\label{eq:g2}
\ee
where we recall that the label $^{(0)}$ on the expectation symbol indicates that we are computing 
it at $U=0$.
Since $\mathcal H^{(0)}_L-\mu\mathcal N_L$ is quadratic in the fermionic creation/annihilation operators, $\media{\cdot}_{\b,L}^{(0)}$ 
can be computed via the fermionic Wick rule, which is completely analogous to the one 
described for the infinite volume and zero temperature truncated expectation after \eqref{eq:3.33},
with the following minor differences: (i) since the expectation in \eqref{eq:g2} is {\it not} truncated, 
after having re-expressed $\VV_L(t_1)\cdots\VV_L(t_n)$ as a linear combination of monomials
of order $4n$ in the creation and annihilation operators, we have to sum over {\it all} possible pairings of these creation/annihilation operators, rather than just on the connected ones; (ii) the finite volume and finite temperature propagator associated with the pair $(\psi^-_{(t,\vec x),\s},\psi^+_{(t',\vec x'),\s'})$ is 
\bea&& g^{\b,L}_{\s,\s'}(t-t',\vec x-\vec x')=\media{{\bf T}\psi^-_{(t,\vec x),\s}\psi^+_{(t',\vec x'),\s'}}_{\b,L}^{(0)}
\\
&&=
\frac1{L^2}\sum_{\vec k\in\BBB_L}e^{-i\vec k(\vec x-\vec x')}\Big[e^{-(t-t')(\hat H^{(0)}(\vec k)-\mu)}\Big(\frac{\mathds 1(t>t')}{1+e^{-\b(\hat H^{(0)}(\vec k)-\mu)}}-
\frac{\mathds 1(t\le t')\, e^{-\b(\hat H^{(0)}(\vec k)-\mu)}}{1+e^{-\b(\hat H^{(0)}(\vec k)-\mu)}}\Big)\Big]_{\s,\s'}\;.\nonumber
\eea
In the following, we denote by  $g^{\b,L}(t,\vec x)$ the matrix whose elements are 
$g^{\b,L}_{\s,\s'}(t,\vec x)$.
Note that, if $0<t<\beta$, then $g^{\b,L}(t-\beta,\vec x)=-g^{\b,L}(t,\vec x)$. 
Therefore, it is natural to extend $g^{\b,L}(t,\vec x)$, which is a priori defined only 
on the time interval $(-\beta,\b)$, to the whole real line, by anti-periodicity in the imaginary time, 
i.e., via the rule 
$g^{\b,L}(t+n\b, \vec x)=(-1)^ng^{\b,L}(t,\vec x)$. The resulting 
extension can be expanded in Fourier series w.r.t. $t$, so that, for all $t\neq n\beta$,
\be\label{eq:hatg_bis}
g^{\beta,L}(t,\vec x) = \frac{1}{\beta L^{2}}\sum_{\substack {k_{0}\in\BBB_\b \\ \vec k\in \mathcal{B}_{L}}} e^{-i\vec k\cdot \vec x-ik_0 t}\, \hat g^{\beta,L}(k_0,\vec k)\ee
with $\mathcal B_\b=\frac{2\pi}{\beta}(\mathbb Z+ \frac{1}{2})$ and 
\be 
\hat g^{\beta,L}(k_0,\vec k) := \frac{1}{-ik_{0} + \hat H^{(0)}(\vec k) - \mu}\;.
\ee
If, instead, $t= n\beta$, then $g^{\b,L}(n\b,\vec x)=(-1)^n\lim_{t\to 0^-}g^{\b,L}(t,\vec x)$.
Note that, by the very definition of the propagator and the canonical anti-commutation relations, 
$g^{\b,L}_{\s,\s'}(0^+,\vec x)-g^{\b,L}_{\s,\s'}(0^-,\vec x)=\d_{\vec x,\vec 0}\d_{\s,\s'}$, so that 
the only discontinuity points of $g^{\b,L}(t,\vec x)$ are $(n\b,\vec 0)$.

In the following we will also need a variant of $g^{\beta,L}(t,\vec x)$, to be denoted by 
$\bar g^{\beta,L}(t,\vec x)$, which coincides with $g^{\beta,L}(t,\vec x)$, $\forall (t,\vec x)\neq (n\b,\vec 0)$, and 
with the arithmetic mean of $g^{\b,L}(0^+,\vec 0)$ and $g^{\b,L}(0^-,\vec 0)$ at the discontinuity points:
\be \bar g^{\beta,L}(\xx)\big|_{\xx=(n\b,\vec x)}=
\frac{g^{\b,L}(0^+,\vec 0)+ g^{\b,L}(0^-,\vec 0)}2.
\ee
The function $\bar g^{\b,L}(\xx)$ is a natural object to introduce, in that it is the limit as $M\to\infty$ of a regularization of $g^{\b,L}(\xx)$
obtained by cutting off the ultraviolet modes $|k_0|>2^M$ in the right side of \eqref{eq:hatg_bis}. More specifically, if we 
take a smooth even compact support function $\chi_0(t)$,
equal to $1$ for $|t|<1$ and equal to $0$ for $|t|>2$, and we define
\be
\bar g^{\beta,L,M}(\xx) = \frac{1}{\beta L^{2}}\sum_{\kk\in \BBB_\b\times\BBB_L} e^{-i\kk\cdot\xx}\chi_0(2^{-M}k_0/\d_\m)\hat g^{\beta,L}(\kk),
\ee
then 
\be \bar g^{\b,L}(\xx)=\lim_{M\to\infty}\bar g^{\b,L,M}(\xx).\ee
These propagators can be used to re-express the formal perturbation theory in \eqref{eq:g2} in terms of the limit 
of a regularized theory with finitely many degrees of freedom, which is advantageous for performing rigorous bounds on the 
convergence of the series. More precisely, we note that \eqref{eq:g2},
as an identity between (a priori formal) power series, can be equivalently 
rewritten as 
\be \frac{\Tr_{\mathcal F}e^{-\beta(\mathcal H_L-\mu\mathcal N_L)}}{\Tr_{\mathcal F}e^{-\beta(\mathcal H^{(0)}_L-\mu\mathcal N_L)}}=
\lim_{M\to\infty}\Big[
1+\sum_{n\ge 1}\frac{(-U)^n}{n!}\int_0^\b dt_1\cdots \int_0^{\b}dt_n\, \bar{\mathbb E}_{\b,L,M}
\Big(\bar\VV_L(t_1)\cdots\bar\VV_L(t_n)\Big)\Big],\label{eq:g9}
\ee
where 
\be \bar\VV_L(t)=\sum_{\vec x,\vec y\in \L_L}\sum_{\s,\s'\in I}\Big(\psi^+_{(t,\vec x),\s}\psi^-_{(t,\vec x),\s}+\frac12\Big)v_{\s\s'}(\vec x-\vec y)
\Big(\psi^+_{(t,\vec y),\s'}\psi^-_{(t,\vec y),\s'}+\frac12\Big)\label{eq:barv}
\ee
and $\bar{\mathbb E}_{\b,L,M}(\cdot)$ acts linearly on normal-ordered polynomials in $\psi^\pm_{(t,\vec x),\s}$, the action 
on a normal-ordered monomial being defined by the fermionic Wick rule with propagator $$\bar{\mathbb E}_{\b,L,M}(\psi^-_{(t,\vec x),\s}\psi^+_{(t',\vec x'),\s'})=\bar g^{\b,L,M}_{\s,\s'}(t-t',\vec x-\vec x').$$
In order to check that the right side of \eqref{eq:g9} coincides order by order with the right side of \eqref{eq:g2}, it is enough to note the following (assume, again without loss of generality, that 
the times $t_1,\ldots, t_n$ are all distinct):
\begin{itemize}
\item all the pairings contributing to $\media{{\bf T}\VV_L(t_1)\cdots\VV_L(t_n)}_{\b,L}^{(0)}$ without {\it tadpoles} (i.e., without contractions of two fields 
at the same space-time point) give the same contribution as the corresponding pairing in 
$\lim_{M\to\infty}\bar{\mathbb E}_{\b,L, M}
\Big(\bar\VV_L(t_1)\cdots\bar\VV_L(t_n)\Big)$, simply because $g^{\b,L}(\xx)=\bar g^{\b,L}(\xx)$, $\forall \xx\neq (\b n, \vec 0)$;
\item in the pairings contributing to $\media{{\bf T}\VV_L(t_1)\cdots\VV_L(t_n)}_{\b,L}^{(0)}$ that contain tadpoles, every tadpole 
corresponds to a factor $\media{\psi^+_{(t,\vec x),\s}\psi^-_{(t,\vec x),\s}}^0_{\b,L}=-g^{\b,L}_{\s,\s}(0^-,\vec 0)$, while the corresponding 
tadpole in $\lim_{M\to\infty}\bar{\mathbb E}_{\b,L, M}\Big(\bar\VV_L(t_1)\cdots\bar\VV_L(t_n)\Big)$ contributes a factor 
$$\lim_{M\to\infty}\bar{\mathbb E}_{\b,L, M}\Big(\psi^+_{(t,\vec x),\s}\psi^-_{(t,\vec x),\s}\Big)=-\bar g^{\b,L}_{\s,\s}(0,\vec 0)=
-\frac12\big[ g^{\b,L}_{\s,\s}(0^+,\vec 0))+g^{\b,L}_{\s,\s}(0^-,\vec 0)\big].$$ The difference between the two is $$-\bar g^{\b,L}_{\s,\s}(0,\vec 0)+
g^{\b,L}_{\s,\s}(0,\vec 0)=-\frac12\big[ g^{\b,L}_{\s,\s}(0^+,\vec 0))-g^{\b,L}_{\s,\s}(0^-,\vec 0)\big]=-\frac12,$$ which is compensated exactly by the  $+\frac12$'s 
appearing in the definition \eqref{eq:barv}.
\end{itemize}

A concise way of rewriting the series in brackets in \eqref{eq:g9} is in terms of Grassmann integrals:
\be 1+\sum_{n\ge 1}\frac{(-U)^n}{n!}\int_0^\b dt_1\cdots \int_0^{\b}dt_n\, \bar{\mathbb E}_{\b,L, M}
\Big(\bar\VV_L(t_1)\cdots\bar\VV_L(t_n)\Big)=\int P_{\le M}(d\Psi)e^{-UV_{\b,L}(\Psi)},
\label{eq:g11}\ee
where $V_{\b,L}(\Psi)$ and $\int P_{\le M}(d\Psi)$ are, respectively, an element of a finite Grassmann algebra, and a linear map from the even part of the same algebra to the real numbers,
defined as follows. 
Let $\BBB_\b^*=\BBB_\b \cap\{k_0\,:\,\c_0(2^{-M}k_0)>0\}$, with $\BBB_\b$ defined after 
\eqref{eq:hatg_bis}, and $\BBB_{\b,L}^*=\BBB_\b^*\times\BBB_L$.
We consider the finite
Grassmann algebra generated by the Grassmann variables
$\{\hat\Psi^\pm_{\kk,\s}\}_{ \kk \in
\BBB_{\b,L}^*}^{\s\in I}$ and we let 
\be V_{L,\b}(\Psi)
=\sum_{\substack{\vec x,\vec y\in\L_L\\ \s,\s'\in I}}\int_{0}^\b dt \,
\Big(\Psi^+_{(t,\vec x),\s}\Psi^-_{(t,\vec x),\s}+\frac12\Big)v_{\s\s'}(\vec x-\vec y)
\Big(\Psi^+_{(t,\vec y),\s'}\Psi^-_{(t,\vec y),\s'}+\frac12\Big),
\ee
where 
\be\Psi_{\xx,\s}^{\pm}=\frac{1}{\b L^2}\sum_{\kk\in\BBB_{\b,L}^*}
e^{\pm i\kk\xx}\hat\Psi^\pm_{\kk,\s}\;.
\label{2.6}\ee
Moreover, $\int P_{\le M}(d\Psi)$ acts on a generic even monomial in the Grassmann variables as follows: 
it gives non zero only if the number of $\hat\Psi^+_{\kk,\s}$ variables is the same as the number
of $\hat \Psi^-_{\kk,\s}$ variables, in which case
\be \int P_{\le M}(d\Psi) \hat \Psi^-_{\kk_1,\s_1}\hat \Psi^+_{\pp_1,\s_1'}\cdots
\hat \Psi^-_{\kk_m,\s_m}\hat \Psi^+_{\pp_m,\s_m'}=\det[C(\kk_i,\s_i;\pp_j,\s_j')]_{i,j=1,\ldots,m},
\ee
where $C(\kk,\s;\pp,\s')=\b L^2 \d_{\kk,\pp}\c_0(2^{-M}k_0/\d_\m)\hat g^{\b,L}_{\s,\s'}(\kk)$.
In particular, 
\be \int P_{\le M}(d\Psi)\Psi_\xx^-\Psi^+_\yy=\bar g^{\b,L,M}(\xx-\yy).\label{eq:g14.b}\ee
If needed, $\int P_{\le M}(d\Psi)$ can be written explicitly in terms of the usual Berezin integral $\int d\Psi$,
which is the linear functional on the Grassmann algebra acting non trivially on a monomial 
only if the monomial is of maximal degree, in which case 
$$\int d\Psi \prod_{\kk\in\BBB_{\b,L}^*}
\prod_{\s\in I}\hat \Psi^-_{\kk,\s}\hat \Psi^+_{\kk,\s}=1.$$ The explicit expression of 
$\int P_{\le M}(d\Psi)$ in terms of $\int d\Psi$ is
\bea &&\int P_{\le M}(d\Psi)\big(\cdot\big) = \frac1{N_{\b,L,M}}\int d\Psi \exp\Big\{-\frac1{\b L^2}
\sum_{\kk\in\BBB_{\b,L}^*}\c_0^{-1}(2^{-M}k_0)
\hat\Psi^{+}_{\kk,\cdot}\big[{\hat g}_\kk^{\b,L}\big]^{-1}\hat\Psi^{-}_{\kk,\cdot}\Big\}\big(\cdot\big),
\nonumber\\
&& \quad {\rm with}\qquad N_{\b,L,M}=\prod_{\kk\in\BBB_{\b,L}^*}[\b L^2 \c_0(2^{-M}k_0/\d_\m)]^{|I|}
\det {\hat g}_\kk^{\b,L}\;,
\label{2.3}\eea
which motivates the appellation ``Gaussian integration'' that is usually given to the reference 
``measure'' $P_{\le M}(d\Psi)$. Because of \eqref{eq:g14.b}, $P_{\le M}(d\Psi)$ is also called the Gaussian integration with propagator $\bar g^{\b,L,M}$.

It is straightforward to check that the definitions above are given in such a way that the two sides 
of \eqref{eq:g11} coincide, order by order in $U$. Note, by the way, that \eqref{eq:g11} 
is a (finite) polynomial in $U$, for every finite $\b,L,M$, simply because the 
Grassmann algebra entering the definition of the right side of \eqref{eq:g11} is finite. 

Summarizing,
\be \frac{\Tr_{\mathcal F}e^{-\beta(\mathcal H_L-\mu\mathcal N_L)}}{\Tr_{\mathcal F}
e^{-\beta(\mathcal H^{(0)}_L-\mu\mathcal N_L)}}=\lim_{M\to\infty}\int P_{\le M}(d\Psi)e^{-UV_{\b,L}(\Psi)},\label{g16}\ee
as an identity between (a priori formal) power series in $U$. In a similar way, one can show (details left to the reader)
that the power series expansion for the truncated multipoint bond current-density correlations can be rewritten as 
\bea && \media{{\bf T}\big(J_{\vec x_1\,\vec y_1}^{\s_1\s'_1}\big)_{t_1};\cdots;\big(J_{\vec x_m\,\vec y_m}^{\s_m\s_m'}\big)_{t_m};n_{\xx_{m+1}}^{\s_{m+1}};\cdots;n_{\xx_{m+n}}^{\s_{m+n}}
}_{\b,L}=\label{g17}\\
&&\qquad =\lim_{M\to\infty}\frac{\partial^{m+n}}{\partial A_{\xx_1,\vec y_1}^{\s_1\s_1'}\cdots \partial\phi^{\s_{m+n}}_{\xx_{m+n}}}
\log\int P_{\le M}(d\Psi) e^{-UV_{\b,L}(\Psi)+(\phi,n)+(A,J)}\Big|_{A=\phi=0}\;,\nonumber
\eea
where 
\bea&& (\phi,n)=\int_0^\b dt \sum_{\vec x,\s}\phi^\s_{(t,\vec x)}\big(
\Psi^+_{(t,\vec x),\s}\Psi^-_{(t,\vec x),\s}+\frac12\big)\;,\\
&& (A,J)=\int_0^\b dt \sum_{\vec x,\vec y}\sum_{\s,\s'}A_{(t,\vec x),\vec y}^{\s\s'}\big[i\Psi^+_{(t,\vec x),\s}H^{(0)}_{\s\s'}(\vec x-\vec y)
\Psi^-_{(t,\vec y),\s'}-i\Psi^+_{(t,\vec y),\s'}H^{(0)}_{\s'\s}(\vec y-\vec x)\Psi^-_{(t,\vec x),\s}\big]\;.\nonumber
\eea
The goal of the incoming discussion is to show that \eqref{g16} and \eqref{g17} are not just identities between formal power series, 
but rather between analytic functions of $U$. Recalling the connection between the total current and the bond current, \eqref{eq:Ji}, it is clear that this will in turn implies the same 
for the multipoint (total) current-density correlations. Therefore, from now on, we shall restrict our attention to the bond (rather than total) current-density correlations. 

In order to prove that  that \eqref{g16} and \eqref{g17} are identities between analytic functions, 
it actually suffices to prove the uniform analyticity in $M$, as $M\to\infty$, 
and the existence of the limit as $M\to\infty$ of the regularized free energy per site and correlations, 
as the following elementary lemma shows. 

\begin{lemma}\lb{p2.1}
Assume that, for any finite $\b$ and $L$, there exists $\e_{\b,L}>0$ such that 
the regularized free energy per site
\be f_{\b,L,M}=-\frac1{\b L^2}\log \int P_{\le M}(d\Psi)e^{-U V_{\b,L}(\Psi)}\ee
and the regularized truncated correlations
\bea && K^{\b,L,M}(\xx_1,\vec y_1,\s_1,\s_1';\ldots;\xx_{m+n},\s_{m+n})=\\
&&\qquad =\frac{\partial^{m+n}}{\partial A_{\xx_1,\vec y_1}^{\s_1\s_1'}\cdots \partial\phi^{\s_{m+n}}_{\xx_{m+n}}}
\log\int P_{\le M}(d\Psi) e^{-UV_{\b,L}(\Psi)+(\phi,n)+(A,J)}\Big|_{A=\phi=0}\nonumber\eea
are analytic functions of $U$ in the domain $D_{\b,L}=\{U\in\mathbb C :\ |U|< \e_{\b,L}\}$, uniformly in $M$ as $M\to\infty$. Moreover, 
assume that in any compact subset of $D_{\b,L}$ the sequences 
$\{f_{\b,L,M}\}_{M\ge 1}$ and $\{K^{\b,L,M}(\xx_1,\vec y_1,\s_1,\s_1';\ldots;\xx_{m+n},\s_{m+n})\}_{M\ge 1}$ converge uniformly as $M\to\infty$.  
Then \eqref{g16} and \eqref{g17} are valid as identities between analytic functions of $U$ in $D_{\b,L}$.
\end{lemma}

\begin{oss} In the following we will prove the assumption of this lemma, and actually much more:
namely, we will prove the analyticity 
of $f_{\b,L,M}$ and $K^{\b,L,M}(\xx_1,\vec y_1,\s_1,\s_1';\ldots;\xx_{m+n},\s_{m+n})$, uniformly in $\b,L,M$ (not just in $M$). We will also prove that 
these functions converge not only as $M\to\infty$,
but also as $L\to\infty$ and $\b\to\infty$, which in turn implies that the limiting correlations 
in the thermodynamic and zero temperature limits are analytic as well, as claimed in Proposition \ref{prp:analyticity}. 
\end{oss}

\0 {\bf Proof of Lemma \ref{p2.1}.} 
Let us start by proving (\ref{g16}), which is equivalent to
\be  \frac{\Tr_{\mathcal F}e^{-\beta(\mathcal H_L-\mu\mathcal N_L)}}{\Tr_{\mathcal F}
e^{-\beta(\mathcal H^{(0)}_L-\mu\mathcal N_L)}}=\lim_{M\to \infty} e^{-\b L^2f_{\b,L,M}}\;.\label{g21}\ee
The first key remark is that, if $\b,L$ are finite, the left side of this equation is an entire function of $U$, as
it follows from the fact that the Fock space generated by the fermion
operators $\psi^\pm_{\vec x,\s}$, with $\vec x\in\L_L$, $\s\in I$, is finite dimensional.
On the other hand, by assumption, $f_{\b,L,M}$ is analytic in 
$D_{\b,L}$ and uniformly 
convergent as $M\to\io$ in every compact subset of $D_{\b,L}$.
Hence, by Vitali's convergence theorem for analytic functions, the limit 
$f_{\b,L}=\lim_{M\to\io}f_{\b,L,M}$ is analytic in $D_{\b,L}$
and its Taylor coefficients coincide with the limits as $M\to\io$
of the Taylor coefficients of $f_{\b,L,M}$. Moreover, by construction, as discussed after \eqref{eq:g9},
the Taylor coefficients of $e^{-\b L^2 f_{\b,L}}$ 
coincide with the Taylor coefficients of the left side of \eqref{g21},
which implies the validity of \eqref{g21} as an identity between analytic functions in $D_{\b,L}$,
simply because the left side is entire
in $U$, the right side is analytic in $D_{\b,L}$ and the Taylor coefficients
at the origin of the two sides are the same. By taking the logarithm at both sides, 
we also find that
$$f_{\b,L}=-\frac1{\b L^2}\log  \frac{\Tr_{\mathcal F}e^{-\beta(\mathcal H_L-\mu\mathcal N_L)}}{\Tr_{\mathcal F}
e^{-\beta(\mathcal H^{(0)}_L-\mu\mathcal N_L)}}$$ 
as an identity between analytic functions in $D_{\b,L}$. In particular, the left side of \eqref{g21}
does not vanish on $D_{\b,L}$.

In order to prove the analogous claim for the correlation functions, we note that 
the truncated correlations
$\media{{\bf T}\big(J_{\vec x_1\,\vec y_1}^{\s_1\s_1'}\big)_{t_1};\cdots;
\big(J_{\vec x_m\,\vec y_m}^{\s_m,\s_m'}\big)_{t_m};n_{\xx_{m+1}}^{\s_{m+1}};
\cdots;n_{\xx_{m+n}}^{\s_{m+n}}
}_{\b,L}$ are linear combination of ratios of entire functions, simply because they are 
linear combinations of products of non-truncated functions, each of which is a ratio of entire functions.
The denominator in these ratios is proportional to a power of the left side of \eqref{g21} that, as 
observed earlier, does not vanish on $D_{\b,L}$. Therefore, the truncated correlations are 
analytic in $D_{\b,L}$, which allow us to repeat the same argument used above for the free energy, to 
conclude the validity of \eqref{g17} as well, as an identity between analytic functions in $D_{\b,L}$.

\subsection{Uniform analyticity of the regularized correlation functions}
\label{appG.2}

In this section, we prove the uniform analyticity of the regularized free energy per site and regularized correlations,
in a domain $D$ independent not only of $M$, but also of $\beta,L$. Later, we will discuss the 
existence of the limit as $M,L,\b\to\infty$ of the regularized functions, thus proving the assumptions of 
Lemma \ref{p2.1}, as well as the existence and analyticity of the infinite volume and zero temperature limits. 
Throughout the proof, $C, C_i, c, c_i, \ldots,$ stand for unspecified constants, independent of $\b,L,M$ and of $\d_\m$, unless specified otherwise.
The key result proved in this section is the following. 

\begin{lemma}\label{lemg.2}
There exists $\e_0=\e_0(\d_\m)>0$ such that the regularized free energy $f_{\b,L,M}$ and correlations $K^{\b,L,M}(\xx_1,\vec y_1,\s_1,\s_1';\ldots;\xx_{m+n},\s_{m+n})$ are 
analytic in the common analyticity domain $D_0=\{U:\ |U|\le \e_0\}$.
Moreover, the regularized correlations are translation invariant and they satisfy the cluster property with faster-than-any-power decay rate, i.e., 
for any collection of integers
$\underline m=\{m_{i,j},m_k\}_{k=1,\ldots,m}^{i,j=1,\ldots,m+n}\ge 0$, there exists a constant $C_{\underline m}=
C_{\underline m}(\d_\m)$ such that 
\be\frac1{\b L^2}\int _{\L_{\b,L}^{m+n}}\!\!\!d\underline\xx\sum_{\underline{\vec y}\in \L_L^m}\big|K^{\b,L,M}(\xx_1,\vec y_1,\s_1,\s_1';\ldots;\xx_{m+n},\s_{m+n})\big|\,
d_{\underline m}(\underline \xx,\underline{\vec y})\le C_{\underline m}.
\label{G9}\ee
Here $\underline \xx=\{\xx_1,\ldots,\xx_{m+n}\}$, $\underline{\vec y}=\{\vec y_1,\ldots,\vec y_m\}$, $\L_{\b,L}=(0,\beta)\times\L_L$,
$\int_{\L_{\b,L}}\!\!d\xx$ is a shorthand for $\int_0^\b dx_0\sum_{\vec x\in\L_L}$, and $d_{\underline m,\underline m'}(\underline\xx,\underline{\vec y})=\prod_{i,j=1}^{m+n}|\xx_i-\xx_j|^{m_{i,j}}\prod_{k=1}^m|\vec y_k-\vec x_k|^{m_k}_L$,
where, if $|x_0|_\b=\min_{n\mathbb Z}|x_0 +n\beta|$ is the distance on the one-dimensional torus of
size $\b$ and $|\vec x|_L$ is the distance on the torus $\L_L$, we denoted $|\xx|=\mathfrak e_0|x_0|_\b+|\vec x|_L$, with $\mathfrak e_0$ the energy scale defined in \eqref{eq:h0}.

\end{lemma}

\0 {\bf Proof of Lemma \ref{lemg.2}.} 
The proof is long and, therefore, we split it into three main steps: we first define the multiscale decomposition of the Grassmann integral, which 
we intend to perform in an iterative fashion; next, we explain in detail how to integrate the first scale; finally, we explain the iterative procedure, whose 
output is conveniently organized in the form of a tree expansion.  

\medskip

{\it Multiscale decomposition.}
In order to prove the analyticity of the regularized free energy and correlations, 
we perform the Grassmann integration in a multiscale fashion, by rewriting the propagator $\bar g^{\b,L,M}$ as a sum of 
smooth ``single scale'' propagators $g^{(h)}$, $h=0,1,\ldots, M$, each decaying faster than any power on a 
specific time scale $\sim 2^h$:
\be\label{eq:g23}
\bar g^{\beta,L,M}(\xx) = \sum_{h=0}^{M} g^{(h)}(\xx),\qquad g^{(h)}(\xx)=\frac{1}{\beta L^{2}}\sum_{\kk\in \BBB_{\b,L}^*} e^{-i\kk\cdot\xx} \,\frac{f_h(k_0)}{-ik_{0} + \hat H^{(0)}(\vec k) - \mu}.
\ee
Here $f_h(k_0)=\c_0(2^{-h}k_0/\d_\m)-\c_0(2^{-h+1}k_0/\d_\m)$ for $h\ge 1$ and $f_{0}(k_0)=\chi_0(k_0/\d_\m)$. For later use, note that 
the single scale propagator $g^{(h)}(\xx)$ satisfies the bound
\be |g^{(h)}(\xx)|\le \frac{C_K}{1+ (2^h \d_\m|x_0|_\b + (\d_\m/\mathfrak e_0)|\vec x|_L)^K}
\;,\lb{B.3}\ee
for all $h,K$ such that $ 0\le h\le M$, $K\ge 0$. In particular, 
\be \|g^{(h)}\|_{1,n}:=\int d\xx\, \|g^{(h)}(\xx)\|\cdot|\xx|^n\le C_n\d_\m^{-3-n}2^{-h}. \lb{cc1}\ee
where $\int d\xx \equiv\int_{\L_{\b,L}} d\xx$ is a shorthand for $\int_0^\b dx_0\sum_{\vec x\in\L_L}$.
If $n=0$, we shall denote $\|g^{(h)}\|_{1}=\|g^{(h)}\|_{1,0}$.
Moreover, $g^{(h)}(\xx)$ admits a Gram decomposition, which will be useful in deriving combinatorially optimal bounds on the generic order 
of perturbation theory: 
\be g^{(h)}_{\s_1,\s_2}(\xx-\yy)=\big(A_{h,\xx,\s_1},B_{h,\yy,\s_2}\big)\equiv\sum_{\s'}\int d\zz\, A_{h,\xx,\s_1}^*(\zz,\s')\cdot B_{h,\yy,\s_2}(\zz,\s')\;,\label{eq:gram}\ee
with
\bea &&
A_{h,\xx,\s}(\zz,\s')=\frac1{\b L^2}\sum_{\kk\in\BBB^*_{\b,L}}  e^{i\kk(\xx-\zz)}\sqrt{f_h(k_0)}\Biggl[
\frac{1}{k_0^2+(\hat H_0(\vec k)-\m)^2}\Biggr]_{\s'\s}
\,,\nonumber\\
&&B_{h,\xx,\s}(\zz,\s')=\frac1{\b L^2}\sum_{\kk\in\BBB^*_{\b,L}} e^{i\kk(\xx-\zz)}\sqrt{f_h(k_0)}\big[ik_0+\hat H_0(\vec k)-\m\big]_{\s'\s} \;,\nonumber\eea
and
\be ||A_{h,\xx,\s}||^2:=(A_{h,\xx,\s},A_{h,\xx,\s})\le C(\d_\m 2^h)^{-3}\;,\qquad
||B_{h,\xx,\s}||^2\le C (\d_\m 2^h)^3\;.\lb{B.5}\ee
The decomposition \eqref{eq:g23} of the propagator allows us to compute the 
regularized Grassmann generating function, 
\be \mathcal W_{M}(\phi,A)=\log \int P_{\le M}(d\Psi) e^{-UV_{\b,L}(\Psi)+(\phi,n)+(A,J)}\;,\ee
in an iterative way, by first integrating 
the degrees of freedom corresponding to $g^{(M)}$, then those corresponding to $g^{(M-1)}$, and so on. 
Technically, we make use of the so-called addition formula for Grassmann Gaussian integrations: if $g_1,g_2$ are two propagators and $g:=g_1+g_2$, then the Gaussian integration $P_g(d\psi)$ with propagator $g$ can be 
rewritten as $P_g(d\psi)=P_{g_1}(d\psi_1)P_{g_2}(d\psi_2)$,
in the sense that for every polynomial $f$
\be 
\int P_g(d\psi) f(\psi)=\int P_{g_1}(d\psi_1)\int P_{g_2}(d\psi_2)
f(\psi_1+\psi_2)\;.\label{fonfo1}
\ee
In our context, we rewrite $P_{\le M}(d\Psi)=\prod_{h=0}^MP_h(d\Psi^{(h)})$, where  
$P_h(d\Psi^{(h)})$ is the Gaussian integration with propagator $g^{(h)}$, so that 
\be\lb{2za} e^{\WW_M(\phi,A)} = \int P_0(d\Psi^{(0)})
\cdots P_{h}(\Psi^{(h)}) e^{-\VV^{(h)}(\Psi^{(\le h)},\phi,A)}, \ee
where $\Psi^{(\le h)}:=\sum_{j=0}^h\Psi^{(j)}$, so that 
\be \VV^{(h)}(\Psi,\phi,A)
=-\log \int P_{h+1}(d\Psi^{(h+1)})
\cdots P_{M}(\Psi^{(M)}) e^{-U V_{\b,L}(\Psi+\Psi^{(h+1)}+\cdots +\Psi^{(M)})+(\phi,n)+(A,J)}.
\label{eq:5.32}\ee
and $\VV^{(M)}(\Psi,\phi,A)=UV_{\b,L}(\Psi)-(\phi,n)-(A,J)$.

\medskip

{\it The first integration step.}
In order to compute the sequence $\VV^{(h)}$ iteratively, let us start by explaining in detail the first step: 
\be \VV^{(M-1)}(\Psi,\phi,A)=-\log\int P_M(d\Psi^{(M)})e^{-\VV^{(M)}(\Psi+\Psi^{(M)},\phi,A)}\;.\label{eq:g34}
\ee
The logarithm in the right side
can be expressed as a series of truncated expectations:
\bea && \label{eq:g34bis}\log\int P_M(d\Psi^{(M)})e^{-\VV^{(M)}(\Psi+\Psi^{(M)},\phi,A)}=\\
&&\qquad =\sum_{s\ge 1}\frac{(-1)^s}{s!}\EE^T_M\big( \underbrace{\VV^{(M)}(\Psi+\Psi^{(M)},\phi,A);\cdots;
\VV^{(M)}(\Psi+\Psi^{(M)},\phi,A)}_{s\ {\rm times}} \big)\,,\eea
where
\be \EE_{M}^T(X_1(\Psi^{(M)});\cdots;X_s(\Psi^{(M)}))=\frac{\dpr^s}{\dpr\l_1\cdots\dpr\l_s}\log\int P_{M}
(d\Psi^{(M)})e^{\l_1X_1(\Psi^{(M)})+\cdots+\l_sX_s(\Psi^{(M)})}\big|_{\l_i=0}\label{2.trun}\,,\ee
and the $X_i$'s are all even elements of the Grassmann algebra generated by the field $\Psi^{(M)}$ 
we are integrating over and by the ``external'' Grassmann field $\Psi$.
The functional $\EE_{M}^T$ is multilinear in its arguments, the action on a collection of monomials being defined by the 
{\it truncated} Wick rule with propagator $g^{(M)}$, which, as already explained above, is similar to the usual fermionic Wick rule,
modulo the extra condition that, if the number $s$ of monomials involved is $\ge 2$, then the pairings one has to sum over 
are only those for which the collection of monomials $X_1$, \ldots, $X_s$
is {\it connected} (this means that for all $\mathcal I\subsetneq\{1,\ldots, s\}$, there exists at least one contracted pair involving one variable in 
the group $\{X_i\}_{i\in \mathcal I}$ and one in $\{X_i\}_{i\in \mathcal I^c}$).

A convenient representation of the truncated expectation, due to 
Battle, Brydges and Federbush \cite{BF,Brydges,BrF}, is the following (for a proof, see, e.g., \cite{GeM,Gi}). 
For a given (ordered) set of indices $P=(f_1,\ldots,f_{p})$, with 
$f_i=(\xx_i,\s_i,\e_i)$, let 
\be \Psi_P:=\Psi^{\e(f_1)}_{\xx(f_1),\s(f_1)}\cdots 
\Psi^{\e(f_{p})}_{\xx(f_{p}),\s(f_{p})}
\;,\label{s5.1}\ee
where $\xx(f_i)=\xx_i$, etc.
It is customary to represent each variable $\Psi^{\e(f)}_{\xx(f),\s(f)}$ as an oriented
half-line, emerging from the point $\xx(f)$ and carrying an arrow, pointing in the direction entering 
or exiting the point, depending on whether $\e(f)$ is equal to $-$ or $+$, respectively; moreover,
the half-line carries the labels $\s(f)\in I$.
Given $n$ sets of indices $P_1,\ldots,P_n$, we can enclose the points $\xx(f)$ belonging 
to the set $P_j$ in a box: in this way, assuming that all the points $\xx(f)$, 
$f\in\cup_iP_i$, are distinct, we obtain $n$ disjoint boxes. Given these definitions, if $\sum_{i=1}^s|P_i|$ is even we can write
\be\EE^T_M(\Psi_{P_1}; \ldots;\Psi_{P_s})=
\sum_{T\in{\bf T}_M}\a_T\prod_{\ell\in T}g_\ell^{(M)}
\int dP_{T}({\bf t}) \det G^{(M)}_{T}({\bf t})\;,\label{s5.3}\ee
where:\begin{itemize}
\item
any element $T$ of the set ${\bf T}_M={\bf T}_M(P_1,\ldots, P_s)$ is a set of lines forming an {\it anchored tree} between
the boxes $P_1,\ldots,P_{s}$, i.e., $T$ is a set
of lines that becomes a tree if one identifies all the points
in the same box; each line $\ell$ corresponds to a pair of half-lines indexed by 
two distinct variables $f,f'\in \cup_iP_i$ such that $\e(f)=-\e(f')$ (i.e., the directions of the two half-lines have to be compatible); if $\ell$ is obtained by contracting $f$ and $f'$, we shall write $\ell=(f,f')$, with the convention 
that $\e(f')=-\e(f)=+$.
\item $\a_T$ is a sign (irrelevant for the subsequent bounds), which depends on the choice of the anchored tree 
$T$;
\item if $\ell=(f,f')$, then $g_\ell^{(M)}$ stands for $g_{\s(f),\s(f')}^{(M)}(\xx(f)-\xx(f'))$;
\item if ${\bf t}=\{t_{i,i'}\in [0,1], 1\le i,i' \le n\}$, then $dP_{T}({\bf t})$
is a probability measure (depending on the anchored tree $T$) with support on a set of ${\bf t}$ such that
$t_{i,i'}=\uu_i\cdot\uu_{i'}$ for some family of vectors $\uu_i\in \RRR^s$ of
unit norm;
\item if $2N=\sum_{i=1}^s|P_{i}|$, then 
$G^{(M)}_{T}({\bf t})$ is a $(N-s+1)\times (N-s+1)$ matrix (depending both on the sets $P_i$ and on the anchored tree $T$), whose
elements are given by $[G^{(M)}_{T}({\bf t})]_{f,f'}=t_{i(f),i(f')}g^{(M)}_{(f,f')}$, where
$f, f'\in\cup_i P_i\setminus \cup_{\ell\in T}\{f^-_\ell,f^+_\ell\}$ (with $\ell=(f^-_\ell,f^+_\ell)$), and 
$i(f)\in\{1,\ldots,s\}$ is the index such that $f\in P_{i(f)}$.
\end{itemize}

If $s=1$ the sum over $T$ is empty, but we can still
use the Eq.(\ref{s5.3}) by interpreting the r.h.s.
as equal to $1$ if $P_{1}$ is empty and equal to $\det G^T({\bf 1})$ otherwise.

In order to use \eqref{s5.3} in \eqref{eq:g34}-\eqref{eq:g34bis}, we first rewrite 
$\VV^{(M)}$ as
\be  \label{eq:g39}\VV^{(M)}(\Psi,\phi,A)=E_M(\phi)+
\sum_{\r=1}^4\,\sum_{\s,\s'\in I}\int d\xx d\yy K^\r_{\s\s'}(\xx,\yy)
\big[\phi^\s_\xx\big]^{\d_{\r,1}}\big[A^{\s\s'}_{\xx,\yy}\big]^{\d_{\r,2}}
\Psi_{P^\r}\;,\ee
where $E_M(\phi)=\frac{\b L^2}4 U\sum_{\s}\n_\s-\frac12\sum_{\s}\int d\xx\, \phi^\s_\xx$, with  
$\n_\s=\sum_{\vec x\in\L_L}\sum_{\s'\in I}v_{\s\s'}(\vec x)$. Moreover, 
$A^{\s\s'}_{\xx,\yy}=A_{\xx,\vec y}^{\s\s'}-A_{\yy,\vec x}^{\s'\s}$, 
\bea && K^1_{\s\s'}(\xx,\yy)=-\d_{\s,\s'}\d(\xx-\yy)\;,\qquad 
K^2_{\s\s'}(\xx,\yy)=-i\d(x_0-y_0)H^{(0)}_{\s\s'}(\vec x-\vec y)\;,\label{eq:5.41}\\
&&K^3_{\s\s'}(\xx,\yy)=U\n_\s \d_{\s,\s'}\d(\xx-\yy)\;,\qquad 
K^4_{\s\s'}(\xx,\yy)=U\d(x_0-y_0)v_{\s\s'}(\vec x-\vec y)\;,\label{eq:5.42}\eea
and
\bea && P^1=P^2=P^3=\big((\xx,\s,+),(\yy,\s',-)\big)\;,\\
&&  P^4=\big((\xx,\s,+),(\xx,\s,-),(\yy,\s',+),(\yy,\s',-)\big)\;.\eea
Plugging \eqref{eq:g39} into \eqref{eq:g34}-\eqref{eq:g34bis}, 
we obtain
\bea && \VV^{(M-1)}(\Psi,\phi,A)=E_M(\phi)-
\sum_{s\ge 1}\frac{(-1)^s}{s!}\hskip-.5truecm\sum_{\substack{\r_1,\ldots,\r_s\\ 
\s_1,\s_1',\ldots,\s_s,\s_s'}}\hskip-.35truecm
\int d\xx_1d\yy_1\cdots d\xx_s\, d\yy_s\times \nonumber\\
&& \times
\big[\prod_{i\,:\,\r_i=1}\phi^{\s_i}_{\xx_i}\big]
\big[\prod_{i\,:\,\r_i=2}A^{\s_i\s_i'}_{\xx_i,\yy_i}\big]
\big[\prod_{i=1}^sK^{\r_i}_{\s_i\s_i'}(\xx_i,\yy_i)\big] 
\EE^T_M\big((\Psi+\Psi^{(M)})_{P^{\r_1}_1};\cdots;
(\Psi+\Psi^{(M)})_{P^{\r_s}_s}\big)\;.\nonumber
\eea
The truncated expectation in the right side can be further rewritten as
\be \EE^T_M\big((\Psi+\Psi^{(M)})_{P^{\r_1}_1};\cdots;
(\Psi+\Psi^{(M)})_{P^{\r_s}_s}\big)=\sum_{P\subseteq \cup_i P^{\r_i}_i}\a_P\Psi_P
\EE^T_M\big(\Psi^{(M)}_{P^{\r_1}_1\setminus Q_1};\cdots;
\Psi^{(M)}_{P^{\r_s}_s\setminus Q_s}\big)\;,\ee
where $\a_P$ is a sign, and $Q_i= P\cap P^{\r_i}_i$, so that,
applying \eqref{s5.3}, we find
\bea &&  \VV^{(M-1)}(\Psi,\phi,A)=E_M(\phi)-
\sum_{s\ge 1}\frac{(-1)^s}{s!}\sum_{\underline\r,\,\underline\s}\int d\underline\xx\,
d\underline\yy\,\big[\prod_{i\,:\,\r_i=1}\phi^{\s_i}_{\xx_i}\big]
\big[\prod_{i\,:\,\r_i=2}A^{\s_i\s_i'}_{\xx_i,\yy_i}\big] \times\nonumber\\
&&\qquad \times
\big[\prod_{i=1}^sK^{\r_i}_{\s_i\s_i'}(\xx_i,\yy_i)\big] 
\sum_{P\subseteq \cup_i P^{\r_i}_i}\!\!\Psi_P\sum_{T\in{\bf T}_M}
\a_{P,T}\prod_{\ell\in T}g_\ell^{(M)}\!
\int dP_{T}({\bf t}) \det G^{(M)}_{T}({\bf t})\;,\label{eq:5.47}\eea
where $\underline\r$, $\underline\s$, $\underline\xx$ and $\underline\yy$ are shorthands for $(\r_1,\ldots,\r_s)$, 
$(\s_1,\s_1',\ldots,\s_s,\s_s')$, $(\xx_1,\ldots,\xx_s)$ and $(\yy_1,\ldots,\yy_s)$, respectively,
and $\a_{P,T}=\a_P\a_T$. Eq.\eqref{eq:5.47} can be equivalently rewritten as 
\bea && \VV^{(M-1)}(\Psi,\phi,A)=E_M(\phi)+\label{eq:5.48}\\
&&+\sum_{n\ge 0}\,\sum_{s_1,s_2\ge 0}\sum_{\underline\s,\underline\e}
\int d\underline\xx d\underline\yy d\underline\zz\,  W^{(M-1)}_{2n,s_1,s_2,\underline\s,\underline\e}(\underline
\xx, \underline\yy,\underline\zz)
\Big[\prod_{i=1}^{s_1}\phi^{\s_i}_{\xx_i}\big]\,\big[\prod_{i=s_1+1}^{s_1+s_2}A^{\s_i\s_i'}_{\xx_i,\yy_i}\big]
\Big[\prod_{i=1}^{2n} \Psi^{\e_i}_{\zz_i,\s_i''}\big]\;,
\nonumber
\eea
with 
\bea&& W^{(M-1)}_{2n,s_1,s_2,\underline\s,\underline\e}(\underline
\xx, \underline\yy,\underline\zz)=\sum_{\substack{s_3\ge 0\\ s_4\ge n-1}}^*\frac{(-1)^{s
-1}}{s_1!s_2!s_3!s_4!}\sum_{\substack{\s_i,\s_i':\\ i>s_1+s_2}}
\int \big[\prod_{i>s_1+s_2}d\xx_{i} d\yy_i \big]\times\label{eq:5.48bis}\\
&&
\times
\big[\prod_{i=1}^{s}K^{\bar\r_i}_{\s_i\s_i'}(\xx_i,\yy_i)\big]
\sum_{\substack{P\subseteq \cup_i P^{\bar\r_i}_i:\\
|P|=2n}}\d(P-P_{ext})
\sum_{T\in{\bf T}_M}
\a_{P,T}\prod_{\ell\in T}g_\ell^{(M)}\!
\int dP_{T}({\bf t}) \det G^{(M)}_{T}({\bf t})\;,\nonumber
\eea
where $s=s_1+s_2+s_3+s_4$, 
the $*$ on the sum indicates the constraint that $s\ge 1$,
and $\bar\r_i$ is equal to 1 if $i\le s_1$, is equal to 2 if $0<i-s_1\le s_2$,
is equal to 3 if $0<i-s_1-s_2\le s_3$, and is equal to 4 otherwise. Moreover, $P_{ext}=((\zz_1,\s_1'',\e_1),\ldots,(\zz_{2n},\s_{2n}'',\e_{2n}))$, and $\d\big(P-P_{ext})$ is a shorthand for the product of delta functions $\prod_{f_i\in P}
\d(\xx(f_i)-\zz_i)\d_{\s(f_i),\s''_i}\d_{\e(f_i),\e_i}$, where the labeling $P=(f_1,\ldots,f_{2n})$
is understood. Note that, in the case that $n=s_1=s_2=0$, in the right side of \eqref{eq:5.48} 
there are neither sums over $\underline\s,\underline\e$ nor integrals over $\underline \xx,\underline\yy,\underline\zz$,
and $ W^{(M-1)}_{0,0,0}$ is a constant, given by \eqref{eq:5.48}, with the understanding that the meaningless factors or sums or integrals should be replaced by one. 

We are finally in the position of proving the analyticity of the integral kernels of $\VV^{(M-1)}$. By using \eqref{eq:5.48} we obtain 
\bea && \frac1{\b L^2}\int d\xx d\yy d\zz \big|W^{(M-1)}_{2n,s_1,s_2,\underline\s,\underline\e}(\underline
\xx, \underline\yy,\underline\zz)\big|\le\label{eq:5.49}\\
&&\qquad \le\sum_{\substack{s_3\ge 0\\ s_4\ge n-1}}^*\frac{|I|^{2s_3+2s_4}}{s_1!s_2!s_3!s_4!}
\Big[\prod_{j=1}^4||K^j||_1^{s_j}\Big]
{2s+2s_4\choose 2n} (C^s s!) ||g^{(M)}||_1\cdot||\det G^{(M)}_{T}||_\infty\;,\nonumber
\eea
where: 
$|I|^{2s_3+2s_4}$ bounds the number of terms in the sum over $\s_i,\s_i'$; 
$||K^j||_1=\sup_{\s,\s'}$ $\int d\xx|K^j_{\s\s'}(\xx,\V0)|$; ${2s+2s_4\choose 2n}$ bounds the number of terms in the sum over $P$; $(C^s s!)$ bounds the number of terms in the sum over $T$. 
Recalling \eqref{cc1} for $n=0$ and the definitions \eqref{eq:5.41}-\eqref{eq:5.42}, from which  
$||K^j||_1\le C|U|^{\d_{j,3}+\d_{j,4}}$, we find that \eqref{eq:5.49} implies 
\be \frac1{\b L^2}\int d\xx d\yy d\zz \big|W^{(M-1)}_{2n,s_1,s_2,\underline\s,\underline\e}(\underline
\xx, \underline\yy,\underline\zz)\big|\le\label{eq:5.50}
\sum_{\substack{s_3\ge 0\\ s_4\ge n-1}}^* C^s |U|^{s_3+s_4}
(\d_\m^{-3}2^{-M})^{s-1}||\det G^{(M)}_{T}||_\infty\;.
\ee
In order to bound $\det G_T^{(M)}$, we use the {\it Gram-Hadamard inequality}, stating
that, if $M$ is a square matrix with elements $M_{ij}$ of the form
$M_{ij}=(A_i,B_j)$, where $A_i$, $B_j$ are vectors in a Hilbert space with
scalar product $(\cdot,\cdot)$, then
\be |\det M|\le \prod_i ||A_i||\cdot ||B_i||\;.\label{s5.6}\ee
where $||\cdot||$ is the norm induced by the scalar product. In our case, $[G_T^{(M)}({\bf t})]_{f,f'}=
\uu_{i(f)}\cdot \uu_{i(f')} (A_{M,\xx(f),\s(f)},B_{M,\xx(f'),\s(f')})$, so that, using 
\eqref{B.5} and recalling that $G_T^{(M)}$ is a $(s_4-n+1)\times(s_4-n+1)$ matrix, 
\be ||\det G^{(M)}_{T}||_\infty\le C^{s_4-n+1}.\label{eq:5.51}\ee
Plugging this last ingredient into \eqref{eq:5.50}, we finally obtain
\bea \frac1{\b L^2}\int d\xx d\yy d\zz \big|W^{(M-1)}_{2n,s_1,s_2,\underline\s,\underline\e}(\underline
\xx, \underline\yy,\underline\zz)\big|&\le&
\sum_{\substack{s_3\ge 0\\ s_4\ge n-1}}^* C^s |U|^{s_3+s_4}
(\d_\m^{-3}2^{-M})^{s-1}\nonumber\\
&\le& C^{n}|U|^{[n-1]_+}(\d_\m^{-3}2^{-M})^{[s_1+s_2+n-2]_+}
\;,\label{eq:5.52}
\eea
where $[\cdot]_+=\max\{\cdot,0\}$ denotes the positive part. Eq.\eqref{eq:5.52} 
proves the analyticity of the kernels of $\VV^{(M)}$ for $U$ small enough, uniformly in $M$ 
(but not in $\d_\m$, in general). 

Moreover, the kernels $W^{(M-1)}_{2n,s_1,s_2,\underline\s,\underline\e}(\underline \xx, \underline\yy,\underline\zz)$ decay faster than any power, on scale $\d_\m^{-1}$, in the relative distances between the coordinates $\xx_i,\yy_i,\zz_i$.
In order to prove this, we multiply the argument of the integral in the left side of \eqref{eq:5.49}
by a product of factors of the form $|\xx_i-\xx_j|^{m_{i,j}}$, or $|\xx_i-\yy_j|^{m'_{i,j}}$, etc.
We denote by $m=\sum_{i,j}(m_{i,j}+m_{i,j}'+\cdots)$ the sum of these exponents.
Again, we use the representation \eqref{eq:5.48}, and we decompose each factor 
``along the anchored tree $T$'', that is we bound it by using  \be|\xx_i-\xx_j|\le 
\sum_{\ell\in T}|\xx(f^-_\ell)-\xx(f^+_\ell)|+\sum_{i=1}^s d_i,\label{eq:anc}\ee where 
$d_i=\max_{f,f'\in P_{T,i}}|\xx(f)-\xx(f')|$ and $P_{T,i}=\cup_{\ell\in T}\{f^-_\ell, f^+_\ell\}\cap 
P^{\bar\r_i}_i$. In this way, the right side of \eqref{eq:5.49} is replaced by a sum of terms, each 
of which is obtained by replacing some of the factors $||K^j||_1$ and $||g^{(M)}||_1$ by 
$||K^j||_{1,n_i}=\sup_{\s,\s'}$ $\int d\xx|K^j_{\s\s'}(\xx,\V0)|\,|\xx|^{n_i}\le C_{n_i}$ and by 
$||g^{(M)}||_{1,n'_i}$, 
respectively. Recall that, by \eqref{cc1}, the dimensional estimate of $||g^{(M)}||_{1,n'_i}$
differs from that of $||g^{(M)}||_{1}$ just by a factor $\d_\m^{-n'_i}$.
Moreover, the total sum of the 
exponents $n_i,n'_i$, etc., equals the exponent $m$ introduced earlier. 
Therefore, the product of the extra factors $\d_{\m}^{-n_i'}$ is smaller than $\d_\m^{-m}$.
All in all, the dimensional estimate on the  kernels $W^{(M-1)}_{2n,s_1,s_2,\underline\s,\underline\e}(\underline \xx, \underline\yy,\underline\zz)$, multiplied by the extra factors $|\xx_i-\xx_i|^{m_{i,j}}$, etc, 
is the same as \eqref{eq:5.52}, up to an extra factor $C_m\d_\m^{-m}$, for all $m\ge 0$.

\medskip

\begin{figure}[ht]
\includegraphics[height=.18\textwidth]{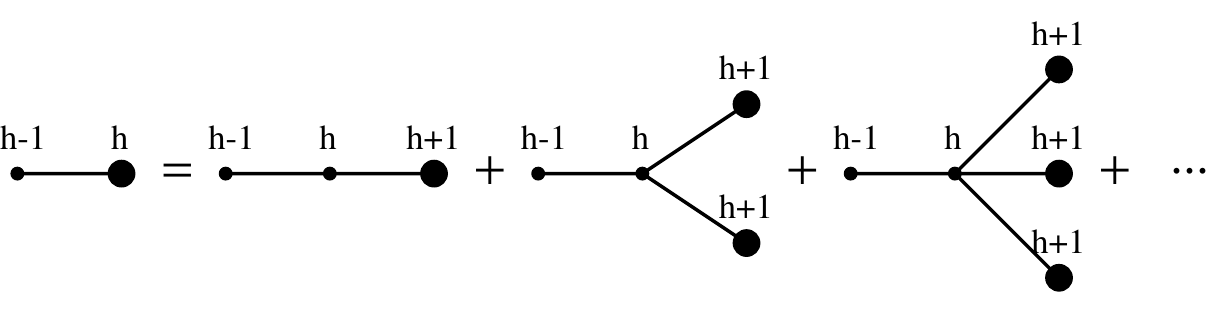}
\caption{The graphical representation of $\VV^{(h-1)}$.\label{tree_h}}
\end{figure}

{\it The iterative integration procedure and the tree expansion.}
We are now in the position of iterating the procedure used above for computing the integral over the scale $M$. By using \eqref{eq:5.32}
and the definition of truncated expectation $\EE^T_h$ (which is the same as \eqref{2.trun}, with $M$ replaced by $h$), we obtain 
\bea && \VV^{(h-1)}(\Psi,\phi,A)=-\log\int P_h(d\Psi^{(h)})e^{-\VV^{(h)}(\Psi+\Psi^{(h)},\phi,A)}=\label{eq:5.55}
\\
&&\hskip3.truecm =\sum_{s\ge 1}\frac{(-1)^s}{s!}\EE^T_h\big( \underbrace{\VV^{(h)}(\Psi+\Psi^{(h)},\phi,A);\cdots;
\VV^{(h)}(\Psi+\Psi^{(h)},\phi,A)}_{s\ {\rm times}} \big)\,.\nonumber
\eea
Eq.(\ref{eq:5.55}) can be graphically represented as in Fig.\ref{tree_h}.
The tree in the left side, consisting of a single horizontal branch, connecting the 
left node 
(called the {\it root} and associated with the {\it scale label} $h-1$)
with a big black dot on scale $h$, represents ${\cal V}^{(h-1)}$. In the right side, 
the term with $s$ final points represents the corresponding term in 
the right side of (\ref{eq:5.55}): a scale label 
$h-1$ is attached to the leftmost node (the root); a scale label 
$h$ is attached to the central node (corresponding to the action of $\EE^T_h$);
a scale label $h+1$ is attached to the $s$ rightmost nodes with the big black dots 
(representing $\VV^{(h)}$). 

\begin{figure}[ht]
\includegraphics[height=.5\textwidth]{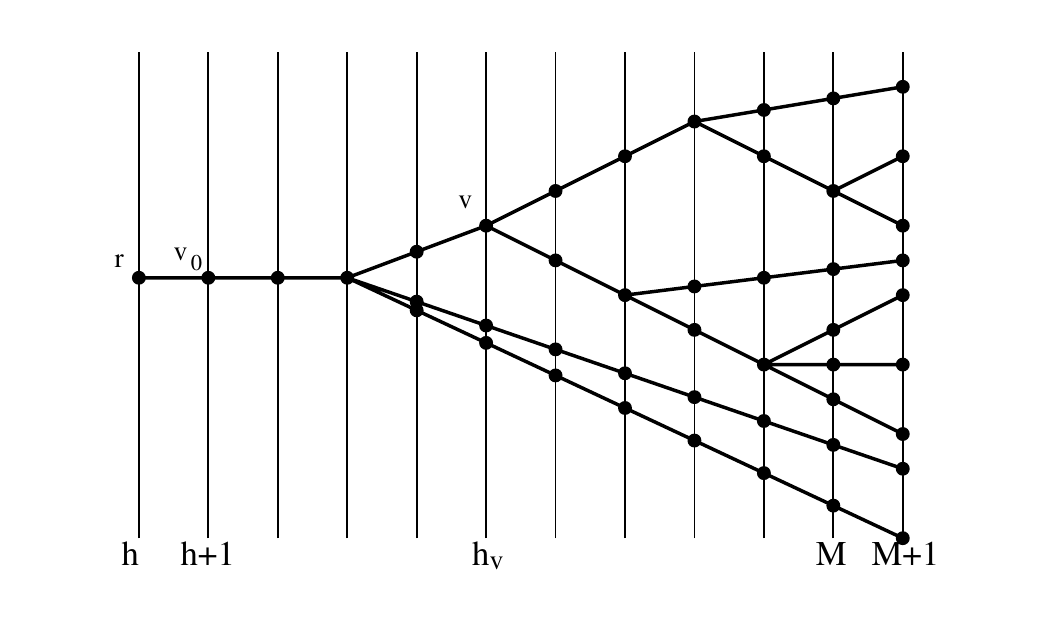}
\caption{A tree $\t\in\widetilde\TT_{M;h,N}$ with $N=9$: 
the root is on scale $h$ and the 
endpoints are on scale $M+1$.\label{fig6.2}}
\end{figure}

Iterating the graphical equation in Fig.\ref{tree_h} up to scale 
$M$, and representing the endpoints on scale $M+1$ as simple dots (rather than big black dots),
we end up with a graphical representation of $\VV^{(h)}$ in terms of {\it Gallavotti-Nicol\`o}
trees {\cite{GN,GaRMP}}, see Fig.\ref{fig6.2}, defined in terms of the following features.

\begin{enumerate}
\item Let us consider the family of all trees which can be constructed
by joining a point $r$, the {\it root}, with an ordered set of $N\ge 1$
points, the {\it endpoints} of the {\it unlabeled tree},
so that $r$ is not a branching point. $N$ will be called the
{\it order} of the unlabeled tree and the branching points will be called
the {\it non trivial vertices}.
The unlabeled trees are partially ordered from the root
to the endpoints in the natural way; we shall use the symbol $<$
to denote the partial order.
Two unlabeled trees are identified if they can be superposed by a suitable
continuous deformation, so that the endpoints with the same index coincide.
It is then easy to see that the number of unlabeled trees with $N$ end-points
is bounded by $4^N$ (see, e.g., \cite[appendix A.1.2]{GeM} for a proof of this fact).
We shall also consider the {\it labeled trees} (to be called
simply trees in the following); they are defined by associating
some labels with the unlabeled trees, as explained in the
following items.
\item We associate a label $0\le h\le M-1$ with the root and we denote by
$\widetilde\TT_{M;h,N}$ the corresponding set of labeled trees with $N$
endpoints. Moreover, we introduce a family of vertical lines,
labeled by an integer taking values in $[h,M+1]$, and we represent
any tree $\t\in\widetilde\TT_{M;h,N}$ so that, if $v$ is an endpoint, it is contained 
in the vertical line with index $h_v=M+1$, while if it is a non trivial vertex, 
it is contained in a vertical line with index
$h<h_v\le M$, to be called the {\it scale} of $v$; the root $r$ is on
the line with index $h$.
In general, the tree will intersect the vertical lines in set of
points different from the root, the endpoints and the branching
points; these points will be called {\it trivial vertices}.
The set of the {\it
vertices} will be the union of the endpoints, of the trivial
vertices and of the non trivial vertices; note that the root is not a vertex.
Every vertex $v$ of a
tree will be associated to its scale label $h_v$, defined, as
above, as the label of the vertical line whom $v$ belongs to. Note
that, if $v_1$ and $v_2$ are two vertices and $v_1<v_2$, then
$h_{v_1}<h_{v_2}$.
\item There is only one vertex immediately following
the root, called $v_0$ and with scale label equal to $h+1$.
\item Given a vertex $v$ of $\t\in\widetilde\TT_{M;h,N}$ that is not an endpoint,
we can consider the subtrees of $\t$ with root $v$, which correspond to the
connected components of the restriction of
$\t$ to the vertices $w\ge v$. If a subtree with root $v$ contains only
$v$ and one endpoint on scale $h_v+1$,
it is called a {\it trivial subtree}.
\item With each endpoint $v$ we associate one of the terms contributing to $\VV^{(M)}(\Psi,\phi,A)$,
see \eqref{eq:g39}. In order to distinguish between the various terms in the right side of \eqref{eq:g39},
we introduce a {\it type label} $\r_v\in\{0,1,2,3,4\}$.
If $\r_v=0$, then we associate the endpoint with a contribution $E_M(\phi)$, while, 
if $1\le \r_v\le 3$, then we associate the endpoint with a contribution 
$K^{\r_v}_{\s_v\s'_v}(\xx_v,\yy_v)
\big[\phi^{\s_v}_{\xx_v}\big]^{\d_{\r_v,1}}\big[A^{\s_v\s'_v}_{\xx_v,\yy_v}\big]^{\d_{\r_v,2}}
\Psi_{I_v}$. 
\end{enumerate}

The field labels attached to the endpoints $v$ of $\t$ are denoted by $I_v$. If $\r_v=0$, then 
$I_v=\emptyset$; if $\r_v=1,2,3$, then $I_v=\big((\xx_v,\s_v,+),(\yy_v,\s'_v,-)\big)$; if $\r_v=4$, then 
$I_v=\big((\xx_v,\s_v,+),(\xx_v,\s_v,-),(\yy_v,\s'_v,+),(\yy_v,\s'_v,-)\big)$. 
Moreover, given any vertex $v\in\t$, we denote by $I_v$ the set of field labels associated with the endpoints following the vertex $v$; given $f\in I_v$, $\xx(f)$, $\s(f)$ and $\e(f)$ denote the
space-time point, the $\s$ index and the $\e$ index of the Grassmann variable with label $f$.
In the following, the ``sum'' over the field labels associated with the endpoints should be understood as $\sum_{\underline\s_{v_0}}
\int d\underline\xx_{v_0}$, where $v_0$ is the leftmost vertex of $\t$, $\underline \s_{v}=\cup_{f\in I_v}\{\s(f)\}$ and 
$\underline \xx_{v}=\cup_{f\in I_v}\{\xx(f)\}$.

\medskip

In terms of trees, the effective potential ${\cal V}^{(h)}$,
$-1\le h\le M$ (with $\VV^{(-1)}$ identified with $\WW_M$), can be written as
\be {\cal V}^{(h)}(\Psi^{(\le h)}) =
\sum_{N=1}^\io\sum_{\t\in\widetilde\TT_{M;h,N}}
\widetilde{\cal V}^{(h)}(\t,\Psi^{(\le h)})\;,\label{B.12}\ee
where, if $v_0$ is the first vertex of $\t$ and $\t_1,\ldots,\t_s$
($s=s_{v_0}$) are the subtrees of $\t$ with root $v_0$,
$\widetilde{\cal V}^{(h)}(\t,\Psi^{(\le h)})$
is defined inductively as:
\be \widetilde{\cal V}^{(h)}(\t,\Psi^{(\le h)})=\frac{(-1)^{s-1}}{s!} \EE^T_{h+1}
\big[\widetilde{\cal V}^{(h+1)}(\t_1,\Psi^{(\le h+1)});\ldots;\widetilde{\cal V}^{(h+1)}
(\t_{s},\Psi^{(\le h+1)})\big]\;.\label{B.12a}\ee
where, if $\t$ is a trivial subtree with root on scale $M$, then $\widetilde{\cal V}^{(M)}(\t,\Psi^{(\le M)})={\cal V}^{(M)}(\Psi^{(\le M)})$ (for lightness of notation, we are dropping the arguments $(\phi,A)$, which are implicitly understood here and in the following).

\begin{figure}[ht]
\includegraphics[height=.5\textwidth]{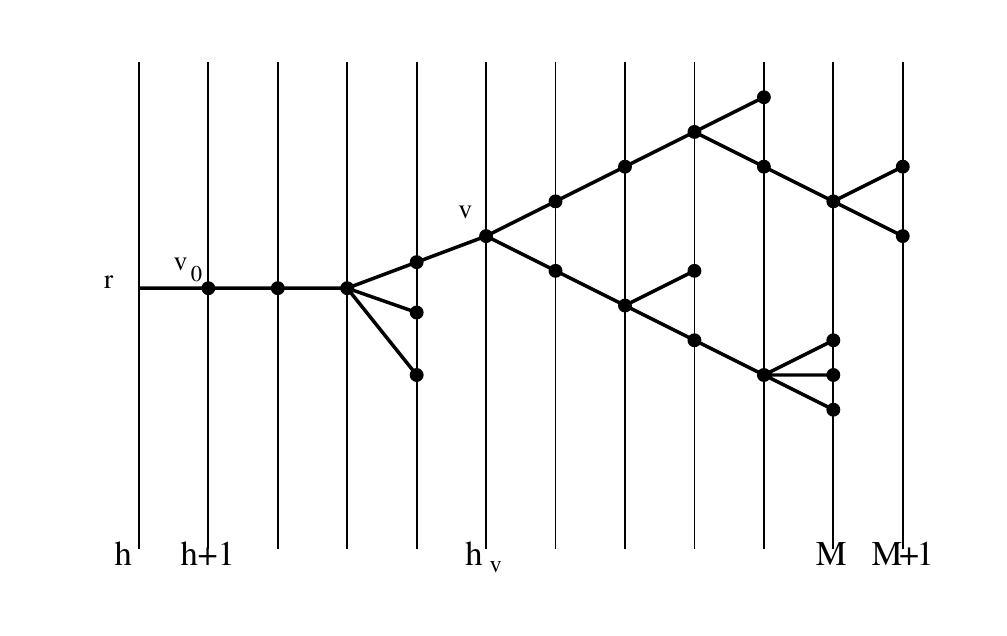}
\caption{A tree $\t\in\TT_{M;h,N}$ with $N=9$: the root is on scale $h$ and the 
endpoints are on scales $\le M+1$.\label{fig6.3}}
\end{figure}

For what follows, it is important to specify the action of the truncated expectations on the 
branches connecting any endpoint $v$ to the closest {\it non-trivial} vertex $v'$
preceding it. In fact, if $\t$ has only one end-point, 
it is convenient to rewrite $\widetilde\VV^{(h)}(\t,\Psi^{(\le h)})=\EE^T_{h+1}\EE^T_{h+2}\cdots
\EE^T_M(\VV(\Psi^{(\le M)}))\equiv \widetilde\VV^{(h)}(\Psi^{(\le h)})$ as: 
\be \widetilde\VV^{(h)}(\Psi^{(\le h)})=\VV^{(M)}(\Psi^{(\le h)})+
\EE^T_{h+1}\cdots\EE^T_M\big(\VV^{(M)}(\Psi^{(\le M)})-
\VV^{(M)}(\Psi^{(\le h)})\big)\;.\label{B.12b}\ee
The second term in the right side can be evaluated explicitly and gives:
\be \EE^T_{h+1}\cdots\EE^T_M\big(\VV^{(M)}(\Psi^{(\le M)})-
\VV^{(M)}(\Psi^{(\le h)})\big)= e_{[h+1,M]}+\sum_{\s,\s'}\int d\xx d\yy\,
k^{[h+1,M]}_{\s\s'}(\xx,\yy)\Psi^+_{\xx,\s}  \Psi^-_{\yy,\s'}\;,\label{B.12by}\ee
where, denoting $g^{[h+1,M]}(\xx)=\sum_{h'=h+1}^Mg^{(h')}(\xx)$, 
\bea &&e_{[h+1,M]}=e_{[h+1,M]}(\phi,A)=-\sum_{\s,\s'}\int d\xx d\yy\,\Big\{
\big[ K^1_{\s\s'}(\xx,\yy)\phi^\s_\xx + K^2_{\s\s'}(\xx,\yy)A^{\s\s'}_{\xx,\yy}+K^3_{\s\s'}(\xx,\yy)\big]\cdot
\nonumber\\ 
&&\cdot g^{[h+1,M]}_{\s'\s}(\V0)+K^4_{\s\s'}(\xx,\yy)\big[g_{\s\s'}^{[h+1,M]}(\xx-\yy)
g_{\s'\s}^{[h+1,M]}(\yy-\xx)-g_{\s\s}^{[h+1,M]}(\V0)g_{\s'\s'}^{[h+1,M]}(\V0)
\big]\Big\}\;,\nonumber\eea
and 
\be k^{[h+1,M]}_{\s\s'}(\xx,\yy)=2 U g_{\s\s'}^{[h+1,M]}((0,\vec x-\vec y))\d(x_0-y_0) \big[v_{\s\s'}(\vec x-\vec y)-
\n_\s\d_{\s\s'}\d(\vec x-\vec y)\big]\,.\label{eq:k}
\ee
Therefore, it is natural to shrink all the branches of $\t\in\widetilde\TT_{M;h,n}$
consisting of a subtree $\t'\subseteq\t$, having root $r'$ on scale $h'\in[h,M]$ 
and only one endpoint on scale $M+1$,  into a trivial subtree, rooted in $r'$ and associated 
with a factor $\widetilde\VV^{(h')}(\Psi^{(\le h')})$, which has the same structure 
as the right side of \eqref{eq:g39}, with $E_M(\phi)$ replaced by $E_{h'}(\phi,A)=E_M(\phi)+
e_{[h'+1,M]}(\phi,A)$, $K^3_{\s\s'}(\xx,\yy)$ replaced by 
$K^{3}_{h'+1;\s\s'}(\xx,\yy):=K^3_{\s\s'}(\xx,\yy)+k^{[h'+1,M]}_{\s\s'}(\xx,\yy)$, and 
$\Psi$ replaced by $\Psi^{(\le h')}$. Note that $k^{[h+1,M]}_{\s\s'}(\xx,\yy)$ is bounded proportionally to $U$, and decays faster 
than any power, {\it uniformly in $M$}, in the sense that
\be \|k^{[h+1,M]}\|_{1,n}=\sup_{\s,\s'}\int d\xx|k^{[h+1,M]}_{\s\s'}(\xx,\V0)|\cdot|\xx|^{n}\le C_{n}2^{-h}|U|,\qquad \forall n\ge 0\;.\label{eq:kh}\ee
In particular, the $({1,n})$-norm of $K^{3}_{h'}$ is bounded uniformly in $h'$ and $M$, proportionally to $|U|$. 
By shrinking all the linear subtrees in the way explained above, we end up with an alternative representation of the
effective potentials, which is based
on a slightly modified tree expansion. The set of modified trees with $N$ endpoints contributing 
to $\VV^{(h)}$ will be denoted by $\TT_{M;h,N}$; every $\t\in\TT_{M;h,N}$ is characterized
in the same way as the elements of $\widetilde\TT_{M;h,N}$, but for two features: 
(i) the endpoints of $\t\in\TT_{M;h,N}$ are not necessarily on scale $M+1$; (ii) every endpoint $v$ 
of $\t$ is attached to a non-trivial vertex on scale $h_v-1$ and is 
associated with the factor $\widetilde\VV^{(h_v-1)}(\Psi^{(\le h_{v}-1)})$. See Fig.\ref{fig6.3}.
In terms of these modified trees, \eqref{B.12a} is changed into 
\be {\cal V}^{(h)}(\Psi^{(\le h)}) =
\sum_{N=1}^\io\sum_{\t\in\TT_{M;h,N}}{\cal V}^{(h)}(\t,\Psi^{(\le h)})\;,\label{B.12az}\ee
where
\be {\cal V}^{(h)}(\t,\Psi^{(\le h)})=\frac{(-1)^{s-1}}{s!} \EE^T_{h+1}
\big[{\cal V}^{(h+1)}(\t_1,\Psi^{(\le h+1)});\ldots;{\cal V}^{(h+1)}
(\t_{s},\psi^{(\le h+1)})\big]\label{B.13}\ee
and, if $\t$ is a trivial subtree with root on scale $k\in[h,M]$, then 
${\cal V}^{(k)}(\t,\Psi^{(\le k)})=\widetilde {\cal V}(\Psi^{(\le k)})$. 

Using its inductive definition Eq.(\ref{B.13}), the right hand side of Eq.(\ref{B.12az}) can be
further expanded (it is a sum of several contributions, differing for the choices of the 
field labels contracted under the action of the truncated expectations $\EE^T_{h_v}$ 
associated with the vertices $v$ that are not endpoints), and in order to describe the resulting 
expansion we need some more definitions (allowing us to distinguish the fields that are 
contracted or not ``inside the vertex $v$").

We associate with any vertex $v$ of the tree a subset $P_v$ of $I_v$,
the {\it external fields} of $v$. These subsets must satisfy various
constraints. First of all, if $v$ is not an endpoint and $v_1,\ldots,v_{s_v}$
are the $s_v\ge 1$ vertices immediately following it (such that, in particular, $h_{v_i}=h_v+1$), then
$P_v \subseteq \cup_i
P_{v_i}$; if $v$ is an endpoint, $P_v=I_v$.
If $v$ is not an endpoint, we shall denote by $Q_{v_i}$ the
intersection of $P_v$ and $P_{v_i}$; this definition implies that $P_v=\cup_i
Q_{v_i}$. The union ${\cal I}_v$ of the subsets $P_{v_i}\setminus Q_{v_i}$
is, by definition, the set of the {\it internal fields} of $v$,
and is non empty if $s_v>1$.
Given $\t\in\TT_{M;h,N}$ and the set of field labels $I_v$ associated with the endpoints $v$ of $\t$, there are many possible choices of the
subsets $P_v$ associated with the vertices that are not endpoints, which are compatible with all the constraints. We
shall denote by ${\cal P}_\t$ the family of all these choices and by ${\bf P}$
the elements of ${\cal P}_\t$. With these definitions, we can rewrite
${\cal V}^{(h)}(\t,\Psi^{(\le h)})$ as:
\bea &&{\cal V}^{(h)}(\t,\Psi^{(\le h)})=\sum_{\underline\s_{v_0}}
\int d\underline\xx_{v_0}
\sum_{{\bf P}\in{\cal P}_\t}K_{\t,\PP}^{(h+1)}\Psi^{(\le h)}_{P_{v_0}}
\;,\label{eq:55.43}\eea
where $K_{\t,\PP}^{(h+1)}$ is defined inductively by
the following equation, which is valid for any $v\in\t$ that is not an endpoint,
\be K_{\t,\PP}^{(h_v)}=\frac{1}{s_v !}
\prod_{i=1}^{s_v} [K^{(h_{v_i})}_{\t_i,{\bf P}_i}]\; \;\EE^T_{h_v}[\Psi^{(h_v)}_{P_{v_1}\setminus 
Q_{v_1}},\ldots,\Psi^{(h_v)}_{P_{v_{s_v}}\setminus
Q_{v_{s_v}}}]\;.\label{eq:5.63}\ee
Here $\t_1,\ldots,\t_{s_v}$ are the subtrees with root $v$, $v_i$ are their leftmost vertices (such that, in particular, $h_{v_i}=h_v+1$),
and ${\bf P}_i=\{P_w, w\in\t_i\}$.
Moreover, if $v_i$ is an endpoint, 
then $K^{(h_{v_i})}_{\t_i,{\bf P}_i}=K_{v_i}$, with 
\be K_{v}=\begin{cases}E_{h_{v}-1}(\phi,A)\hskip4.82truecm {\rm if}\quad \r_{v}=0\;,\\
K^{\r_{v}}_{h_v;\s_{v}\s_{v}'}(\xx_{v},\yy_{v})\big[\phi^{\s_v}_{\xx_v}\big]^{\d_{\r_v,1}}\big[A^{\s_v\s'_v}_{\xx_v,\yy_v}\big]^{\d_{\r_v,2}}\quad {\rm if}\quad \r_{v}>0\;,
\end{cases}\ee
where $K^{\r_{v}}_{h_v;\s_{v}\s_{v}'}$ should be identified with $K^{\r_{v}}_{\s_{v}\s_{v}'}$ in the case
that $\r_v=1,2,4$.
Combining \eqref{B.12az} with \eqref{eq:55.43} and \eqref{eq:5.63}, and using the
determinant representation of the truncated expectation, see (\ref{s5.3}),
we finally get:
\be  {\cal V}^{(h)}(\Psi^{(\le h)}) =E_h(\phi,A)+
\sum_{N=1}^\io\sum_{\t\in\TT_{M;h,N}}^*
\sum_{\underline\s_{v_0}}
\int d\underline\xx_{v_0}
\sum_{{\bf P}\in{\cal P}_\t}\sum_{T\in {\bf T}}
W_{\t,\PP,T}^{(h)}(\underline\xx_{v_0},\underline\s_{v_0}) \Psi^{(\le h)}_{P_{v_0}}
\;,\label{B.14}\ee
where the $*$ on the sum over $\t$ indicates the constraint that there are no endpoints of type 
$0$, and ${\bf T}$ is the set of the tree graphs on $\underline{\xx}_{v_0}$ obtained by putting together 
an anchored tree graph $T_v$ for each non-trivial vertex $v$ and by adding a line (which is by
definition the only element of $T_v$) for the couple of space-time points belonging
to the set $\underline{\xx}_v$ for each endpoint $v$. Moreover, 
\be
W_{\t,\PP, T}(\xx_{v_0},\underline\s_{v_0}) =\a_T\Big[\prod_{v\  {\rm e.p.}} K_{v}\Big]
\prod_{\substack{v\ {\rm not}\\ {\rm e.p.}}}\frac{1}{s_v!} \int
dP_{T_v}({\bf t}_v)\;{\rm det}\, G^{(h_v)}_{T_v}({\bf t}_v)
\prod_{\ell\in T_v} g_{\ell}^{(h_v)}\;,\label{eq:5.66}\ee
where $\a_T$ is a sign and $G^{(h_v)}_{T_v}({\bf t}_v)$ is a matrix analogous to the one defined
after \eqref{s5.3}, with $g^{(M)}$ replaced by $g^{(h_v)}$. Note that $W_{\t,\PP, T}$ depends on $M$
only 
through: (i) the choice of the scale labels, and (ii) the (weak) $M$-dependence of the endpoints $v$ of type $\r_v=3$, whose value is $K^3_{h_v;\s_v\s_v'}= K^3_{\s_v\s_v'}+k^{[h_v,M]}_{\s_v\s_v'}$,
with $k^{[h_v,M]}_{\s_v\s_v'}$ as in \eqref{eq:k}.
From \eqref{B.14} and \eqref{eq:5.66} we see that $\VV^{(h)}(\Psi)$ can be rewritten as
in \eqref{eq:5.48}, with $M-1$ replaced by $h$, and 
\bea && W^{(h)}_{2n,s_1,s_2,\underline\s,\underline\e}(\underline
\xx, \underline\yy,\underline\zz)=\sum_{N\ge 1}\sum_{\t\in \TT_{M;h,N}}^{**}
\sum_{\underline \s_{v_0}}\int d\underline\xx_{v_0}\!\!
\sum_{\substack{{\bf P}\in\mathcal P_\t:\\
|P_{v_0}|=2n}}\!\!\d(I_{v_0}^1-I^1_{ext})\d(I_{v_0}^2-I^2_{ext})\d\big(P_{v_0}-P_{ext})
\times\nonumber\\
&&\qquad \times
\Big[\prod_{v\  {\rm e.p.}} K^{\r_v}_{h_v;\s_v\s_v'}(\xx_v,\yy_v)\big]\sum_{T\in{\bf T}}\a_T
\prod_{\substack{v\ {\rm not}\\ {\rm e.p.}}}\frac{1}{s_v!} \int
dP_{T_v}({\bf t}_v)\;{\rm det}\, G^{(h_v)}_{T_v}({\bf t}_v)
\prod_{\ell\in T_v} g_{\ell}^{(h_v)}\label{eq:5.67}
\;,\eea
where the $**$ on the sum over $\t$ indicates the constraint that $\t$ has $s_1$ endpoints of type $1$, $s_2$ of type $2$, and no endpoints of type $0$. Note also that, in order for $|P_{v_0}|$ to be equal to 
$2n$, the number of endpoints of type $3$ and $4$ must be $\ge n-1$, that is $N\ge s_1+s_2+n-1$. Moreover, 
$I^1_{ext}=\big((\xx_1,\s_1),\ldots,(\xx_{s_1},\s_{s_1})\big)$,
$I^2_{ext}=\big((\xx_{s_1+1},\yy_{s_1+1},\s_{s_1+1},\s_{s_1+1}'),\ldots,(\xx_{s_2},\yy_{s_2},\s_{s_2},\s_{s_2}')\big)$, $P_{ext}=\big((\zz_1,\s_1'',\e_1),\ldots,(\zz_{2n},\s_{2n}'',\e_{2n})\big)$,
and the functions $\d(I_{v_0}^1-I^1_{ext})$, etc, are shorthands of products of delta functions, in the same sense as $\d(P-P_{ext})$ in \eqref{eq:5.48bis}. Using the explicit expression \eqref{eq:5.67},
we obtain a bound analogous to \eqref{eq:5.49}:
\bea
&& \frac1{\b L^2}\int d\xx d\yy d\zz \big|W^{(h)}_{2n,s_1,s_2,\underline\s,\underline\e}(\underline
\xx, \underline\yy,\underline\zz)\big|\le\label{eq:5.68}\\
&& \le\sum_{\substack{N\ge 1:\\ N\ge s_1+s_2+n-1}} C^N\sum_{\t\in \TT_{M;h,N}}^{**}
\sum_{\substack{{\bf P}\in\mathcal P_\t:\\
|P_{v_0}|=2n}}
\Big[\prod_{v\ {\rm e.p.}}
\|K^{\r_v}\|_1\Big]\sum_{T\in{\bf T}}\Big[\prod_{\substack{v\ {\rm not}\\ {\rm e.p.}}}\frac{1}{s_v!}
\big\|{\rm det}\, G^{(h_v)}_{T_v}\big\|_\infty
\prod_{\ell\in T_v}
\|g^{(h_v)}_\ell\|_1\Big]
\;.\nonumber
\eea
Now: (i) the contribution of the endpoints is bounded as $\|K^{\r_v}\|_1\le C|U|^{\d_{\r_v,3}+\d_{\r_v,4}}$,  (ii)
the $1$-norm of the propagators is bounded as in \eqref{cc1}, that is $\|g^{(h_v)}_\ell\|_1\le C\d_\m^{-3}2^{-h}$, and (iii) the determinant, recalling the Gram representation of the propagator \eqref{eq:gram}, 
can be bounded by using the Gram--Hadamard inequality (\ref{s5.6}) in a way analogous to \eqref{eq:5.51}, that is 
\be \|{\rm det} G^{(h_v)}_{T_v}\|_{\infty} \le
C^{\sum_{i=1}^{s_v}|P_{v_i}|-|P_v|-2(s_v-1)}\;,\label{2.54}\ee
where $v_1,\ldots,v_{s_v}$ are the vertices immediately following $v$ on $\t$. 
Plugging these bounds into \eqref{eq:5.68}, and using the fact that 
$\sum_{v\ {\rm not.}\ {\rm e.p.}}(\sum_{i=1}^{s_v}|P_{v_i}|-|P_v|)\le 4(N-s_1-s_2)$,
we obtain 
\be \sum_{\substack{N\ge 1:\\ N\ge s_1+s_2+n-1}} C^N|U|^{N-s_1-s_2}\sum_{\t\in \TT_{M;h,N}}^{**}
\sum_{\substack{{\bf P}\in\mathcal P_\t:\\
|P_{v_0}|=2n}}
\sum_{T\in{\bf T}}\Big[\prod_{\substack{v\ {\rm not}\\ {\rm e.p.}}}\frac{1}{s_v!}
(C\d_\m^{-3}2^{-h_v})^{s_v-1}\Big]
\;.\label{B.18}\ee
Using the following relation, which can be easily proved by induction,
\be\sum_{\substack{v\ {\rm not}\\ {\rm e.p.}}}h_v(s_v-1)=h(N-1)+
\sum_{\substack{v\ {\rm not}\\ {\rm e.p.}}}(h_v-h_{v'})(n(v)-1)\;,\label{B.18rt}\ee
where $v'$ is the vertex immediately preceding $v$ on $\t$
and $n(v)$ the number of endpoints following $v$ on $\t$, 
we find that Eq.(\ref{B.18}) can be rewritten as
\be
\sum_{\substack{N\ge 1:\\ N\ge s_1+s_2+n-1}} \sum_{\t\in \TT_{M;h,N}}^{**}
\sum_{\substack{{\bf P}\in\mathcal P_\t:\\
|P_{v_0}|=2n}}
\sum_{T\in{\bf T}}C^N\d_\m^{-3(N-1)}|U|^{N-s_1-s_2}2^{-h(N-1)}
\Big[\prod_{\substack{v\ {\rm not}\\ {\rm e.p.}}}\frac{1}{s_v!}
2^{(h_v-h_{v'})(n(v)-1)}\Big]
\;.\label{B.18tris}\ee
where, by construction, if $N>1$, then $n(v)>1$ for any vertex $v$ of $\t\in\TT_{M;h,N}$
that is not an endpoint (simply because every endpoint $v$ 
of $\t$ is attached to a non-trivial vertex on scale $h_v-1$, see the discussion after \eqref{eq:k} 
and item (ii) after \eqref{eq:kh}). If $N=1$, the only tree contributing to the sum in 
\eqref{B.18tris} is trivial, with four possible type labels attached to the endpoint. The corresponding contribution to \eqref{B.18tris} is (const.)$|U|^{\d_{s_1+s_2,0}}$. The contribution to \eqref{B.18tris}
from the terms with $N\ge 2$ can be bounded as follows: first of all, the number of terms in $\sum_{T\in {\bf T}}$ is bounded by
$C^N\prod_{v\ {\rm not}\ {\rm
e.p.}} s_v!$ (see, e.g.,  \cite[appendix A.3.3]{GeM}); moreover, $|P_v|\le 4n(v)$ and $n(v)-1\ge \max\{1,\frac{n(v)}2\}$, so that 
$n(v)-1\ge \frac12+\frac{|P_v|}{16}$, and, therefore,
\bea&&
\frac1{\b L^2}\int d\xx d\yy d\zz \big|W^{(h)}_{2n,s_1,s_2,\underline\s,\underline\e}(\underline
\xx, \underline\yy,\underline\zz)\big|\le\sum_{\substack{N\ge 1:\\ N\ge s_1+s_2+n-1}}
C^N\d_\m^{-3(N-1)}|U|^{N-s_1-s_2}2^{-h(N-1)} \times\nonumber\\
&&\hskip3.truecm\times \sum_{\t\in \TT_{M;h,N}}^{**}
\big(\!\!\!\!\prod_{v\ {\rm not}\ {\rm e.p.}}2^{-\frac12(h_v-h_{v'})}\big)
\sum_{\substack{{\bf P}\in\mathcal P_\t:\\
|P_{v_0}|=2n}}
\big(\prod_{\substack{v\ {\rm not}\\ {\rm e.p.}}}
2^{-|P_v|/16}\big)\;.\label{s6.boh} \eea
Now, the sums over $\t$ and $\PP$ in the second line can be both bounded by (const.)$^N$, see 
\cite[Lemma A.2 in appendix A.1 and appendix A.6.1]{GeM}, which implies the uniform analyticity 
of the kernels of the effective potentials on scale $h$, for all $-1\le h<M$, provided $U$ is small enough, namely $|U|\le ({\rm const.})\d_\m^3$. Note that the 
regularized free energy and correlation functions are nothing but the constant part and the kernels of the effective potential with $h=-1$. Therefore, the regularized free energy is analytic in $U$, uniformly in $\b,L,M$. Similarly, the regularized correlation functions are uniformly analytic and satisfy \eqref{G9}, uniformly in $\b,L,M$, for $\underline m=\underline 0$ and $|U|$ small enough. 
The proof of \eqref{G9}
for general choices of $\underline m$ follows similarly, by combining the previous strategy 
with the idea of decomposing the factors $|\xx_i-\xx_j|$ along the tree $T$, as in \eqref{eq:anc} and following discussion. 
{This concludes the proof of (\ref{G9}) and of Lemma \ref{lemg.2}}. \qed

\subsection{Proof of Proposition \ref{prp:analyticity}}\label{sec5c}

We are left with proving the existence of the limit as $\b,L,M\to\infty$ of the regularized free energy and 
correlation functions. In order to prove it, we show that these regularized functions form a Cauchy 
sequence. Let us start by showing that, for fixed $\b,L$, and $M'>M$, for all $0<\theta<1$, there exists 
$C_\theta>0$ such that 
\be \| K^{\b,L,M}_{m,n}-K^{\b,L,M'}_{m,n}\|_{1,r}\le C_\theta 2^{-\theta M}\;,\label{eq:2-m}\ee
where 
\be \| K^{\b,L,M}_{m,n}\|_{1,r}=
\frac1{\b L^2}\sup_{\underline\s}\sup_{\substack{\underline{m}:\\
|\underline m|=r}} \int _{\L_{\b,L}^{m+n}}\!\!\!d\underline\xx\sum_{\underline{\vec y}\in \L_L^m}\big|K^{\b,L,M}(\xx_1,\vec y_1,\s_1,\s_1';\ldots;\xx_{m+n},\s_{m+n})\big|\,
d_{\underline m}(\underline \xx,\underline{\vec y})\;.\label{eq:norm}\ee
As already remarked above, the regularized correlation function are the kernels of the 
effective potential on scale $-1$. Therefore, both $K^{\b,L,M}$ and $K^{\b,L,M'}$ 
can be expressed in terms of the tree expansion described above. As already remarked after \eqref{eq:5.66}, 
the expansions for $K^{\b,L,M}$ and $K^{\b,L,M'}$ differ among each other only because of: (i) the choice of the scale labels (the trees contributing to 
$K^{\b,L,M}$, resp. $K^{\b,L,M'}$, have endpoints on scales $\le M+1$, resp. $\le M'+1$);
(ii) the dependence on the ultraviolet cutoff of the endpoints of type $3$, whose 
value is $K^3_{h_v;\s_v\s_v'}= K^3_{\s_v\s_v'}+k^{[h_v,M]}_{\s_v\s_v'}$ in the trees 
contributing to $K^{\b,L,M}$, and similarly for $K^{\b,L,M'}$. This means that the difference 
$K^{\b,L,M}-K^{\b,L,M'}$ can be expressed as a sum over trees whose root is on scale 
$-1$ and:  (A) either there is at least one endpoint on scale $>M+1$, or (B) there is one endpoint of type $3$ associated with a difference $k^{[h_v,M]}_{\s_v\s_v'}-k^{[h_v,M']}_{\s_v\s_v'}=k^{[M+1,M']}_{\s_v\s_v'}$. 

The contributions from the case (A) can be bounded as in \eqref{B.18tris}, with $h=-1$ and 
the extra constraint that there is at least one endpoint on scale $>M+1$. This means that the 
factor $\prod_{\substack{v\ {\rm not}\\ {\rm e.p.}}}
2^{(h_v-h_{v'})(n(v)-1)}$ is smaller than $2^{-M}$. The idea is then to split this term into two factors, 
in the form $\big[\prod_{\substack{v\ {\rm not}\\ {\rm e.p.}}}
2^{\theta(h_v-h_{v'})(n(v)-1)}\big]\times \big[\prod_{\substack{v\ {\rm not}\\ {\rm e.p.}}}
2^{(1-\theta)(h_v-h_{v'})(n(v)-1)}\big]$. The first factor is smaller than $2^{-\theta M}$, while the 
sum over the scale and field labels of the second factor can be bounded exactly in the same away as 
it was explained after \eqref{s6.boh}.

Concerning case (B), it is enough to note that the norm of 
$k^{[M+1,M']}_{\s_v\s_v'}$ is proportional to $2^{-M}$, see \eqref{eq:kh}, which implies that the overall 
contribution from these trees is smaller  than the norm of $K^{\b,L,M}$ by a factor $2^{-M}$. 

In conclusion, we obtain \eqref{eq:2-m}. By {Vitali}'s uniform convergence theorem for analytic functions, we conclude that the limit as $M\to\infty$ of {any weighted integral of the regularized correlations
(with weights growing at most polynomially at large space-time differences, and the integral normalized by $1/(\b L^2)$)} is analytic, and its Taylor coefficients are the $M\to\infty$ limit of the coefficients of the regularized correlations. {Analogously, one proves the same for the correlation functions at fixed space-time positions.} Moreover, the same argument is valid for the limit as $\b,L\to\infty$, see \cite[appendix D]{GM} for a thorough discussion of this limit. Of course, the same claims are valid for the regularized free energy, too. 

Finally, the statement of proposition \ref{prp:analyticity} follows from the remark that 
that the correlation functions in momentum space can be expressed as the Fourier 
transforms of their space-time counterparts, and that their derivatives of order 
$r$ are controlled by the $(1,r)$ norms \eqref{eq:norm} of the space-time correlation functions, which are finite and bounded uniformly in $\b,L,M$, as we just proved. 
\qed

\section{Reconstruction of the real-time Kubo formula}\label{sec:reconstr}

In this section we prove theorem \ref{thm:realkubo}, which says that the Kubo formula (\ref{eq:kubo}), which can be expressed as an imaginary-time integral of the current-current correlation, can be analytically continued to real times. In other words, we rigorously prove the validity of the {\it Wick rotation} for the Kubo conductivity of the class of systems under investigation. In the language of Quantum Field Theory, we prove a {\it reconstruction theorem} for the conductivity matrix of weakly interacting gapped fermionic systems. 

\bigskip

The proof of theorem \ref{thm:realkubo} is based on the existence of the real-time correlation functions in the infinite volume and zero temperature limits, as well as on the decay of the complex time correlations, as summarized in the following proposition.

\begin{prp}{\bf [Properties of the current-current correlations at complex times]}\label{prp:realJJ}
Let $J_{i}(z) := e^{z\mathcal{H}_{L}} J_{i} e^{-z\mathcal{H}_{L}}$, $z\in\mathbb{C}$,
with $J_i$ defined in \eqref{eq:currdefgen}--\eqref{eq:defbondcurr} and following lines.
Under the same assumptions as theorem \ref{thm:main}, and if in addition $U\in\mathbb R$, the following is true.
\begin{itemize}
\item{(i)} Let $z\in \mathbb{C}^{+} = \{ z\in\mathbb{C} \mid {\rm{Re}}\,z>0 \}$. Then, the limit
\be
\lim_{\beta\to\infty}\lim_{L\to\infty}\frac1{L^2} \langle J_{i}(z) J_{j}(0) \rangle_{\beta, L} =: \langle\!\!\langle J_{i}(z) J_{j}(0) \rangle\!\!\rangle_{\infty}
\ee
exists, and it is analytic in $z\in \mathbb{C}^{+}$. Moreover, it decays faster than any power in ${\rm Re}\,z$, i.e., 
\be\label{eq:boundCS}
\big|\langle\!\!\langle J_{i}(z) J_{j}(0) \rangle\!\!\rangle_{\infty}\big|\leq  \frac{C_{M}}{1 + ({\d_\m}{\rm Re}\, z)^{M}}\;,\qquad z\in \mathbb{C}^{+}\;,
\ee
for all $M\ge 0$ and a suitable $C_{M}>0$.

\item{(ii)} Let $t\in \mathbb{R}$. Then, the limit
\be\label{eq:limitrealtime}
\lim_{\b\to\infty} \lim_{L\to\infty} \frac{1}{L^{2}} \big\langle \big[ J_{i}(it), J_{j}(0) \big] \big\rangle_{\beta, L}=:\langle\!\!\langle \big[ J_{i}(it), J_{j}(0) \big]\rangle\!\!\rangle_\infty
\ee
exists and is finite, uniformly in $t$.
\end{itemize}
\end{prp}

\noindent{\bf Proof of proposition \ref{prp:realJJ}.} To begin with, let us prove item (i). The starting point is to notice that the positive temperature, finite volume current-current correlation
\be
\langle J_{i}(z) J_{j}(0) \rangle_{\beta, L} = \frac{\Tr\, e^{-\beta (\mathcal{H}_{L} - \mu \mathcal{N}_{L})}\, e^{z \mathcal{H}_{L}} J_{i}\,e^{-z\mathcal{H}_{L}} J_{j}}{\Tr\, e^{-\beta (\mathcal{H}_{L} - \mu \mathcal{N}_{L})}}
\ee
is {\it entire} in $z$. By the Cauchy-Schwarz inequality and the cyclicity of the trace, recalling that $U\in\mathbb R$, we get:
\bea\label{eq:CS}
\big| \langle J_{i}(z) J_{j}(0) \rangle_{\beta, L}  \big| &\leq& \big| \langle J_{i}(z/2) J_{i}(z/2)^{\dagger} \rangle_{\beta, L} \big|^{1/2} \big| \langle J_{j}(-z/2)^{\dagger} J_{j}(-z/2) \rangle_{\beta,L} \big|^{1/2}\\
&=&\big| \langle J_{i}( \text{Re}\,z) J_{i}(0) \rangle_{\beta,L}\big|^{1/2}\,\big| \langle J_{j}(\text{Re}\, z) J_{j}(0) \rangle_{\beta,L}\big|^{1/2}\;.\nonumber
\eea
Thus, for $0 \leq \text{Re}\, z < \beta$, the right side of (\ref{eq:CS}) can be estimated via (the proof of)
proposition \ref{prp:analyticity}. This implies, in particular, that, if $0\le t<\b$, the imaginary-time correlation 
function $L^{-2}\langle J_{i}(t) J_{i}(0) \rangle_{\beta,L}$ decays faster than any power in $|t|_\b=\min_{n\mathbb Z}|t +n\beta|$, uniformly in $\b,L$ (see Lemma \ref{lemg.2} and section \ref{sec5c}). 
Therefore, \eqref{eq:CS} implies that, for every fixed $z\in \overline{\mathbb{C}^{+}}$, there exists $\beta_{0}$ such that, for $\beta > \beta_{0}$, 
\be\frac1{L^2}
\big| \langle J_{i}(z) J_{j}(0) \rangle_{\beta, L}  \big| \leq \frac{C_{M}}{1 + ({\d_\m}\text{Re}\, z)^{M}}\;,
\label{eq..4.9}
\ee 
for all $M\ge 0$ and a suitable $C_{M}>0$, independent of $\beta, L, z$. Moreover, the proof of 
proposition \ref{prp:analyticity} implies that the limit
\be
\lim_{\beta \to\infty}\lim_{L\to\infty}\frac1{L^2} \langle J_{i}(t) J_{j}(0) \rangle_{\beta, L}
\ee
exists, for all $t\ge 0$. Therefore, by Vitali's theorem on the convergence of holomorphic functions, 
we conclude that the limit
\be
\lim_{\beta\to\infty}\lim_{L\to\infty}\frac1{L^2} \langle J_{i}(z) J_{j}(0) \rangle_{\beta, L} =: \langle\!\!\langle J_{i}(z) J_{j}(0) \rangle\!\!\rangle_{\infty} 
\ee
exists and is analytic in $z$ in the whole open right half-plane $\mathbb{C}^{+}$. Moreover, the convergence to the limit is uniform on any compact subset of $\mathbb C^+$. By \eqref{eq..4.9},
the limit satisfies  (\ref{eq:boundCS}), which concludes the proof of item (i). 

Let us now prove (ii). By using the translational invariance of the Gibbs state, we rewrite (recalling
\eqref{eq:currbond}-\eqref{eq:defbondcurr}):
%
\be
\frac{1}{L^{2}}\big\langle \big[ J_{i}(it), J_{j}(0)\big] \big\rangle_{\beta, L} =-\frac14\sum_{\vec x,\vec y,\vec z\in \L_{L}} \sum_{\s_1,...,\s_4\in I}( \vec x_{\s_1}-\vec y_{\s_2} )_i(\vec 0_{\s_3}-\vec z_{\s_4} )_j
\big\langle \big[ \big(J_{\vec x\,\vec y}^{\s_1\s_2}\big)(it), J_{\vec0\,\vec z}^{\s_3\s_4}\big] \big\rangle_{\beta, L}\;,
\label{eq:LR1}\ee
where, again, $A(it)=e^{i\mathcal H_L t}Ae^{-i\mathcal H_L t}$.
Now, the summand in the right side of \eqref{eq:LR1} is absolutely summable, uniformly in $\b,L$.
This can be proven using {\it Lieb-Robinson bounds}; see, e.g., \cite{LR, Nachtergaele1,Nachtergaele2} for a derivation of these bounds for quantum spin systems and \cite{BP} for an extension to fermionic 
systems. 
The key result is (see, e.g., \cite[Theorem 2.1]{Nachtergaele1}, or \cite[Theorem 3.1]{BP}): 
\be\label{eq:LRbound}
\big\| \big[ \big(J_{\vec x\,\vec y}^{\s_1\s_2}\big)(it), J_{\vec0\,\vec z}^{\s_3\s_4}\big] \big\| \leq \|J_{\vec x\,\vec y}^{\s_1\s_2}\|\,\|J_{\vec 0\,\vec z}^{\s_3\s_4}\|\frac{C_Me^{v|t|}}{\big[1+{\rm{dist}}(\{\vec x,\vec y\},\{\vec 0,\vec z\})\big]^M}\;,
\ee
for all $M\ge 0$ and suitable constants $C_M$ and $v$, independent of $\beta, L,t$.
Here $\|\cdot\|$ denotes the operator norm. 
By using the fact that $\|J_{\vec x\,\vec y}^{\s_1\s_2}\|\le 2\|H_{\s_1\s_2}(\vec x-\vec y)\|$, which decays faster than any power in $|\vec x-\vec y|$,
we see that the  sum in the right side of (\ref{eq:LR1}) is absolutely convergent, uniformly in $\beta, L$.

Therefore, in order to prove the existence of the limit of \eqref{eq:LR1} as $\b,L\to\infty$, it is enough to prove, term by term, 
the existence of the limit of the summands in the right side. The proof of this fact is a straightforward consequence of the existence of the infinite volume dynamics (see \cite{Nachtergaele1, Nachtergaele2, BP}) 
and of the existence of the $\beta,L\to\infty$ limit of the Gibbs state. In appendix C, we reproduce this proof; that is, we prove that the limit $\lim_{\b\to\infty}\lim_{L\to\infty}\big\langle \big[ \big(J_{\vec x\,\vec y}^{\s_1\s_2}\big)(it), J_{\vec0\,\vec z}^{\s_3\s_4}\big] \big\rangle_{\beta, L}$ exists, thus concluding the proof of the existence of the limit in item (ii). The uniform boundedness of the limit is a consequence of \eqref{eq:CS}-\eqref{eq..4.9}. \qed

\bigskip

We are now in the position of proving theorem \ref{thm:realkubo}.

\bigskip

\noindent{\bf Proof of theorem \ref{thm:realkubo}.} We start from the very definition of imaginary-time conductivity
\eqref{eq:kubo}, that is $\bar\s_{ij}(U)= -\lim_{\omega\to 0^{+}} (A\o)^{-1}\big[ \widehat K_{ij}(\omega,\vec 0) - \widehat K_{ij}(0,\vec 0) \big]$, where 
\be  \widehat K_{ij}(\omega,\vec 0)=\lim_{\b\to\infty}\lim_{L\to\infty}\frac1{\b L^2}\int_0^\b dt \int_0^{\b}
dt' e^{-i\o (t-t')}\media{{\bf T} J_{i}(t)\, J_{j}(t')}_{\b,L}.\ee
Note that there is no semicolon between the two current operators in the right side (that is, the 
expectation is {\it un}truncated): the reason is that the Gibbs average of $J_i$ vanishes, 
simply because $J_i$ is proportional to the commutator of $\mathcal H_L$ with $X_i$. 
With some abuse of notation, we denoted by the same symbol the frequency $\o$ in the 
two sides of the equation. However, it should be recalled that the (Matsubara) frequency in the right side is an integer multiple of $2\p/\b$, i.e., it should be understood as being equal to $\o_n=(2\p/\b)n$,
with $\o_n\to\o$ as $\b\to\infty$.

We start by analyzing and suitably rewriting $\widehat K_{ij}(\omega,\vec 0)$, with $\o>0$. 
By the cyclicity of the trace and the fact that $\o_n$ is an integer multiple of $(2\p/\b)$, we can rewrite 
\be \frac1\b\int_0^\b dt \int_0^{\b}
dt' e^{-i\o_n (t-t')}\media{{\bf T} J_{i}(t)\, J_{j}(t')}_{\b,L}=\int_{-\b/2}^{\b/2} dt\,
 e^{-i\o_n t}\media{{\bf T} J_{i}(t)\, J_{j}(0)}_{\b,L},\ee
so that 
\be  \widehat K_{ij}(\omega,\vec 0)=\lim_{\b\to\infty}\lim_{L\to\infty}\frac1{ L^2}\int_{-\b/2}^{\b/2} dt\,  e^{-i\o_n t}\media{{\bf T} J_{i}(t)\, J_{j}(0)}_{\b,L}.\ee
Recalling that $L^{-2}\media{{\bf T} J_{i}(t)\, J_{j}(0)}_{\b,L}$ decays faster than any power in $\|t\|_\b$,
uniformly in $\b,L$, 
and that it converges as $\b,L\to\infty$, we find that, for any $T>0$,
\be  \widehat K_{ij}(\omega,\vec 0)=\int_{-T}^{T} dt\,e^{-i\omega t} \langle\!\!\langle {\bf T}\,J_{i}(t) J_{j} (0)\rangle\!\!\rangle_{\infty} + R_{T}(\omega)\;, 
\ee
where 
\be\label{eq:R1}
|R_{T}(\omega)| \leq \frac{C_{M}}{1 + ({\d_\m}T)^M}\,,
\ee
for all $M\ge 0$ and a suitable $C_M>0$, independent of $T$ and $\o$. Therefore,
\be \widehat K_{ij}(\omega,\vec 0)=
\lim_{T\to\infty} \int_{-T}^{T} dt\,e^{-i\omega t} \langle\!\!\langle {\bf T}\,J_{i}(t) J_{j}(0)\rangle\!\!\rangle_{\infty} \;.\label{eq:R1dert}
\ee
We rewrite:
\be\label{eq:JJTT}
\int_{-T}^{T} e^{-i\omega t} \langle\!\!\langle {\bf T}\,J_{i}(t) J_{j}(0)\rangle\!\!\rangle_{\infty} = \int_{0}^{T} e^{-i\omega t} 
\langle\!\!\langle J_{i}(t) J_{j}(0) \rangle\!\!\rangle_{\infty}+
\int_{-T}^{0} e^{-i\omega t} \langle\!\!\langle J_{j}(0) J_{i}(t) \rangle\!\!\rangle_{\infty} \;.
\ee
We study the two integrals in the right side separately, starting from the first. Recall that the integrand is analytic in
$\mathbb C^+$: therefore, by Cauchy theorem, the integral along any closed path in $\mathbb C^+$ is identically zero. 
We choose the closed path consisting of the union of: the segment $[\e,T]$ on the real line ($\e$ being a small positive number, to be eventually sent to zero), directed from left to right; 
the quarter circle of radius $T-\e$ centered in $\e$, connecting  the point $T$ with the point 
$-i(T-\e)+\e$, in the clockwise direction; and the vertical segment connecting $-i(T-\e)+\e$ 
with $\e$, in the upwards direction. We thus rewrite:
\bea\label{eq:proofrk3}
\int_{0}^{T} dt\, e^{-i\omega t} \langle\!\!\langle J_{i}(t) J_{j}(0)\rangle\!\!\rangle_{\infty} &=& 
\lim_{\e\to 0}\Big[-i\int_{-T + \e}^{0} dt\, e^{\omega(t-i\e)} \langle\!\!\langle 
J_{i}(it+\e) J_{j}(0)\rangle\!\!\rangle_{\infty} \\
&+&i(T-\e) \int_{-\p/2}^0 d\theta\, 
e^{i\theta}e^{-i\omega (\e+(T-\e)e^{i\theta})} \langle\!\!\langle J_{i}\big(\e+(T-\e)e^{i\theta}\big) J_{j}(0)
\rangle\!\!\rangle_{\infty}\Big] \;.\nonumber\eea
%
Recalling \eqref{eq:boundCS}, we can bound the term in the second line by:
\be T \lim_{\e\to 0} \int_{-\p/2}^0 d\theta\, 
e^{\omega (T-\e)\sin\theta}\frac{C_M}{1+\big[{\d_\m}(\e+(T-\e)\cos\theta)\big]^M}\le
\frac\p4C_MT\Big[e^{-\o T/\sqrt2}+\frac1{1+({\d_\m}T/\sqrt2)^M}\Big]\;,
\ee 
which tends to zero faster than any power as $T\to\infty$, for every $\o>0$. Repeating the same argument for the 
second integral in the right side of \eqref{eq:JJTT}, and plugging the result back into \eqref{eq:R1dert},
we find that, for every $\o>0$,
\be \widehat K_{ij}(\o,\vec 0)=-i\lim_{\e\to0}
\int_{-\infty}^0dt\, e^{\o t}\Big[ \langle\!\!\langle 
J_{i}(it+\e) J_{j}(0)\rangle\!\!\rangle_{\infty}- \langle\!\!\langle J_{j}(0)
J_{i}(it-\e) \rangle\!\!\rangle_{\infty}
\Big]\;.\ee
By adding and subtracting the expression in square brackets at $\e=0$, we get
\be \widehat K_{ij}(\o,\vec 0)=-i
\int_{-\infty}^0dt\, e^{\o t}\langle\!\!\langle 
\big[ J_{i}(it), J_{j}(0)\big]\rangle\!\!\rangle_{\infty}+ \lim_{\e\to0} R(\o,\e)\;,
\label{roe0}\ee
where we used item (ii) of proposition \ref{prp:realJJ}, and we defined:
\be R(\o,\e)=\lim_{\b\to\infty}\lim_{L\to\infty}\frac{-i}{L^2}\int_{-\infty}^0\!\!dt\, e^{\o t} \langle 
\big[J_{i}(it+\e)-J_i(it)\big] J_{j}(0)-J_{j}(0)
\big[J_{i}(it-\e)-J_i(it)\big] \rangle_{\b,L}
\;.\label{roe}\ee
The term $ \langle 
\big[J_{i}(it+\e)-J_i(it)\big] J_{j}(0)\rangle_{\b,L}$ can be bounded by rewriting it as:
\bea  \langle 
\big[J_{i}(it+\e)-J_i(it)\big] J_{j}(0)\rangle_{\b,L} &=& \int_{0}^{\e} ds\, \Big\langle \frac{d}{ds} e^{(it + s)\mathcal{H}_{L}} J_{i} e^{-(it + s)\mathcal{H}_{L}} J_{j} \Big\rangle_{\b,L}\nn\\
&=& \int_{0}^{\e} ds\, \big\langle \big[ \mathcal{H}_{L}, J_{i}(it+s) \big] J_{j} \big\rangle_{\b,L}\;.
\eea
By proceeding as in the proof of proposition \ref{prp:realJJ}, via the analogues of 
\eqref{eq:CS}-\eqref{eq..4.9}, we obtain 
\be \big| \langle 
\big[J_{i}(it+\e)-J_i(it)\big] J_{j}(0)\rangle_{\b,L}\big|\le L^{2} C\e\;,\ee 
with $C>0$ independent of $\b,L,\e$. The same estimate is valid for $\langle J_{j}(0)
\big[J_{i}(it-\e)-J_i(it)\big] \rangle_{\b,L}$. Plugging these estimates back into 
\eqref{roe}, we find that $|R(\o,\e)|\le C\e/\o$, so that, using \eqref{roe0}, we finally get that, for all $\o>0$, 
\be \widehat K_{ij}(\o,\vec 0)=-i
\int_{-\infty}^0dt\, e^{\o t}\langle\!\!\langle 
\big[ J_{i}(it), J_{j}(0)\big]\rangle\!\!\rangle_{\infty}\;,\label{fin}\ee
which is our final expression for $\widehat K_{ij}(\o,\vec 0)$, with $\o>0$.

Concerning $\widehat K_{ij}(0,\vec 0)$, in order to rewrite it conveniently, we use \eqref{eq:WI1prop}
with $n=2$ and $\a_2=j$ in the thermodynamic and zero temperature limits, which reads
(denoting $\pp=(\o,\vec p)$)
\be i\o \widehat K_{0j}(\pp)+p_1\widehat K_{1j}(\pp)+p_2\widehat K_{2j}(\pp)=\widehat S_j(\pp)\;.\ee
If we derive this expression with respect to $p_i$ and then set $\pp=\V0$, we obtain 
\be \widehat K_{ij}(0,\vec 0)=\frac{\partial \widehat S_j}{\partial p_i}(\V0)=
- \langle\!\!\langle \big[[ \mathcal{H},X_{i}], X_{j} \big] \rangle\!\!\rangle_{\infty}\;.\label{finn}\ee
Using  \eqref{fin} and \eqref{finn} in \eqref{eq:kubo}, we finally recognize that $\bar\s_{ij}(U)=$\eqref{eq:kubo0}, as desired.
\qed

\bigskip

{\bf Acknowledgements.} 
We would like to thank Gian Michele Graf and Gianluca Panati for interesting
discussions on the topological aspects of condensed matter physics, and Giuseppe Benfatto for useful comments.
We gratefully acknowledge financial support from: the A*MIDEX project (n$^o$
ANR-11-IDEX-0001-02) funded by the ``Investissements d'Avenir'' French
Government program, managed by the French National Research Agency (A.G.);
the PRIN National Grant {\it Geometric and analytic theory of Hamiltonian systems in finite and infinite dimensions} (A.G. and V. M.); the NCCR SwissMAP (M.P.).

\appendix

\section{The non-interacting conductivity}\label{app:kubo}

In this appendix we reproduce the well known result that, in the absence of interactions, and under the gap condition \eqref{gap}, 
the Kubo conductivity (\ref{eq:kubo}) reduces to the usual formula for the Chern number: 
\be\label{eq:AS2}
\bar\sigma_{ij} (0)= i \int_{\mathcal{B}} \frac{d\vec k}{(2\pi)^{2}}\, \Tr\, P_{-}(\vec k) [\partial_{k_i} P_{-}(\vec k), \partial_{k_j} P_{-}(\vec k)] \;,
\ee
where $P_-(\vec k)$ is the projection onto the filled bands, defined after \eqref{gap}. In light of theorem \ref{thm:realkubo}, the same is true for $\s_{ij}(0)$.

\medskip

Our starting point consists in rewriting the infinite volume limit of the current operator defined in \eqref{eq:currdefgen}--\eqref{eq:defbondcurr} in Fourier space: 
\be\label{eq:eqcond00}
\vec J = i\int_\BBB\frac{d\vec k}{|\BBB|}\sum_{\s,\s'\in I} \hat \psi^{+}_{\vec k,\s} \Big[\big( i\nabla_{\vec k} +\vec r_{\s'}-\vec r_\s\big)\hat H_{\s\s'}^{(0)}(\vec k) \Big] \hat \psi^{-}_{\vec k,\s'} 
\ee
The term $\vec r_{\s'}-\vec r_\s$ can be reabsorbed by conjugating the Bloch Hamiltonian and the fermionic fields with a suitable unitary transformation: if 
we define
$U(\vec k) = \text{diag}\big(e^{i\vec k\cdot \vec r_1}\;,\; \cdots \;,\; e^{i\vec k\cdot \vec r_{|I|}}\big)$,
$\widetilde H^{(0)}(\vec k) = U(\vec k)\hat H^{(0)}(\vec k) U(\vec k)^{\dagger}$, $\widetilde \psi^{-}_{\vec k} = 
U(\vec k) \hat \psi^{-}_{\vec k}$, and $\widetilde \psi^{+}_{\vec k} =\hat \psi^{+}_{\vec k} U(\vec k)^{\dagger}$,
then 
the current in \eqref{eq:eqcond00} can be rewritten as:
\be
\vec J = -\int_\BBB\frac{d\vec k}{|\BBB|}\sum_{\s,\s'\in I} \widetilde\psi^{+}_{\vec k,\s}\big[ \nabla_{\vec k} 
\widetilde H^{(0)}_{\s\s'}(\vec k)\big] \widetilde \psi^{-}_{\vec k,\s'}\;.
\ee
Its imaginary-time evolution, $\vec J_t$, is obtained by replacing  
$\widetilde \psi^{\pm}_{\vec k,\s}$ by its imaginary-time evolution, $\widetilde \psi^{\pm}_{(t,\vec k),\s}$. 
Note that $J_{i,t}$ is the same as (the infinite volume limit of) $\tilde J_{i,(t,\vec p)}\big|_{\vec p=\vec 0}$, where 
$\tilde J_{i,(t,\vec p)}$ was defined in \eqref{eq:Ji}.

The conductivity matrix \eqref{eq:kubo} at $U=0$ can then be re-expressed as:
\bea\label{eq:eqcondprelim}
\bar\sigma_{ij}(0)& = &- \frac{1}{A}\lim_{\o\to 0}\frac{\partial}{\partial\o} \int_{\mathbb R}dt\, e^{-i\o t} 
\int_\BBB\frac{d\vec k}{|\BBB|}
 \int_\BBB\frac{d\vec k'}{|\BBB|}\sum_{\s_1, ..., \s_4}\times\\
 &&\times
\big\langle \mathbf{T} \; \widetilde\psi^{+}_{(t,\vec k),\s_1}\partial_{k_i} 
\widetilde H^{(0)}_{\s_1\s_2}(\vec k)\, \widetilde \psi^{-}_{(t,\vec k),\s_2}\; ;\; 
 \widetilde\psi^{+}_{(0,\vec k'),\s_3}\partial_{k_j} 
\widetilde H^{(0)}_{\s_3\s_4}(\vec k')\,\widetilde \psi^{-}_{(0,\vec k'),\s_4}
 \big\rangle^{(0)}\;,\nn
\eea
where $\media{\cdot}^{(0)}=\lim_{\b\to\infty}\lim_{L\to\infty}\media{\cdot}^{(0)}_{\b,L}$.
The expectation in the second line can be evaluated via the Wick rule, so that 
\be\label{eq:appwick}
\bar\sigma_{ij}(0)= \frac{-i}{A}\int_\BBB\frac{d\vec k}{|\BBB|} \int_{\mathbb R}\,dt\, t\,
\Tr \big\{ \widetilde g(-t,\vec k)\, \partial_{k_{i}} \widetilde H^{(0)}(\vec k)\, \widetilde g(t, \vec k)\, \partial_{k_{j}} \widetilde H^{(0)}(\vec k)\big\}
\ee
where the trace  is over the $\s$ indices, and 
\bea\label{eq:proptilde}
\widetilde g(t,\vec k)= e^{-t ( \widetilde H^{(0)}(\vec k) - \mu )} \Big[ \mathds{1}(t>0)\widetilde P_+(\vec k)
-\mathds{1}(t\le 0)\widetilde P_-(\vec k)\Big]\;,
\eea
with $\widetilde P_-(\vec k)=U(\vec k)P_-(\vec k)U(\vec k)^\dagger$, and $\widetilde P_+(\vec k)=1-\widetilde P_-(\vec k)$. Plugging \eqref{eq:proptilde} into \eqref{eq:appwick}, and noting that 
$A|\BBB|=(2\p)^2$, we find
\bea
&&\bar\sigma_{ij}(0)= i\int_\BBB\frac{d\vec k}{(2\p)^2}\, \Bigl[\,  \int_0^\infty\,dt\, t\,
\Tr \big\{ e^{t\widetilde H^{(0)}(\vec k) }\widetilde P_-(\vec k)\, \partial_{k_{i}} \widetilde H^{(0)}(\vec k)\, e^{-t\widetilde H^{(0)}(\vec k) }\widetilde P_+(\vec k)\, \partial_{k_{j}} \widetilde H^{(0)}(\vec k)\big\}\nn\\
&&\hskip2.truecm +\int_{-\infty}^0\,dt\, t\,
\Tr \big\{ e^{t\widetilde H^{(0)}(\vec k) }\widetilde P_+(\vec k)\, \partial_{k_{i}} \widetilde H^{(0)}(\vec k)\, e^{-t\widetilde H^{(0)}(\vec k) }\widetilde P_-(\vec k)\, \partial_{k_{j}} \widetilde H^{(0)}(\vec k)\big\}\Big]
\nn\\
&&\hskip1.15truecm \equiv  i\int_\BBB\frac{d\vec k}{(2\p)^2}\, \big[
\Sigma_{ij}(\vec k)-\Sigma_{ji}(\vec k)\big]\;,\label{Sigmaij}
\eea
where $\Sigma_{ij}(\vec k)= \int_0^\infty\,dt\, t\, f_{ij}(t,\vec k)$, with 
\be\label{fij} f_{ij}(t,\vec k)=
\Tr \big\{ e^{t\widetilde H^{(0)}(\vec k) }\widetilde P_-(\vec k)\, \partial_{k_{i}} \widetilde H^{(0)}(\vec k)\, e^{-t\widetilde H^{(0)}(\vec k) }\widetilde P_+(\vec k)\, \partial_{k_{j}} \widetilde H^{(0)}(\vec k)\big\}\;.\ee
Note that $f_{ij}(t,\vec k)$ decays exponentially to zero as $t\to\infty$, uniformly in $\vec k$, due to the presence of the projectors 
in the trace and to the gap condition. Now, the key observation is that 
\be f_{ij}(t,\vec k)=\partial_t^2F_{ij}(t,\vec k)\,,\quad {\rm with}\quad F_{ij}(t,\vec k)=\Tr \big\{ e^{t\widetilde H^{(0)}(\vec k) }\widetilde P_-(\vec k)\, \partial_{k_{i}} \widetilde P_-(\vec k)\, e^{-t\widetilde H^{(0)}(\vec k) } \partial_{k_{j}} \widetilde P_-(\vec k)\big\}\;,\label{true}\ee
and $F_{ij}(t,\vec k)$ decays exponentially to zero as $t\to\infty$, uniformly in $\vec k$. Let us first show how this 
identity implies \eqref{eq:AS2}, and let us then come back to its proof. 
In light of \eqref{true}, we can rewrite 
$$\Sigma_{ij}(\vec k)=\int_0^\infty\,dt\, t\, \partial_t^2F_{ij}(t,\vec k)=F_{ij}(0,\vec k)=\Tr \big\{\widetilde P_-(\vec k)\, \partial_{k_{i}} \widetilde P_-(\vec k)\,  \partial_{k_{j}} \widetilde P_-(\vec k)\big\}\;.$$
Plugging this back into \eqref{Sigmaij}, we find that $\bar\s_{ij}(0)$ is equal to the same expression 
as \eqref{eq:AS2}, with $P_-(\vec k)$ replaced by $\widetilde P_-(\vec k)$.
In order to see that we can drop the tilde, note that  
\be
\partial_{k_{i}} \widetilde P_-(\vec k) = U(\vec k) \partial_{k_{i}} P_{-}(\vec k) U(\vec k)^{\dagger} + 
U(\vec k) [ A_{i}, P_{-}(\vec k) ] U(\vec k)^{\dagger}
\ee
where $A_{i} = U(\vec k)^\dagger\partial_{k_{i}}U(\vec k)=i\, 
 \text{diag}\big((\vec r_1)_i\;,\; \cdots \;,\; (\vec r_{|I|})_i\big)$.
 Using this formula, we find that 
\be\label{eq:getridoftilde} \Tr\, \widetilde P_{-}(\vec k) [ \partial_{k_{i}} \widetilde P_{-}(\vec k),  \partial_{k_{j}} \widetilde P_{-}(\vec k)  ] = \Tr\, P_{-}(\vec k) [ \partial_{k_{i}} P_{-}(\vec k),  \partial_{k_{j}} P_{-}(\vec k)  ] + \text{total derivative}\;,\ee
so that the integral over the Brillouin zone of the left side is the same as the integral 
of $ \Tr\, P_{-}(\vec k) [ \partial_{k_{i}} P_{-}(\vec k),  \partial_{k_{j}} P_{-}(\vec k)  ]$, and we thus get  (\ref{eq:AS2}).

We are left with proving \eqref{true} and that $F_{ij}(t,\vec k)$ decays to zero as $t\to\infty$. For this purpose, we rewrite (dropping for notational simplicity the arguments of $\widetilde H^{(0)}$
and $\widetilde P_\pm$)): 
\be \partial_{k_i}\widetilde H^{(0)}=\sum_{\a=\pm}\Big(
\partial_{k_i}\widetilde P_\a\,  \widetilde H^{(0)} 
\widetilde P_\a+\widetilde P_\a \partial_{k_i}\widetilde H^{(0)} 
\widetilde P_\a+\widetilde P_\a \widetilde H^{(0)} 
\partial_{k_i}\widetilde P_\a\Big)\;.\ee
Plugging this identity into \eqref{fij} we find
\bea  f_{ij}(t,\vec k)&=&
\Tr \big\{ e^{t\widetilde H^{(0)} }\widetilde P_-\, \big(\partial_{k_{i}}\widetilde P_+\,  \widetilde H^{(0)}+\widetilde H^{(0)}\partial_{k_{i}}\widetilde P_-\big)
 e^{-t\widetilde H^{(0)} } \widetilde P_+\big(\partial_{k_{j}}\widetilde P_-\,  \widetilde H^{(0)}+\widetilde H^{(0)}\partial_{k_{j}}\widetilde P_+
\big)
\big\}\nn\\
&=&
-\Tr \big\{ \partial_t e^{t\widetilde H^{(0)} }\widetilde P_-\, \partial_{k_{i}}\widetilde P_+\, \partial_t e^{-t\widetilde H^{(0)} } \widetilde P_+\partial_{k_{j}}\widetilde P_-\big\}
 +
 \Tr \big\{ e^{t\widetilde H^{(0)} }\widetilde P_-\, \partial_{k_{i}}\widetilde P_+\,  \partial_t^2
 e^{-t\widetilde H^{(0)} } \widetilde P_+\partial_{k_{j}}\widetilde P_+
\big\}\nn\\
&&+
\Tr \big\{ \partial_t^2e^{t\widetilde H^{(0)} }\widetilde P_-\, \partial_{k_{i}}\widetilde P_-
 e^{-t\widetilde H^{(0)} } \widetilde P_+\partial_{k_{j}}\widetilde P_-\big\}
 -
 \Tr \big\{ \partial_te^{t\widetilde H^{(0)} }\widetilde P_-\, \partial_{k_{i}}\widetilde P_-
 \partial_te^{-t\widetilde H^{(0)} } \widetilde P_+\partial_{k_{j}}\widetilde P_+
\big\}\;.\nn
\eea
Using the fact that $\partial_{k_i}\widetilde P_+=-\partial_{k_i}\widetilde P_-$,
this is easily recognized to be equal to 
\be f_{ij}(t,\vec k)=\partial_t^2
 \Tr \big\{ e^{t\widetilde H^{(0)} }\widetilde P_-\, \partial_{k_{i}}\widetilde P_-
e^{-t\widetilde H^{(0)} } \widetilde P_+\partial_{k_{j}}\widetilde P_-
\big\}\ee
which is the same as \eqref{true}, simply because $(\partial_{k_{i}}\widetilde P_-)
\widetilde P_+=\widetilde P_-\partial_{k_{i}}\widetilde P_-$. Note that, writing 
$F_{ij}(t,\vec k)= \Tr \big\{ e^{t\widetilde H^{(0)} }\widetilde P_-\, \partial_{k_{i}}\widetilde P_-
e^{-t\widetilde H^{(0)} } \widetilde P_+\partial_{k_{j}}\widetilde P_-
\big\}$, it is apparent that $F_{ij}(t,\vec k)$ decays exponentially to zero as $t\to\infty$,
thanks to the projectors under the trace sign and to the gap condition.

\section{The Haldane model}\label{sec:haldane}

An interesting model that falls into the general class of two-dimensional systems studied in this paper
is the interacting version of the {\it Haldane model} \cite{H1}, 
which describes fermions hopping on the hexagonal lattice, exposed to a suitable external magnetic field. For simplicity, we neglect the spin degrees of freedom.
Let $\L_{L}$ be the triangular lattice, generated by the basis vectors
\be
\vec \ell_{1} = \frac{1}{2}(3, -\sqrt{3})\,,\qquad  \vec \ell_{2} = \frac{1}{2}(3, \sqrt{3})\;.
\ee
The reciprocal lattice $\L_{L}^{*}$ of $\L_{L}$ is the triangular lattice generated by the vectors
\be
\vec G_{1} = \frac{2\pi}{3}(1, -\sqrt{3})\;,\qquad \vec G_{2} = \frac{2\pi}{3}(1, \sqrt{3})\;.
\ee
The hexagonal lattice where the electrons hop on can be thought of as the union of two translates
of $\L_L$, denoted by $\L_L^{(A)}\equiv \L_L$ and $\L_L^{(B)}=\L_L+(1,0)$. 
The creation and annihilation operators associated with the sites of $\L_L^{(A)}\cup\L_L^{(B)}$ 
are denoted by $\psi^\pm_{\vec x,\s}$, with $\vec x\in \L_L$ and  $\s\in\{A,B\}\equiv I$: the operators $\psi^\pm_{\vec x,A}$ create or annihilate a particle at $\vec x\in \L_L^{(A)}\equiv \L_L$, while 
$\psi^\pm_{\vec x,B}$ create or annihilate a particle at $\vec x+(1,0)\in \L_L^{(B)}$. In the notation of 
section \ref{sec:modelgen}, this corresponds to choosing the displacement vectors as $\vec r_A=\vec 0$ and $\vec r_B=(1,0)$.

\begin{figure}[hbtp]
\centering
\def\svgwidth{180pt}
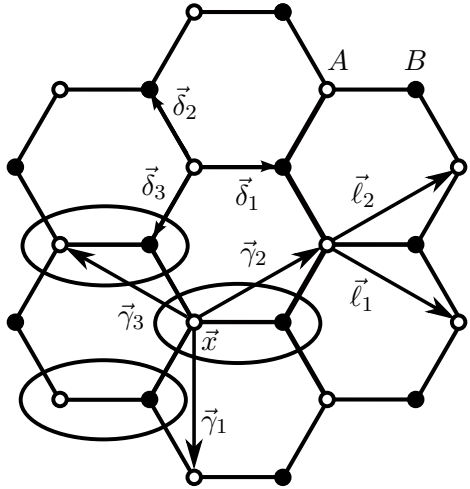
\caption{The honeycomb lattice of the Haldane model. The empty dots belong to $\L^{(A)}_{L}$, while the black dots belong to $\L^{(B)}_{L}$. The ovals encircle the two sites of the fundamental cell,
labeled by the position of the empty dot, i.e., of the site of the $A$ sublattice. The two pairs 
of creation and annihilation operators associated with the two sites of the fundamental cell $\vec x$ 
are denoted by $\psi^\pm_{\vec x,A}$ and $\psi^\pm_{\vec x,B}$.
The nearest neighbor 
vectors $\vec \d_i$, with $\vec \d_1=\vec r_B$, $\vec \d_2=\vec r_B-\vec \ell_1$ and 
$\vec \d_3=\vec r_B-\vec\ell_2$ 
are shown explicitly, together with the next-to-nearest neighbor vectors $\vec\g_i$, and the two basis vectors $\vec \ell_{1,2}$ of $\L_L$.
}\label{fig:haldane}
\end{figure}

The interacting Haldane model is described by the Hamiltonian \eqref{eq:defham}, 
where the non-interacting part is 
\bea\label{eq:H0}
\mathcal{H}^{(0)}_{L} &=& -t_{1}\sum_{\vec x\in \L_{L}}\big[ \psi^{+}_{\vec x,A}\psi^{-}_{\vec x,B} + \psi^{+}_{\vec x,A}\psi^{-}_{\vec x-\vec\ell_{1},B} + \psi^{+}_{\vec x,A}\psi^{-}_{\vec x - \vec\ell_{2},B} + h.c. \big] \nn\\
&&-t_{2}\sum_{\vec x\in \L_{L}}\sum_{\substack{\alpha = \pm \\ j=1,2,3}} \big[ e^{i\alpha\phi}\psi^{+}_{\vec x,A}\psi^{-}_{\vec x + \alpha\vec \g_{j},A} + e^{-i\alpha\phi}\psi^{+}_{\vec x,B}\psi^{-}_{\vec x + \alpha\vec\g_{j},B} \big]\nn\\
&& + W\sum_{\vec x\in \L_{L}}\big[ \psi^{+}_{\vec x,A}\psi^{-}_{\vec x,A} - \psi^{+}_{\vec x,B}\psi^{-}_{\vec x,B} \big],
\eea
where $\vec \g_{1} = \vec\ell_{1} - \vec\ell_{2}$, $\vec \g_{2} = \vec\ell_{2}$,
$\vec\g_{3} = -\vec\ell_{1}$. See Fig.\ref{fig:haldane}.

For definiteness, we assume that $t_{1} > 0$ and $t_{2} > 0$. The 
term proportional to $t_1$ describes nearest neighbor hopping on the hexagonal lattice. The term 
proportional to $t_2$ describes next-to-nearest neighbor hopping,
with the complex phases $e^{\pm i\phi}$ modeling the effect of an external magnetic field, orthogonal 
to the plane of the sample, with {\it zero net flux} through the hexagonal cell. Finally, the term 
proportional to $W$ describes a staggered potential, favoring the occupancy of the $A$ or $B$ 
sublattice, depending on whether $W$ is negative or positive. 

Let us compute explicitly the Bloch Hamiltonian and the Bloch bands associated with $\mathcal H^{(0)}_L$: the Bloch Hamiltonian 
is
\be
\hat H^{(0)}(\vec k) = \begin{pmatrix} -2t_{2}\cos\phi\, \alpha_{1}(\vec k) + m(\vec k) & -t_{1}\Omega^*(\vec k) \\ -t_{1}\Omega(\vec k) & -2t_{2}\cos\phi\alpha_{1}(\vec k) - m(\vec k) \end{pmatrix}
\ee
with
\begin{equation}\label{eq:Haldane2}
\begin{split}
&\alpha_{1}(\vec k) = \sum_{j=1,2,3}\cos(\vec k\cdot \vec \gamma_{i})\;,\qquad \alpha_{2}(\vec k) = \sum_{j=1,2,3} \sin(\vec k\cdot \vec \gamma_{i})\;,\\
&m(\vec k) = W - 2t_{2}\sin\phi\, \alpha_{2}(\vec k)\;,\qquad \Omega(\vec k) = 1 + e^{-i\vec k\cdot \vec\ell_{1}} + e^{-i\vec k\cdot \vec\ell_{2}}\;.
\end{split}
\end{equation}
The corresponding energy bands are 
\begin{equation}\label{eq:bandshaldane}
\e_{\pm}(\vec k) = -2t_{2}\cos\phi\, \alpha_{1}(\vec k) \pm \sqrt{m(\vec k)^{2} + t_{1}^{2}|\Omega(\vec k)|^{2}}\;.
\end{equation}
To make sure that the energy bands do not overlap, we assume that $t_2/t_1<1/3$.
The two bands can only touch  at the Fermi points $\vec k_{F}^{\pm} = \big( \frac{2\pi}{3}, \pm \frac{2\pi}{3\sqrt{3}} \big)$, which are the two zeros of $\Omega(\vec k)$.
The condition that the two bands touch indeed
at $\vec k_F^\o$, with $\o\in\{+,-\}$, is that $m_\o=0$, with 
\be
m_{\o} := m(\vec k_{F}^{\o}) = W +\o 3\sqrt{3}\,t_{2}\sin\phi\;.
\ee
The critical line of the non-interacting Haldane model (i.e., the line in parameter space where the model becomes massless) is then $\{(\phi,W):  W=\pm 3\sqrt3\, t_2\sin \phi\}$. The complement of the critical line consists of four disconnected regions in the $(\phi,W)$ plane, denoted by $\mathcal R_{1}=
\{m_+,m_->0\}$, $\mathcal R_{2}=
\{m_+>0>m_-\}$, $\mathcal R_{3}=
\{m_+<0<m_-\}$, and $\mathcal R_{4}=
\{m_+,m_-<0\}$.

Our theorem applies in the complement of the critical line, non-uniformly in the distance from it. 
It tells us
that, if the Fermi level is chosen inside the gap and $U$ is small enough, then the 
interacting conductivity is {\it equal} to the non-interacting one, which is \cite{H1,KK}
\be\label{eq:checkhaldane}
\bar\sigma_{11}(0) = \bar\sigma_{22}(0) = 0\;,\qquad \bar\sigma_{12}(0) = -\bar\sigma_{21}(0) = \frac{1}{4\p}\big[ \text{sign}(m_{-}) - \text{sign}(m_{+}) \big]\;.
\ee
For completeness, let us derive this formula. The starting point is  (\ref{eq:AS2}),
which immediately implies the vanishing of the longitudinal conductivity. 
In order to compute $\bar\s_{12}(0)$ from (\ref{eq:AS2}) we need an expression for the projector $P_-(\vec k)$, which can be computed from the {\it Bloch function} $u_-$:
\be
u_{-}(\vec k) = \frac{1}{N(\vec k)} \begin{pmatrix}\sqrt{ m(\vec k)^{2} + t_{1}^{2}|\O(\vec k)|^{2} }-  m(\vec k)  \\ t_{1}\O(\vec k)  \end{pmatrix}\;,\ee
where $N(\vec k)= \Big[2\sqrt{m(\vec k)^{2} + t_{1}|\O(\vec k)|^{2}}\, \Big(   \sqrt{m(\vec k)^{2} + t_{1}^{2}|\O(\vec k)|^{2}}-m(\vec k)  \Big)\Big]^{1/2}$.
The Bloch functions are defined only up to a phase. For instance, another possible choice 
for the Bloch function of the negative band is 
\be\label{B.10}
v_{-}(\vec k) = \frac{1}{N'(\vec k)} \begin{pmatrix} t_{1}\O^*(\vec k) \\ \sqrt{m(\vec k)^{2} + t_{1}^{2}|\O(\vec k)|^{2}}+m(\vec k)  \end{pmatrix} \ee
with $N'(\vec k)=\Big[2\sqrt{m(\vec k)^{2} + t_{1}|\O(\vec k)|^{2}}\, \Big(   \sqrt{m(\vec k)^{2} + t_{1}^{2}|\O(\vec k)|^{2}}+m(\vec k)  \Big)\Big]^{1/2}$. The two functions are related by a phase, namely,
$v_-(\vec k)=\frac{\O^*(\vec k)}{|\O(\vec k)|}u_-(\vec k)\equiv e^{i\h(\vec k)}u_-(\vec k)$. Note that, if $(\phi,W)\in\mathcal R_1$ (resp. 
$(\phi,W)\in\mathcal R_4$), 
then $v_-$ (resp. $u_-$) is real analytic in $\vec k$ over the whole Brillouin zone $\BBB$. If $(\phi,W)\in\mathcal R_2$ (resp. 
$(\phi,W)\in\mathcal R_3$), then neither $u_-$ nor $v_-$ are analytic over the whole $\BBB$: $u_-$ is singular at 
$\vec k_F^+$ (resp. $\vec k_F^-$) and $v_-$ is singular at $\vec k_F^-$ (resp. $\vec k_F^+$).
Of course, in any of the regions $\mathcal R_i$, the projector $P_-(\vec k)$ is independent 
of the specific choice of the Bloch function, and is analytic over the whole $\BBB$.

If $(\phi,W)\in\mathcal R_1$, recalling that $v_-$ is analytic over the whole Brillouin zone, 
we write $P_-=|v_-\rangle\langle v_-|$, and we thus find
\be \Tr\, P_{-}(\vec k) [\partial_{k_1} P_{-}(\vec k), \partial_{k_2} P_{-}(\vec k)]=
\langle \partial_{k_1} v_{-}(\vec k), \partial_{k_2} v_{-}(\vec k) \rangle - \langle \partial_{k_2} v_{-}(\vec k), \partial_{k_1} v_{-}(\vec k) \rangle\;.\ee
Integrating over $\BBB$ we get zero, which proves that $\bar\s_{12}(0)=0$, for all $(\phi,W)\in \mathcal R_1$. The same argument, with $v_-$ replaced by $u_-$, shows that $\bar\s_{12}(0)=0$, for all $(\phi,W)\in \mathcal R_4$. 

If $(\phi,W)\in\mathcal R_2$, recalling that $u_-$ is singular at 
$\vec k_F^+$ and $v_-$ is singular at $\vec k_F^-$, we write:
$P_-=|v_-\rangle\langle v_-|$, if $\vec k\in \BBB_+$, and $P_-=|u_-\rangle\langle u_-|$, if $\vec k\in \BBB_-$, where $\BBB_\pm=\{\vec k\in\BBB: \|\vec k-\vec k_F^\pm\|<\|\vec k-\vec k_F^\mp\|\}$,
and $\|\vec q\|=\min_{n_1,n_2}|\vec q+n_1\vec G_1+n_2\vec G_2|$ is the norm on the torus $\BBB$. 
Note that $\BBB=\overline{(\BBB_+\cup\BBB_-)}$. We thus get
\bea \bar\s_{12}(0)&=&i\Big[\int_{\BBB_+}\frac{d\vec k}{(2\p)^2}\Big(
\langle \partial_{k_1} v_{-}(\vec k), \partial_{k_2} v_{-}(\vec k) \rangle - \langle \partial_{k_2} v_{-}(\vec k), \partial_{k_1} v_{-}(\vec k) \rangle\Big)\label{B.12xyz}\\
&&+\int_{\BBB_-}\frac{d\vec k}{(2\p)^2}\Big(
\langle \partial_{k_1} u_{-}(\vec k), \partial_{k_2} u_{-}(\vec k) \rangle - \langle \partial_{k_2} u_{-}(\vec k), \partial_{k_1} u_{-}(\vec k) \rangle\Big)\Big]\;.\nn
\eea
By Stokes' theorem, this can be re-expressed as
\be
\bar\sigma_{12}(0) = \frac{1}{(2\pi)^2}\oint_{\partial \mathcal{B_+}} \big[ \vec{\mathcal{V}}(\vec k) - \vec{\mathcal{U}}(\vec k) \big]\cdot d\vec k\;,
\ee
where the integration path is run counterclockwise. Moreover, $\vec{\mathcal{V}}(\vec k) = \langle v_{-}(\vec k), i\nabla_{\vec k} v_{-}(\vec k) \rangle$ and $\vec{\mathcal{U}}(\vec k) = \langle u_{-}(\vec k), i\nabla_{\vec k} u_{-}(\vec k)\rangle$ are the {\it Berry connections} of $v_{-}$ and $u_{-}$, respectively. 
Recalling that $v_-=e^{i\h}u_-$ (see the lines after \eqref{B.10}), we get 
\be
\bar\sigma_{12}(0) = -\frac{1}{(2\pi)^2} \int_{\partial B_+} \nabla_{\vec k}\eta(\vec k) \cdot d\vec k = -\frac1{2\p}\;,
\ee
which is the same as \eqref{eq:checkhaldane}. The same argument, with $v_-$ replaced by $u_-$, shows that $\bar\s_{12}(0)=1/(2\pi)$, for all $(\phi,W)\in \mathcal R_3$.

\section{Infinite volume dynamics}\label{app:LR}

In this appendix, we prove the existence of the thermodynamic and zero temperature limits of real-time correlations, as stated in section \ref{sec:reconstr}, see in particular the 
discussion after \eqref{eq:LRbound}. The proof is a simple adaptation of \cite{Nachtergaele1, BP}, the only difference being the choice of boundary conditions (periodic, rather than free).
We consider two bounded operators $A, B$ on the fermionic Fock space, even in the fermionic operators, with supports $X$ and $Y$, respectively, independent of $L$. We shall think the torus $\L_L$ as a subset of $\L$ `centered' at the barycenter of $X$ and $Y$, to be denoted $\vec z_0$. In this way, the `boundary' of $\L_L$ is more and more far from $X$ and $Y$ as $L\to\infty$. Periodic boundary conditions are
enforced by properly choosing (in an $L$-dependent way), the local potentials contributing to $\mathcal{H}_{L}-\m \mathcal N_L$, see section \ref{sec2.b}.
For notational convenience, we rename these potentials $\Phi^L_X$, via the following: $\mathcal{H}_{L}-\m\mathcal N = \sum_{X\subset \L_{L}} \Phi^{L}_{X}$.
We also drop the vector symbol from the elements of $\L$. For any fixed $X$ (at a fixed distance from the barycenter $z_0$) and for $L\geq R'$,
\be\label{eq:diffPHI}
\| \Phi^{L}_{X} - \Phi^{R'}_{X} \|\to 0\qquad \text{as $R'\to\infty$.}\;
\ee
The main result of this appendix is that the following limit exists: 
\be\label{eq:claim}
\lim_{\beta\to\infty}\lim_{L\to\infty} \big\langle A_{L}(it) B(0) \big\rangle_{\beta,L}
\ee
for all $t\in \mathbb{R}$ and with $A(it) = e^{i\mathcal{H}_{L}t}Ae^{-i\mathcal{H}_{L}t}$. 

It is convenient to introduce the following norm: 
\be\label{eq:decay}
\| \Phi^{L} \|_{M,\L_L} := \sup_{x,y\in \L_{L}} \sum_{\substack{ X \ni x,y \\ X\subset \L_{L}}} \frac{\| \Phi^{L}_{X} \|}{F_{M}(d_{L}(x,y))}\;,\qquad 
F_{M}(d_{L}(x,y)) := \frac{1}{(1 + d_{L}(x,y))^{M}}\;,
\ee
with $d_L$ the distance on the torus $\L_L$: $d_{L}(x,y) := \inf_{n\in \L} |x - y + nL|$.
Notice that, by the assumptions on $\mathcal H_L$, the potentials satisfy the bound $\|\Phi^{L} \|_{M,\L_L}\le C_{M}$, for all $M\ge 0$ and suitable $C_M>0$, independent of $L$. 

Let $L\ge R'\ge R$. In order to prove \eqref{eq:claim}, we rewrite
\bea
 \big\langle A_{L}(it) B(0) \big\rangle_{\beta,L} &\equiv&  \big\langle A_{L,L}(it) B(0) \big\rangle_{\beta,L} \nn\\
 &=& \big\langle A_{R',R}(it) B(0) \big\rangle_{\beta,L} + \big\langle \big(A_{L,L}(it) - A_{L,R}(it)\big) B(0) \big\rangle_{\beta,L}\nn\\
 && + \big\langle \big(A_{L,R}(it) - A_{R',R}(it)\big) B(0) \big\rangle_{\beta,L}\;,\label{eq:c.4}
\eea
where $A_{L,R}(it)$ is the operator evolved with the dynamics generated by 
\be
\mathcal{H}_{L,R} := \sum_{Z\subset \L_{R}} \Phi^{L}_{Z}\;,
\ee
that is, 
\be
A_{L,R}(it) := e^{i\mathcal{H}_{L,R}t} A e^{-i\mathcal{H}_{L,R}t}\;.
\ee
By \cite[Eq.(2.28)]{Nachtergaele1}, 
\be\label{eq:NOS}
\big\| A_{L,L}(it) - A_{L,R}(it) \big\| \leq  \sup_{x\in X}\sum_{y\in \L_{L}\setminus \L_{R}}\frac{C_{X,M}(t)}{(1 + d_{L}(x,y))^{M}}\;,\qquad \forall M\in\mathbb{N},\;
\ee
with $C_{X,M}(t)$ independent of $L$ (exponentially growing with $t$, as $t\to\infty$). 
In particular, 
\be\label{eq:NOSind}
\big\| A_{L,L}(it) - A_{L,R}(it) \big\| \leq \e(R)\,,\quad \text{for some $\e(R)\to 0$ as $R\to\infty$}.
\ee
Moreover, the difference $A_{L,R}(it) - A_{R',R}(it)$ can be bounded as follows:
\bea
\big\| A_{L,R}(it) - A_{R',R}(it)  \big\| &=& \big\| A - e^{-i\mathcal{H}_{L,R}t}e^{i\mathcal{H}_{R',R}t} A e^{-i\mathcal{H}_{R',R}t} e^{i\mathcal{H}_{L,R}t} \big\| \nn\\
&\leq& \int_{0}^{t} ds\, \big\| \big[\mathcal{H}_{L,R} - \mathcal{H}_{R',R}\,, e^{i\mathcal{H}_{R',R}s} A e^{-i\mathcal{H}_{R',R}s} \big] \big\|\nn\\
&\leq & \int_{0}^{t}ds\, \sum_{Z\subset \L_{R}} \big\| \big[ \Phi^{L}_{Z} - \Phi^{R'}_{Z}, A_{R',R}(is) \big] \big\|\nn\\
&\leq& \int_{0}^{t}ds\, C(s)\big\|A\big\|\, \sum_{Z\subset \L_{R}} \big\| \Phi^{L}_{Z} - \Phi^{R'}_{Z} \big\| \sum_{w\in Z}\sum_{x\in X} F_{M}(d_{R'}(w,x))\;.
\nn\eea
where in the last step we used the Lieb-Robinson bound (see \cite[Theorem 2.1]{Nachtergaele1}, or \cite[Theorem 3.1]{BP}). We now use 
$\sum_{Z\subset \L_R}\sum_{w\in Z} \cdots \le \sum_{z,w\in \L_{R}}\sum_{\L_R\supset Z\ni z,w}\cdots$, so that
\bea\label{eq:LL'2}
&&\big\| A_{L,R}(it) - A_{R',R}(it)  \big\|\le 
 \nn\\
&&\quad \leq  C'(t) \sum_{z\in \L_{R}}\sum_{w\in \L_{R}} F_{M}(d_{R'}(z,w)) \Big[ \sup_{z,w\in \L_{R}}\frac{\sum_{\L_R\supset Z\ni z,w}\| \Phi^{L}_{Z} - \Phi^{R'}_{Z} \| }{F_{M}(d_{R'}(z,w))} \Big] \sum_{x\in X} F_{M}(d_{R'}(w,x)).\nn\eea
Now, using (\ref{eq:diffPHI}) and the fact that the norm in (\ref{eq:decay}) is bounded uniformly in $L$, the sup in square brackets is smaller than a suitable $\e(R,R')$, with $\e(R,R')\to 0$ as $R'\to\infty$.
Moreover, $\sum_{w\in \L_{R}} F_{M}(d_{R'}(z,w))F_{M}(d_{R'}(w,x))\le (\const.)F_M(d_{R'}(z,x))$, so that 
\be \big\| A_{L,R}(it) - A_{R',R}(it)  \big\| \leq C''(t) \sum_{z\in \L_{R}}\sum_{x\in X} F_{M}(d_{R'}(z,x))\, \e(R,R') \to 0 \qquad \text{as $R'\to \infty$.}
\ee
We now plug (\ref{eq:NOSind}), (\ref{eq:LL'2}) into \eqref{eq:c.4}, thus getting 
\be\label{eq:bound0}
\Big|\big\langle A_{L}(it) B(0) \big\rangle_{\beta,L} - \big\langle A_{R',R}(it) B(0) \big\rangle_{\beta,L} \Big| \leq \widetilde \e(R,R'),
\ee
with $\widetilde\e(R,R') \to 0$ in the limit $R'\to\infty$, then $R\to \infty$.
 Also, it is easy to see that the limit $\lim_{\beta,L\to\infty} \big\langle A_{R',R}(it) B(0) \big\rangle_{\beta,L}$ exists, for every fixed $R',R$. In fact, using the boundedness of the fermionic operators and the fact that $\| \mathcal{H}_{R',R} \|\leq CR^{2}$,
\be
A_{R',R}(it) = \sum_{n\geq 0} \frac{t^{n}}{n!}\text{ad}^{n}_{\mathcal{H}_{R',R}}(A)\;,\qquad \| \text{ad}^{n}_{\mathcal{H}_{R',R}}(A) \|\leq \|A\|(2C)^{n}R^{2n}\;, 
\ee
where $\text{ad}^{n}_{\mathcal{H}_{R',R}}(A)$ is the $n$-fold commutator of $A$ with $\mathcal{H}_{R',R}$, and $C$ is a constant independent of $R',R$. Therefore, the existence of the limit $\lim_{\beta,L\to\infty}\big\langle A_{R',R}(it) B(0) \big\rangle_{\beta,L}=:\big\langle A_{R',R}(it) B(0) \big\rangle$ follows from the existence of the (time-independent) limit $\lim_{\beta,L\to\infty} \big\langle \text{ad}^{n}_{\mathcal{H}_{R',R}}(A) B(0) \big\rangle_{\beta,L} $ for all $n$, which can be proved along the lines of the proof in section \ref{app:analyt}.

We now let $L\to\infty$ in \eqref{eq:bound0}, so that 
\bea
-\widetilde \e(R,R') &\leq&  \liminf_{\beta,L\to\infty} \big\langle A_{L}(it) B(0) \big\rangle_{\beta,L} - \big\langle A_{R',R}(it) B(0) \big\rangle
\le\\
&\le&   \limsup_{\beta,L\to\infty} \big\langle A_{L}(it) B(0) \big\rangle_{\beta,L} - \big\langle A_{R',R}(it) B(0) \big\rangle \le \widetilde \e(R,R') \,,
\eea
that is:
\be
  \limsup_{\beta,L\to\infty} \big\langle A_{L}(it) B(0) \big\rangle_{\beta,L} -\widetilde \e(R,R')\le  \big\langle A_{R',R}(it) B(0) \big\rangle
\le \liminf_{\beta,L\to\infty} \big\langle A_{L}(it) B(0) \big\rangle_{\beta,L}+ \widetilde \e(R,R').\nn
\ee
Therefore, letting $R',R\to\infty$, we find that the liminf and limsup coincide, as desired. 
\qed

\end{document}